\documentclass[aps,twocolumn,prc,superscriptaddress,showpacs,nofootinbib,floatfix,amssymb,amsfonts,amsmath]{revtex4-1}

\usepackage{graphicx}
\usepackage{dcolumn}
\usepackage{bm}

\usepackage{amsmath}    
\usepackage{amsfonts}   
\usepackage{amssymb}
\usepackage{graphicx}   

\begin{document}
\title{Global performance of covariant energy density functionals:
ground state observables of even-even nuclei and the estimate
of theoretical uncertainties}

\author{S.\ E.\ Agbemava}
\affiliation{Department of Physics and Astronomy, Mississippi
State University, MS 39762}

\author{A.\ V.\ Afanasjev}
\affiliation{Department of Physics and Astronomy, Mississippi
State University, MS 39762}

\author{D.\ Ray}
\affiliation{Department of Physics and Astronomy, Mississippi
State University, MS 39762}

\author{P. Ring}
\affiliation{Fakult\"at f\"ur Physik, Technische Universit\"at M\"unchen,
 D-85748 Garching, Germany}

\date{\today}

\begin{abstract}
Covariant density functional theory (CDFT) is a modern theoretical
tool for the description of nuclear structure phenomena. The
current investigation aims at the global assessment of the
accuracy of the description of the ground state properties
of even-even nuclei. We also estimate {\it theoretical
uncertainties} defined here as the spreads of predictions within four
covariant energy density functionals (CEDF) in known regions of
the nuclear chart and their propagation towards the neutron
drip line. Large-scale axial relativistic Hartree-Bogoliubov (RHB)
calculations are performed for all $Z\leq 104$ even-even nuclei
between the two-proton and two-neutron drip lines with four modern
covariant energy density functionals such as NL3*, DD-ME2, DD-ME$\delta$
and DD-PC1. The physical observables of interest include the
binding energies, two-particle separation energies, charge quadrupole
deformations, isovector deformations, charge radii, neutron skin
thicknesses and the positions of the two-proton and two-neutron
drip lines. The predictions for the two-neutron drip line are also
compared  in a systematic way with the ones obtained in
non-relativistic models. As an example, the data set of the
calculated properties of even-even nuclei obtained with DD-PC1 CEDF
is provided as Supplemental Material with this article at
\cite{Sup}.
\end{abstract}


\maketitle

\section{Introduction}
\label{Introduc}

Density functional theories (DFT) are extremely useful for
the microscopic description of quantum mechanical many-body systems.
They map the complicated $N$-body systems on an effective systems
of $N$ uncorrelated single particles. They have been applied
with great success for many years in Coulombic systems~\cite{KS.65,KS.65a},
where they are, in principle, exact and where the functional can
be derived without any phenomenological adjustments directly from
the Coulomb interaction. In nuclear physics the situation is much
more complicated:

Nuclei are self-bound systems with translational invariance.
Because of the large spin-orbit interaction spin degrees
of freedom play an important role and cannot be neglected. There
are also isospin degrees of freedom and many open shell nuclei
are superfluid systems. In addition, there are strong
indications~\cite{Ring2012_PST150-014035}
that an optimal description of nuclei should be relativistic.
As a consequence, the single-particle wave functions form at
each point in $r$-space a spinor of dimension 4, or 8
(with superfluidity) or 16 (in the relativistic case).

The bare nuclear force is usually adjusted to scattering data. This
requires additional assumptions and is connected with additional uncertainties
such as the off-shell behavior. As compared to the Coulomb force, the
two-body part of the nuclear force is extremely strong at short distances
and has a relatively short range. There are convincing indications that it
contains, on the non-relativistic level, an important three-body part.

Despite all these restrictions, in the last forty years non-relativistic
and relativistic (covariant) density functional theories have been
developed and successfully applied to the description of a variety
of nuclear phenomena \cite{BHP.03,LNP.641,VALR.05,Niksic2011_PPNP66-519}
with great
success. All of these applications are based on phenomenological
parametrizations of the underlying density functionals. Usually
the form of these functionals is determined by arguments of symmetry
and simplicity and the remaining set of parameters is fitted to
experimental data in finite nuclei, such as binding energies, radii
etc. Only recently there were attempts to reduce the number of
phenomenological parameters by using information from {\it ab-initio}
calculations for non-relativistic
\cite{Fayans1998_JETPL68-169,Baldo2008_PLB663-390,Drut2010_PPNP64-120}
and for relativistic \cite{DD-PC1,DD-MEdelta} functionals. It is clear,
however, that the required accuracy of a few hundred keV for the binding
energies, i.e. in heavy nuclei an accuracy of 10$^{-4}$ and below,
can, in foreseeable future, only be achieved by additional fine tuning
of a few extra phenomenological parameters.

Among these nuclear DFT's, covariant density functional theory
is one of most attractive since covariant energy density
functionals exploit basic properties of QCD at low energies,
such as symmetries and the separation of scales~\cite{LNP.641}. They
provide a consistent treatment of the spin degrees of freedom, they
include the complicated interplay between the large Lorentz scalar
and vector self-energies induced on the QCD level by the in-medium changes
of the scalar and vector quark condensates~\cite{CFG.92}. In addition,
these functionals include {\it nuclear magnetism} \cite{KR.89}, i.e. a
consistent description of currents and time-odd mean fields important
for odd-mass nuclei \cite{AA.10}, the excitations with unsaturated spins,
magnetic moments \cite{HR.88} and nuclear rotations \cite{AR.00,TO-rot}.
Because of Lorentz invariance no new adjustable parameters are required for
the time-odd parts of the mean fields. Of course, at present, all
attempts to derive these functionals directly from the bare
forces~\cite{BT.92,HKL.01,SOA.05,HSR.07} do not reach the required
accuracy. However, in recent years modern phenomenological covariant
density functionals have been derived~\cite{DD-ME2,DD-PC1,DD-MEdelta}
which provide an excellent description of ground
and excited states all over the nuclear chart ~\cite{VALR.05,NVR.11}
with a high predictive power. Modern versions of these forces
derive the density dependence of the vertices from state-of-the-art
ab-initio calculations and use only the remaining few parameters for
a fine tuning of experimental masses in finite spherical
~\cite{DD-MEdelta} or deformed~\cite{DD-PC1} nuclei.

The theoretical description of ground state properties of nuclei
is important for our understanding of their structure. It is
also important for nuclear astrophysics, where we are facing the
problem of an extrapolation to the nuclei with large isospin. Many
of such nuclei will not be studied experimentally even with the next
generation of facilities, or forever. Thus, it is important to to
answer two questions, first, how well the existing nuclear EDF's describe
available experimental data, and second, how well do they extrapolate
to the region of unknown nuclei.

Unfortunately, even the answer on the first question is not
possible for the majority of nuclear EDF's since their
global performance is not known. This is especially true
for covariant energy density functionals. Very few of them
were confronted with experimental data on a global scale.
Even the new generation of CEDF's such as NL3* \cite{NL3*},
DD-ME2~\cite{DD-ME2}, DD-ME$\delta$ \cite{DD-MEdelta} and DD-PC1
\cite{DD-PC1}, which were fitted during last decade, have not
passed this critical test. This is because only limited sets
of nuclei, usually located in the region of nuclei used in
the fitting protocol, were confronted with calculations.
Thus, it is not known how well they describe ground state
properties on a global scale and what are their strong and
weak points in that respect.

The answer on the question ``How well a given CEDF extrapolates
towards neutron-rich nuclei?'' is intimately connected with the
answer to the first question. This is because one can estimate
its reliability for the description of nuclei far away from
the region of known nuclei only by assessing its global
performance on existing experimental data. Of course, a good
performance in known nuclei is only a necessary condition and
one has to be very careful with extrapolations of models where
this good performance has only been achieved with a large number
of phenomenological parameters. It is one of the essential advantages
of relativistic models that covariance reduces the number of
parameters considerably.

It was suggested in Refs.\ \cite{RN.10,Eet.12,Dobaczewski2014_arXiv1402.4657}
to use the methods of information theory and to define 
the uncertainties in the EDF parameters.
These uncertainties come from the selection of the form
of EDF as well as from the fitting protocol details, such as the selection
of the nuclei under investigation, the physical observables, or the
corresponding weights. Some of them are called {\it statistical errors}
and can be calculated from a statistical analysis during the fit,
others are systematic errors, such as for instance the form of the EDF
under investigation. On the basis of these statistical errors and under
certain assumptions on the independence of the form of many EDF's one
hopes to be able to deduce in this way {\it theoretical error bars} for the prediction of physical
observables~\cite{RN.10,Eet.12,Dobaczewski2014_arXiv1402.4657}.
It is very difficult to perform the analysis of statistical errors
on a global scale since the properties of transitional and deformed
nuclei have to be calculated repeatedly for different variations of
original CEDF. Thus, such statistical analysis has been performed
mostly for spherical nuclei \cite{RN.10,KENBGO.13}
or selected isotopic chains of deformed nuclei \cite{Eet.12}.

Although such an analysis has its own merits, at present, it does not
allow to fully estimate theoretical uncertainties in the description of
physical observables. This is because they originate not only
from the uncertainties in model parameters, but also from the
definition and the limitations of the model itself, in particular,
from an insufficient form of the nuclear energy density functional.
The later uncertainties are very difficult to estimate.
As a consequence, any analysis of theoretical uncertainties (especially, for
extrapolations to neutron-rich nuclei) contains a degree of
arbitrariness related to the choice of the model and fitting
protocol.

Thus, in the given situation we concentrate mostly on the
uncertainties related to the present choice of energy density
functionals which can be relatively easily deduced globally.
We therefore define {\it theoretical systematic uncertainties} for a
given physical observable via the spread of theoretical predictions
within the four CDEF's
\begin{equation}
\Delta O(Z,N)=|O_{\rm max}(Z,N)-O_{\rm min}(Z,N)|
\label{eq:TSUC}
\end{equation}
where $O_{\rm max}(Z,N)$ and $O_{\rm min}(Z,N)$ are the largest and smallest
values of the physical observable $O(Z,N)$ obtained with
the four employed CEDF's for the $(Z,N)$ nucleus. In the following
we use the word {\it spread} for these theoretical
systematic uncertainties for the CEDF's.
 Three different classes of the CEDF's are used for this purpose
(see Sec.\ \ref{CEDF}). Note that these {\it theoretical
uncertainties} are only spreads of physical observables due
to a very small number of functionals and, thus, they are only
a crude approximation to the {\it systematic theoretical errors}
discussed in Ref.~\cite{Dobaczewski2014_arXiv1402.4657}. As in
the case of present Skyrme functionals, the different covariant
functionals do not form an independent statistical ensemble.
Their number is very small and they are all based on a very
similar form. For example, no tensor terms are present in the
relativistic case and simple power laws are used  for the density
dependence  in the Skyrme DFT. The parameters of these functionals
are fitted according to similar protocols including similar types
of physical observables such as binding energies and radii.

Thus, there are two main goals of the current manuscript. First
is the assessment of global performance of the state-of-the art
CEDF's. In future
it will allow to define the strategies for new fits of CEDF's.
The second goal is to estimate differences in the description of
various physical observables on a global scale and especially
in the regions of unknown nuclei.

The manuscript is organized as follows. The four state-of-the-art
covariant energy density functionals and the details of their
fitting protocols are discussed in Sec.\ \ref{CEDF}. Sec.\
\ref{RHB-eq} describes the solutions of the relativistic Hartree-Bogoliubov
equations. The treatment of the pairing interaction and the selection of
its strength are considered in Sec.\ \ref{Sec-pairing}. We report on the
results for masses (binding energies) and two-particle separation energies
in Secs.\ \ref{B-energies} and \ref{Sep-energies},
respectively. Sec.\ \ref{proton-drip} contains a discussion of
the two-proton drip line and the accuracy of its description in model
calculations. The predictions for the two-neutron drip line,
an analysis of sources for uncertainties of its definition
and a comparison of two-neutron drip line predictions of covariant
and non-relativistic DFT's are presented in Sect.\ \ref{Two-neu-drip-sec}.
Calculated charge quadrupole and hexadecapole deformations and
isovector quadrupole deformations are considered in Sec.\ \ref{Sec-def}.
Charge radii and neutron skin thicknesses are discussed in Sec.\
\ref{Sec-radii}. Note that theoretical uncertainties of relevant physical
observables are discussed in each of the Secs.\ \ref{B-energies},
\ref{Sep-energies}, \ref{proton-drip},
\ref{Two-neu-drip-sec}, \ref{Sec-def} and \ref{Sec-radii}.
Finally, Sec.\ \ref{Concl} summarizes the results of our
work.

\section{Covariant energy density functionals}
\label{CEDF}

Three classes of covariant density functional models are used
throughout this manuscript: the nonlinear meson-nucleon coupling
model (NL), the density-dependent meson-exchange model (DD-ME) and the
density-dependent point-coupling model (DD-PC). The main differences
between them lay in the treatment of the range of the interaction
and in the density dependence. The interaction in the first two classes
has a finite range that is determined by the mass of the mesons.
For fixed density it is of Yukawa type and the range is given by the
inverse of the meson masses. For large meson masses, i.e. for small
ranges, the meson propagator can be expanded in terms of this range.
In zero'th order we obtain $\delta$-forces and in higher order derivative
terms. This leads to the third class of density functionals,
the point coupling models. It is well known from the non-relativistic
Skyrme functionals that pure $\delta$-forces are not able to describe properly
at the same time nuclear binding energies and radii. One needs at least
one derivative term in the isoscalar-scalar channel because the
$\sigma$-mass is considerably smaller than the masses of the
other mesons.

For realistic calculations the density dependence is very
important. It is taken into account by non-linear meson-couplings
in the NL-models and by an explicit density dependence of
the coupling constants in the other two cases, i.e. by density
dependent meson-nucleon vertices in the DD-ME and DD-PC
models.

Each of these classes is represented in the current
manuscript by the covariant energy density functionals (CEDF)
considered to be state-of-the-art, i.e. by NL3* \cite{NL3*} for the NL-models,
by DD-ME2 \cite{DD-ME2} and DD-ME$\delta$ \cite{DD-MEdelta}
for the DD-ME models, and by
DD-PC1 \cite{DD-PC1} for the point coupling models.

In the meson-exchange models \cite{NL3*,DD-ME2,DD-MEdelta}, the nucleus is
described as a system of Dirac nucleons interacting via the exchange of
mesons with finite masses leading to finite-range interactions. The starting
point of covariant density functional theory (CDFT) for these two models
is a standard Lagrangian density~\cite{GRT.90}
\begin{align}
\label{lagrangian}%
\mathcal{L}  &  =\bar{\psi}\left[%
\gamma\cdot(i\partial-g_{\omega}\omega-g_{\rho
}\vec{\rho}\,\vec{\tau}-eA)-m-g_{\sigma}\sigma-g_{\delta}\vec{\tau}\vec{\delta}
\right]%
\psi\nonumber\\
&+\frac{1}{2}(\partial\sigma)^{2}-\frac{1}{2}m_{\sigma}^{2}\sigma^{2}%
+\frac{1}{2}(\partial\vec{\delta})^{2}-\frac{1}{2}m_{\delta}^{2}\vec{\delta}^{2}
\nonumber\\
&-\frac{1}{4}\Omega_{\mu\nu}\Omega^{\mu\nu}+\frac{1}{2}m_{\omega}^{2}\omega^{2}
-\frac{1}{4}{\vec{R}}_{\mu\nu}{\vec{R}}^{\mu\nu}+\frac{1}{2}m_{\rho}^{2}\vec{\rho}^{\,2}\\
&-\frac{1}{4}F_{\mu\nu}F^{\mu\nu}\nonumber
\end{align}
which contains nucleons described by the Dirac spinors $\psi$ with
the mass $m$ and several effective mesons characterized by the
quantum numbers of spin, parity, and isospin. They create effective
fields in a Dirac equation, which corresponds to the Kohn-Sham
equation~\cite{KS.65} of non-relativistic density functional theory.
The Lagrangian (\ref{lagrangian}) contains as parameters the meson
masses $m_{\sigma}$, $m_{\omega}$, $m_{\delta}$, and $m_{\rho}$ and
the coupling constants $g_{\sigma}$, $g_{\omega}$, $g_{\delta}$,
and $g_{\rho}$. $e$ is the charge of the protons and it vanishes
for neutrons.

This linear model has first been introduced by Walecka~\cite{Wal.74,SW.86}.
It has failed, however, to describe the surface properties of realistic nuclei.
In particular, the resulting incompressibility of infinite nuclear matter
is much too large~\cite{BB.77} and nuclear deformations are too small~\cite{GRT.90}. Therefore,
Boguta and Bodmer~\cite{BB.77} introduced a density dependence via a non-linear
meson coupling replacing the term $\frac{1}{2}m_{\sigma}^{2}\sigma^{2}$
in Eq. (\ref{lagrangian}) by
\begin{equation}
U(\sigma)~=~\frac{1}{2}m_{\sigma}^{2}\sigma^{2}+\frac{1}{3}g_{2}\sigma
^{3}+\frac{1}{4}g_{3}\sigma^{4}.
\end{equation}
The nonlinear meson-coupling models are represented by the parameter
set NL3* \cite{NL3*} (see Table \ref{tab1}), which is a modern version of
the widely used parameter set NL3 \cite{NL3}. Both contain no $\delta$-meson.
Apart from the fixed values for the masses $m$, $m_\omega$ and $m_\rho$, there
are six phenomenological parameters $m_\sigma$, $g_\sigma$, $g_\omega$,
$g_\rho$, $g_2$, and $g_3$ which have been fitted in Ref. \cite{NL3*} to
a  set experimental data in spherical nuclei: 12 binding energies,
9 charge radii, and 4 neutron skin thicknesses.

The density-dependent meson-nucleon coupling model has an explicit
density dependence for the meson-nucleon vertices. There are no non-linear
terms for the $\sigma$ meson, i.e. $g_2 = g_3 =0$.
For the form of the density dependence the Typel-Wolter ansatz \cite{TW.99}
has been used:
\begin{equation}
 g_{i}(\rho) = g_i(\rho_{\rm sat})f_i(x) \quad {\rm for} \quad i=\sigma, \omega, \delta, \rho
\end{equation}
where the density dependence is given by \cite{TW.99,DD-ME2,DD-MEdelta}
\begin{equation}\label{fx}
 f_i(x)=a_i\frac{1+b_i(x+d_i)^2}{1+c_i(x+e_i)^2}.
\end{equation}
\textit{x} is defined as the ratio between the baryonic density $\rho$
at a specific location and the baryonic density at saturation
$\rho_{\rm sat}$ in symmetric nuclear matter. The parameters in Eq.\
(\ref{fx}) are not independent, but constrained as follows:
$f_i({\it x}=1)=1$, $f_{\sigma}^{''}({\it x}=1)=f_{\omega}^{''}({\it x}=1)$,
and $f_{i}^{''}({\it x}=0)=0$. In addition, the following constraints
$d_{\sigma}=e_{\sigma}$ and $d_{\omega}=e_{\omega}$ are used. These
constraints reduce the number of independent parameters for the
density dependence. The density-dependent meson-nucleon coupling
model is represented here by the CEDF's DD-ME2 \cite{DD-ME2} and
DD-ME$\delta$ \cite{DD-MEdelta}. The selection of DD-ME$\delta$ in this
class is motivated by the desire to understand the role of the extra ($\delta$)
meson. Note that in the case of DE-ME2 we have no $\delta$-meson and
the density dependence of Eq.\ (\ref{fx}) is used only for the
$\sigma$ and $\omega$ mesons. For the $\rho$ meson
we have an exponential density dependence
\begin{equation}
 f_\rho(x)=\exp(-a_\rho(x-1)).
\end{equation}
in DD-ME2.

 There is an important difference between the functional NL3* and other
three functionals considered in this investigation. NL3*, as all
older non-linear meson coupling functionals
like NL1~\cite{NL1}, NL3~\cite{NL3}, or TM1~\cite{Sugahara1994_NPA579-557},
have no non-linearities in the isovector channel. Therefore, in infinite
nuclear matter, the isovector fields are proportional to the isovector
density, which are given by $N-Z$. This leads to a very stiff symmetry
energy as a function of the density and to relatively
large values for the symmetry energy $J$ and its slope $L$ at saturation
(see Table \ref{tab-nuclear-matter}). $J$ is particularly large in NL1. The fits
of other above-mentioned non-linear meson coupling
functionals have tried to reduce this value. However, because of
the stiffness of the linear ansatz this is possible only to a certain extent.
Although these functionals are very successful for static CDFT close to the
valley of stability~\cite{NL3*}, their common feature is that the neutron skin
thicknesses are larger than those of successful Skyrme EDF's
and DD CEDF's (see Sec.\ \ref{Sec-radii} for more details). The
majority of experimental estimates of the neutron skin thickness based on hadronic
probes favor lower values for this quantity. However, these experimental values
strongly depend on model assumptions. Only the central value of the neutron skin
thickness obtained in the recent PREX~\cite{PREX.12} experiment is in agreement with
CEDF's linear in the isovector channel. This experiment is, however, characterized
by large statistical errors. On the other hand, the information on the symmetry
energy (for more details concerning the present status of our
knowledge on the symmetry energy in nuclei see Ref.~\cite{symmetryenergy})
from ab-initio calculations and from isovector excitations such as the
Giant Dipole Resonance (GDR) indicate clearly that one needs a density dependence
in the isovector channel~\cite{Niksic2002_PRC66-024306}, as we have it in the
CEDF's DD-ME2 DD-ME$\delta$, or DD-PC1.

\begin{figure*}[ht]
\includegraphics[width=8.7cm,angle=0]{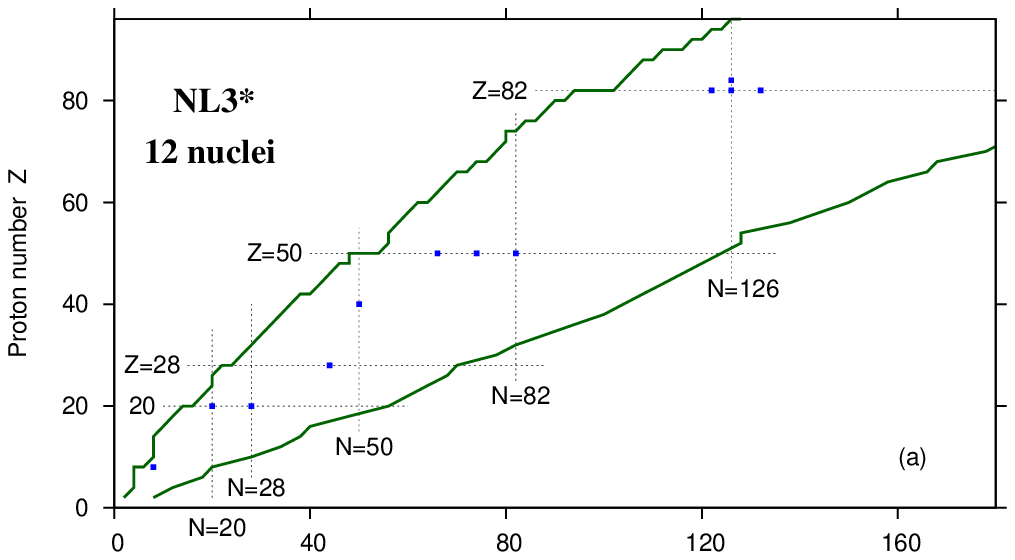}
\includegraphics[width=8.7cm,angle=0]{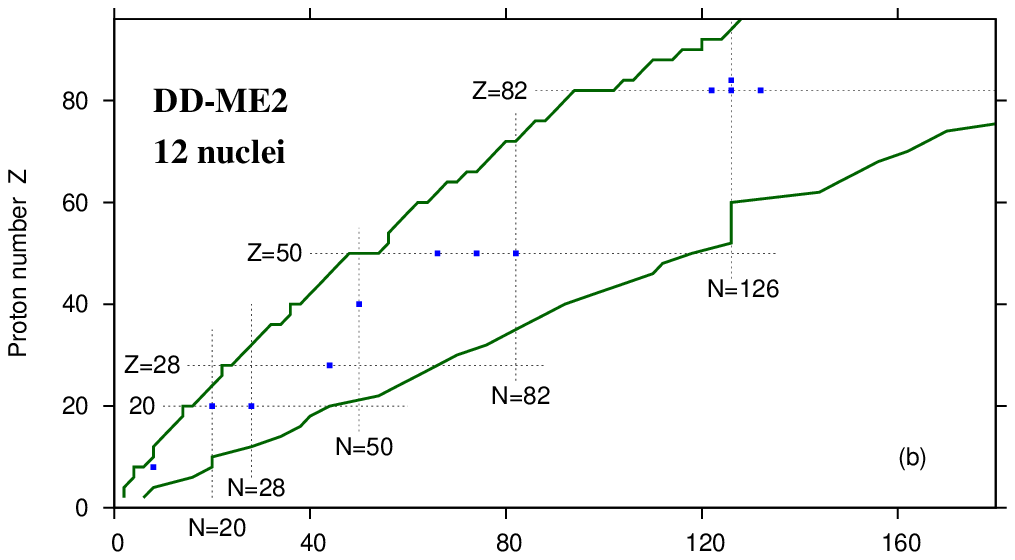}
\includegraphics[width=8.7cm,angle=0]{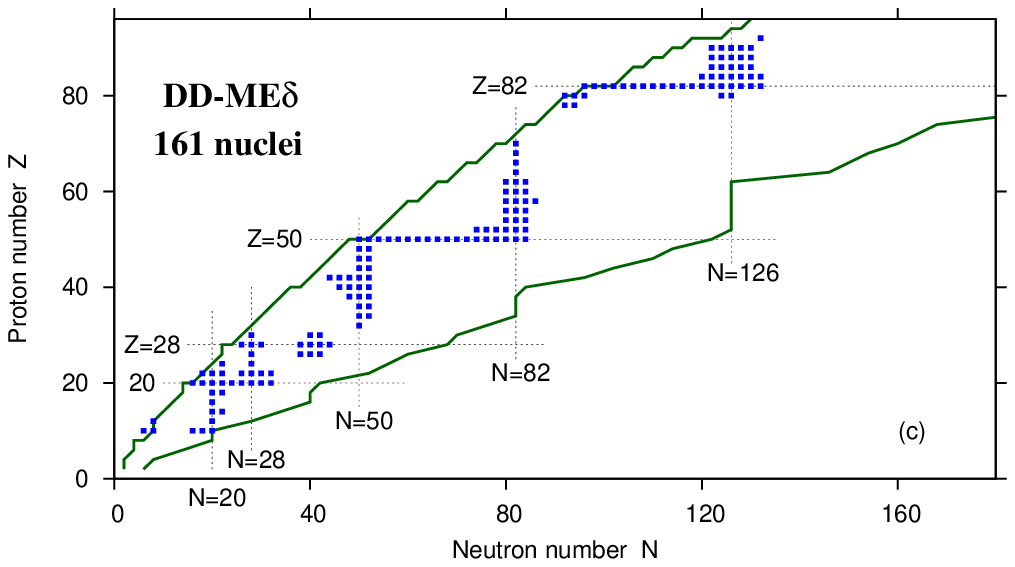}
\includegraphics[width=8.7cm,angle=0]{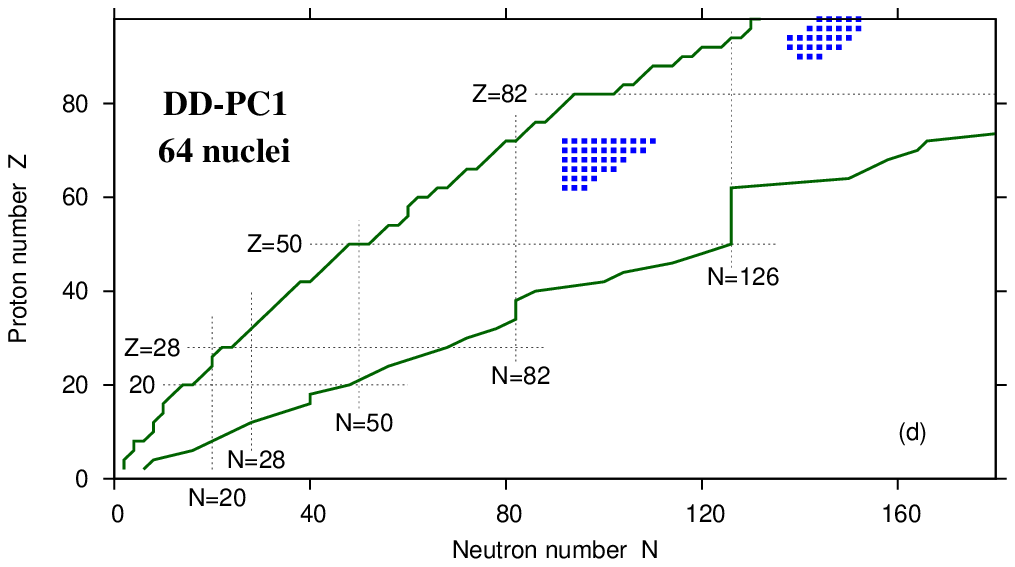}
\caption{(Color online) The nuclei (solid squares), shown
in the $(N,Z)$ plane, which were used in the fit of indicated
CDFT parametrizations. Their total number is shown below the
parametrization label. Magic shell closures are shown by
dashed lines.}
\label{Par-fit}
\end{figure*}

For the functional DD-ME2 \cite{DD-ME2} the masses $m$, $m_\omega$ and
$m_\rho$ are kept at fixed values.
As discussed above the density dependence of the coupling constants $f_i(x)$ $i=\sigma,\omega,\rho$
is given by four independent parameters. Therefore, together with the
four parameters $m_\sigma$, $g_\sigma(\rho_{\rm sat})$, $g_\omega(\rho_{\rm sat})$,
and $g_\rho(\rho_{\rm sat})$ DD-ME2 contains eight independent parameters which have been fitted
in Ref. \cite{DD-ME2} to a set experimental data in spherical nuclei:
12 binding energies, 9 charge radii, and 3 neutron skin thicknesses.

The functional DD-ME$\delta$ \cite{DD-MEdelta} differs from the earlier DD-ME functionals
in the fitting strategy. It tries to use only a minimal number of free
parameters adjusted to the data in finite nuclei and to use ab-initio calculations
to determine the density dependence of the meson-nucleon vertices. Relativistic
ab-initio calculations~\cite{HKL.01,SOA.05} show clearly
that the isovector scalar self-energy, i.e. the field of the $\delta$-meson, is
not negligible.
Therefore, the functional DD-ME$\delta$ differs also
from the other functionals by including the $\delta$-meson, which leads
to a different effective Dirac mass for protons and neutrons:
\begin{equation}
m^*_{n,p} =m+g_\sigma\sigma \pm g_\delta\delta.
\end{equation}
As a consequence, the splittings of the spin-orbit doublets with large
orbital angular momentum {\it l} are slightly different in the models with
and without $\delta$-meson. However, this effect is too small to be seen
in present experiments~\cite{DD-MEdelta}. All the other effects
of the $\delta$-meson on experimental isovector properties of nuclear structure
at densities below and slightly above saturation can be completely absorbed
by a renormalization of the $\rho$-meson-nucleon vertex~\cite{DD-MEdelta}.
Therefore, successful phenomenological CEDF's do not need to include
the $\delta$-meson. However, the effects of the $\delta$-meson are
important for a proper description of the nuclear equation of state (EoS)
at higher densities (see Ref.\ \cite{DD-MEdelta} and references given there)
which play a role in heavy-ion reactions and in astrophysics.

In the earlier parameters sets DD-ME1 \cite{Niksic2002_PRC66-024306} and
DD-ME2 \cite{DD-ME2} all eight independent parameters were adjusted to experimental
data in finite nuclei, whereas for DD-ME$\delta$ only the four independent
parameters $m_\sigma$, $g_\sigma(\rho_{\rm sat})$, $g_\omega(\rho_{\rm sat})$,
and $g_\rho(\rho_{\rm sat})$ have been adjusted  to experimental data in finite nuclei.
This data set includes 161 binding energies  and 86 charge
radii of spherical nuclei.
The parameter $g_\delta(\rho_{\rm sat})$ and the density dependence $f_i(x)$ have been
fitted to parameter-free {\it ab-initio} calculations of infinite nuclear matter of various
densities, as for instance the equations of state (EoS) for symmetric nuclear matter and
pure neutron matter, and the difference of the effective Dirac masses $m^*_p-m^*_n$.
Thus, the functional DD-ME$\delta$ is the most microscopically justified CEDF among
those used in this investigation.

\begin{table}[ptb]
\caption{ The parameters of the NL3*, DD-ME2 and
DD-ME$\delta$ CEDF's. The masses are given in MeV
and the dimension of $g_2$ in NL3* is fm$^{-1}$.
All other parameters are dimensionless. Note that
$g_{\sigma}=g_{\sigma}(\rho_{\rm sat})$,
$g_{\omega}=g_{\omega}(\rho_{\rm sat})$,
$g_{\delta}=g_{\delta}(\rho_{\rm sat})$
and $g_{\rho}=g_{\rho}(\rho_{\rm sat})$
in the case of the DD-ME2 and DD-ME$\delta$
CEDF's}.
\label{tab1}
\begin{center}%
\begin{tabular}
[c]{|c|c|c|c|}\hline
Parameter    &  NL3*      &  DD-ME2   & DD-ME$\delta$ \\\hline
$m$          &  939       &    939    &   939         \\
$m_{\sigma}$ &  502.5742  &  550.1238 &  566.1577     \\
$m_{\omega}$ &  782.600   &  783.000  &  783.00       \\
$m_{\delta}$ &            &           &    983.0      \\
$m_{\rho}$   &  763.000   &  763.000  &    763.0      \\
$g_{\sigma}$ &  10.0944   &  10.5396  &   10.3325     \\
$g_{\omega}$ &  12.8065   &  13.0189  &   12.2904     \\
$g_{\delta}$ &            &           &   7.152       \\
$g_{\rho}$   &  4.5748    &   3.6836  &   6.3128      \\
$g_{2}$      &  -10.8093  &           &               \\
$g_{3}$      &  -30.1486  &           &               \\
$a_{\sigma}$ &            &  1.3881   &  1.3927       \\
$b_{\sigma}$ &            &  1.0943   &  0.1901       \\
$c_{\sigma}$ &            &  1.7057   &  0.3679       \\
$d_{\sigma}$ &            &  0.4421   &  0.9519       \\
$e_{\sigma}$ &            &  0.4421   &  0.9519       \\
$a_{\omega}$ &            &  1.3892   &  1.4089       \\
$b_{\omega}$ &            &  0.9240   &  0.1698       \\
$c_{\omega}$ &            &  1.4620   &  0.3429       \\
$d_{\omega}$ &            &  0.4775   &  0.9860       \\
$e_{\omega}$ &            &  0.4775   &  0.9860       \\
$a_{\delta}$ &            &           &  1.5178       \\
$b_{\delta}$ &            &           &  0.3262       \\
$c_{\delta}$ &            &           &  0.6041       \\
$d_{\delta}$ &            &           &  0.4257       \\
$e_{\delta}$ &            &           &  0.5885       \\
$a_{\rho}$   &            &  0.5647   &  1.8877       \\
$b_{\rho}$   &            &           &  0.0651       \\
$c_{\rho}$   &            &           &  0.3469       \\
$d_{\rho}$   &            &           &  0.9417       \\
$e_{\rho}$   &            &           &  0.9737       \\
\hline
\end{tabular}
\end{center}
\end{table}

The Lagrangian for the density-dependent point coupling model \cite{NHM.92,DD-PC1}
is given by
\begin{align}
\label{Lag-pc}%
\mathcal{L}  &  =\bar{\psi}\left(i\gamma \cdot \partial-m\right)\psi%
-\frac{1}{4}F_{\mu\nu}F^{\mu\nu} - e\bar\psi\gamma \cdot A\psi\nonumber\\
&  -\frac{1}{2}\alpha_S(\rho)\left(\bar{\psi}\psi\right)\left(\bar{\psi}\psi\right)%
-\frac{1}{2}\alpha_V(\rho)\left(\bar{\psi}\gamma^{\mu}\psi\right)\left(\bar{\psi}\gamma_{\mu}\psi\right)\\%
&-\frac{1}{2}\alpha_{TV}(\rho)\left(\bar{\psi}\vec\tau\gamma^{\mu}\psi\right)\left(\bar{\psi}\vec\tau\gamma_{\mu}\psi\right)%
-\frac{1}{2}\delta_S\left(\bar{\psi}\psi\right)\Box \left(\bar{\psi}\psi\right).\nonumber
\end{align}
It contains the free-nucleon part, the coupling of the proton to the electromagnetic field,
and the point coupling interaction terms. The derivative term with the D'Alembert operator $\Box$
accounts for the leading effects of finite-range interaction which are important in nuclei.
In analogy with meson-exchange models, this model contains isoscalar-scalar (S), isoscalar-vector (V)
and isovector-vector (TV) interactions. The coupling constants $\alpha_i(\rho)$ are density dependent.

\begin{table}[ptb]
\caption{The parameters of the DD-PC1 CEDF}.
\label{tab2}
\begin{center}%
\begin{tabular}
[c]{|c|c|}\hline Parameter &  DD-PC1 \\\hline
$m$ &   939\\
$a_{S}$ & -10.04616\\
$b_{S}$ & -9.15042\\
$c_{S}$ & -6.42729\\
$d_{S}$ &  1.37235\\
$a_{V}$ &  5.91946\\
$b_{V}$ &  8.86370\\
$d_{V}$ &  0.65835\\
$b_{TV}$&  1.83595\\
$d_{TV}$&  0.64025\\
\hline
\end{tabular}
\end{center}
\end{table}

In the present work the Lagrangian (\ref{Lag-pc}) is represented by the parametrization DD-PC1 \cite{DD-PC1} given in
Table \ref{tab2}. The following ansatz is used for the functional form of the couplings:
\begin{equation}
\alpha_i (\rho) = a_i + (b_i + c_i x)e^{-d_i x},\quad {\rm for~} i=S,V,TV
\end{equation}
where $x=\rho/\rho_{\rm sat}$ denotes the nucleon density in units of the saturation
density of symmetric nuclear matter. In the isovector
channel a pure exponential dependence is used, i.e. $a_{TV}=0$ and $c_{TV}=0$. The
remaining set of 10 constants, $a_S$, $b_S$, $c_S$, $d_S$, $a_V$, $b_V$, $c_V$, $d_V$,
$b_{TV}$, and $d_{TV}$ that control the strength and density dependence
of the interaction Lagrangian, was adjusted in a multistep
parameter fit exclusively to the experimental masses of 64
axially deformed nuclei.

The fitting protocols used for the derivation of the various
CEDF's differ in the amount and the type of experimental data.
Fig.\ \ref{Par-fit} shows the nuclei which were used in the fits of
the different CEDF's. NL3*, DD-ME2 and DD-ME$\delta$ CEDF were
fitted to spherical nuclei, while DD-PC1 to deformed nuclei
in the rare-earth and actinide regions. Only 12 spherical nuclei
were used in the fitting protocols of NL3* and DD-ME2. On
the contrary, the fits of other CEDF's rely on more extensive
sets of experimental data (161 spherical nuclei in the DD-ME$\delta$
CEDF and 64 deformed nuclei in the DD-PC1 CEDF). In all these fitting 
protocols, the binding energies were used. In addition, the charge 
radii were employed in the fitting of NL3*, DD-ME2 and DE-ME$\delta$. 
In contrast to non-relativistic models, no single-particle information 
has been used in the fits. The number of independent parameters in 
the NL3*, DD-ME2, DD-ME$\delta$ and DD-PC1 CEDF is 6, 8, 14, and 10, 
respectively. Note, however, that in the case of DD-ME$\delta$, only 
the 4 parameters are  fitted to the properties of finite nuclei and 
additional 10 parameters are fitted to pseudo-data obtained from 
{\it ab initio} calculations of nuclear matter.

\section{Solution of the RHB-equations}
\label{RHB-eq}

Pairing correlations play an important role in all open
shell nuclei. On the mean field level they are taken into
account by Bardeen-Cooper-Schrieffer (BCS) or Hartree-Fock-Bogoliubov
(HFB) theory and in the relativistic case by Relativistic Hartree-Bogoliubov
(RHB) theory \cite{Kucharek1991_ZPA339-23,Ring1996_PPNP37-193,CRHB}.
Therefore, density functional theory in nuclei always
has to go beyond the simple density functional theory used
in most of the DFT applications in Coulombic systems, where
the energy depends only on the normal single particle density
$\rho$. Nuclear energy density functionals depend on two
densities, the normal density
\begin{equation}
\label{Eq:rho}
\rho^{}_{n_1 n_2}= \langle\Phi|c^\dag_{n_2} c^{}_{n_1} |\Phi\rangle,
\end{equation}
and the anomalous density
\begin{equation}
\label{Eq:kappa}
\kappa^{}_{n_1 n_2}= \langle\Phi|c^{}_{n_2} c^{}_{n_1}|\Phi\rangle.
\end{equation}
usually called the pairing tensor. $|\Phi\rangle$ is the RHB
wave function, a generalized Slater determinant~\cite{RS.80} and,
therefore, the density $\rho$ as well as $\kappa$ depend on the pairing
correlations. In particular, the density matrix $\rho$ is no longer
a projector on the subspace of occupied states:
\begin{equation}
\rho^2-\rho = \kappa\kappa^*.
\end{equation}
In the relativistic form the nuclear energy functional is usually given by
\begin{equation}
\label{Eq:ERHB}
E_{RHB}[\rho,\kappa]=E_{RMF}[\rho]+E_{pair}[\kappa],
\end{equation}
where $E_{RMF}[\rho]$ has the same functional form as the CEDF's
discussed in the last section, but it is now a functional of the density
$\rho$ in Eq.~(\ref{Eq:rho}) depending on the RHB wave function
$|\Phi\rangle$. The pairing energy\footnote{The details for the
treatment of pairing are presented in Sec.\ \ref{Sec-pairing}}.
is given by
\begin{equation}
\label{Eq:pairing-energy}
E_{pair}[\kappa] = \frac{1}{4}\sum_{n^{}_1n^{}_2,n_1^{\prime}n_2^\prime}
\kappa^\ast_{n^{}_1n^{}_2} \langle n^{}_1 n^{}_2\vert V^{pp}\vert n_1^{\prime}n_2^\prime\rangle\kappa^{}_{n_1^{\prime}n_2^\prime}%
\end{equation}
The Dirac equation for fermion fields $\psi({\bm r})$ is now replaced
by the RHB equation. In the present manuscript, the RHB framework with finite range
pairing and its separable limit are used for a systematic study of ground state properties
of all even-even nuclei from the proton to neutron drip line. It has the proper coupling to
the continuum at the neutron drip line and, therefore, it allows a correct description
of weakly bound nuclei close to the neutron drip line. Even nuclear halo phenomena can be
described by this method, if a proper basis is used, such as the coordinate
space~\cite{Meng1996_PRL77-3963,ZPX.13} or a Woods-Saxon basis~\cite{Li-Lulu2012_PRC85-024312}.

The RHB equations for the fermions are given by \cite{CRHB}
\begin{eqnarray}
\begin{pmatrix}
  \hat{h}_D-\lambda  & \hat{\Delta} \\
 -\hat{\Delta}^*& -\hat{h}_D^{\,*} +\lambda
\end{pmatrix}
\begin{pmatrix}
U({\bm r}) \\ V({\bm r})
\end{pmatrix}_k
= E_k
\begin{pmatrix}
U({\bm r}) \\ V({\bm r})
\end{pmatrix}_k,
\end{eqnarray}
Here, $\hat{h}_D$ is the Dirac Hamiltonian for the nucleons with mass
$m$; $\lambda$ is the chemical potential defined by the constraints on
the average particle number for protons and neutrons;
$U_k ({\bm r})$ and $V_k ({\bm r})$ are quasiparticle Dirac
spinors~\cite{Kucharek1991_ZPA339-23,Ring1996_PPNP37-193,CRHB} and
$E_k$ denotes the quasiparticle energies. The Dirac Hamiltonian
\begin{equation}
\label{Eq:Dirac0}
\hat{h}_D = \boldsymbol{\alpha}(\boldsymbol{p}-\boldsymbol{V}) + V_0 + \beta (m+S).
\end{equation}
contains an attractive scalar potential
\begin{eqnarray}
S(\bm r)=g_\sigma\sigma(\bm r),
\label{Spot}
\end{eqnarray}
a repulsive vector potential
\begin{eqnarray}
V_0(\bm r)~=~g_\omega\omega_0(\bm r)+g_\rho\tau_3\rho_0(\bm r)+e A_0(\bm r),
\label{Vpot}
\end{eqnarray}
and a magnetic potential
\begin{eqnarray}
\bm V(\bm r)~=~g_\omega\bm\omega(\bm r)
+g_\rho\tau_3\bm\rho(\bm r)+e\bm A(\bm r).
\label{Vmag}
\end{eqnarray}
The last term breaks time-reversal symmetry and induces currents.
Time-reversal symmetry is broken when the time-reversed
orbitals are not occupied pairwise. This takes place in odd-mass
nuclei \cite{AA.10}. In the Dirac equation, the space-like components
of the vector mesons $\bm\omega(\bm r)$ and $\bm\rho(\bm r)$ have the
same structure as the space-like component $\bm A(\bm r)$ generated
by the photons. Since $\bm A(\bm r)$ is the vector potential of the
magnetic field, by analogy the effect due to presence of the vector
field $\bm V(\bm r)$ is called {\it nuclear magnetism} \cite{KR.89}.
It affects the properties of odd-mass nuclei \cite{AA.10}. Thus,
the spatial components of the vector mesons are properly taken into
account for such nuclei. This is done only for the study of
odd-even mass staggerings in Sec.\ \ref{Sec-pairing} as it has
been successfully done earlier for the studies of single-particle
\cite{A250,AS.11} and pairing \cite{AO.13} properties of deformed
nuclei. Nuclear magnetism, i.e. currents and time-odd mean fields,
plays no role in the studies of even-even nuclei. The systematic
investigations of such nuclei are performed within the axial RHB
computer code outlined below. As the absolute majority of nuclei are
known to be axially and reflection symmetric in their ground states,
we consider only axial and parity-conserving intrinsic states and
solve the RHB-equations in an axially deformed harmonic oscillator
basis~\cite{GRT.90,RGL.97}.

We have developed a parallel version of the axial RHB computer code starting
from a considerably modified version of the computer code DIZ \cite{RGL.97}.
This code is based on an expansion of the Dirac spinors and the meson fields
in terms of harmonic oscillator wave functions with cylindrical symmetry.
The calculations are performed by successive diagonalizations using the method
of quadratic constraints \cite{RS.80}. The parallel version allows simultaneous
calculations for a significant number of nuclei and deformation points in each
nucleus. For each nucleus, we minimize
\begin{equation}
E_{RHB} + \frac{C_{20}}{2} (\langle\hat{Q}_{20}\rangle-q_{20})^2
\end{equation}
where $E_{RHB}$ in Eq.~(\ref{Eq:ERHB}) is the total energy  and
$\langle\hat{Q}_{20}\rangle$ denotes the expectation value of
the mass quadrupole operator,
\begin{equation}
\hat{Q}_{20}=2z^2-x^2-y^2
\end{equation}
$q_{20}$ is the constrained value of the multipole moment, and
$C_{20}$ the corresponding stiffness constant~\cite{RS.80}.
In order to provide the convergence to the exact value
of the desired multipole moment we use the method suggested in
Ref.~\cite{BFH.05}. Here the quantity $q_{20}$ is replaced by the
parameter $q_{20}^{eff}$, which is automatically modified during
the iteration in such a way that we obtain
$\langle\hat{Q}_{20}\rangle = q_{20}$ for the converged solution.
This method works well in our constrained
calculations.

For each nucleus the potential energy curve is calculated in a
large deformation range from $\beta_2=-0.4$ up to $\beta_2=1.0$
by means of the constraint on the quadrupole moment $q_{20}$.
The lowest in energy minimum is defined from  the potential energy
curve. Then, unconstrained calculations are performed in this
minimum and the correct ground state configuration and its
energy are determined. This procedure is especially important
for the cases of shape coexistence.

The truncation of the basis is performed in such a way that all states
belonging to the major shells up to $N_F = 20$ fermionic shells for
the Dirac spinors and up to $N_B = 20$ bosonic shells for the meson
fields are taken into account. In constrained calculations, the
deformation of the basis is selected in such a way that it corresponds to
the desired deformation of the converged solution. The Coulomb field is
determined by integrating over the Greens function~\cite{RGL.97}.
The comparison with the results obtained with $N_F=26$ and
$N_B=26$ clearly shows that this truncation scheme provides sufficient
numerical accuracy for the description of weakly bound nuclei in the
vicinity of the neutron drip line and of superheavy nuclei. This is even
more true for the  nuclei in the vicinity of $\beta$-stability line and
for the nuclei with masses $A\leq 260$ away from neutron drip line.

It has been found in axial reflection-symmetric calculations for
superheavy nuclei with $Z\geq 106$ that the superdeformed minimum is
frequently lower in energy than the normal deformed one \cite{BBM.04,AAR.12}.
As long as triaxial~\cite{AAR.12} and octupole~\cite{BBM.04,AAR.12}
deformations are not included, this minimum is stabilized by the
presence of an outer fission barrier. Including such deformations,
however, it often turns out that this minimum becomes a saddle point,
unstable against fission \cite{BBM.04,AAR.12}. Since these deformations
are not included in the present calculations, we restrict our consideration
to nuclei with $Z\leq 104$. The investigation of ground state properties
of superheavy $Z\geq 106$ nuclei is inevitable connected with the studies
of fission barriers; such investigations are currently in progress and
their results will be reported in a forthcoming manuscript \cite{AARR.14}.
Of course, in the nuclear chart there exist also a small number of
nuclei with stable octupole or triaxial deformations not considered here
which we have to leave for future investigations.

\section{The effective pairing interaction}
\label{Sec-pairing}

The pair field $\hat{\Delta}$ in RHB theory is given by
\begin{eqnarray}
\hat{\Delta} \equiv \Delta_{ n^{}_1 n^{}_2}~=~\frac{1}{2}\sum_{n_1^{\prime}n_2^\prime}
\langle n^{}_1 n^{}_2\vert V^{pp}\vert n_1^{\prime}n_2^\prime\rangle\kappa^{}_{n_1^{\prime}n_2^\prime}%
\label{gap}
\end{eqnarray}
It contains the pairing tensor  $\kappa$ of Eq.(\ref{Eq:kappa})
\begin{equation}
\kappa = V^{*}U^{T}
\label{kappa}
\end{equation}
and the effective interaction $V^{pp}$ in the particle-particle channel.

\begin{figure*}[h]
\centering
\includegraphics[width=16.0cm]{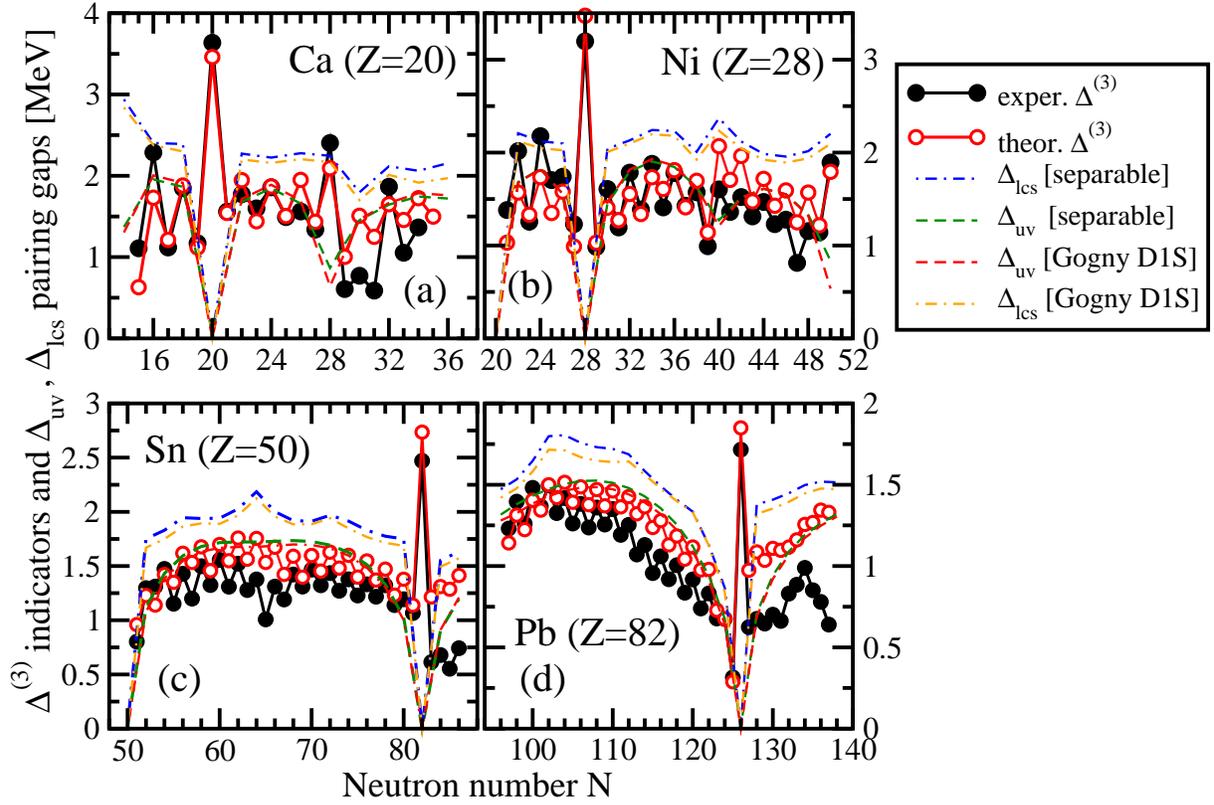}
\caption{(Color online) Experimental and calculated neutron
three-point indicators $\Delta^{(3)}(N)$ and calculated
pairing gaps $\Delta_{\rm uv}$ and $\Delta_{\rm lcs}$ as a function of
the neutron number $N$. Theoretical $\Delta^{(3)}(N)$
indicators shown by open red circles are derived from calculated
binding energies of odd- and even-even nuclei; they are obtained
in RHB calculations with the CEDF  NL3* and the Gogny force D1S of Eq. (\ref{Eq:Vpp})
in the pairing channel. The calculated pairing gaps $\Delta_{\rm uv}$ and
$\Delta_{\rm lcs}$ are shown by lines. They are calculated in even-even nuclei
with the Gogny force D1S (labeled as 'Gogny D1S') and its separable
approximation in Eq. (\ref{Eq:TMR}) (labeled as 'separable' in the figure).}
\label{Delta3N-NL3s}
\end{figure*}

In the literature on nuclear density functional theory several types of effective pairing forces
$V^{pp}$ have been used. The most simple force is the seniority force of Kerman~\cite{Kerman1961_APNY12-300}
with constant pairing matrix elements $G$. For problems with time-reversal symmetry the
corresponding pairing matrix $\Delta$ in Eq. (\ref{gap}) is proportional to unity
for this force and RHB theory is equivalent to RMF + BCS.
This force is widely used, but is has many limitations, e.g. correlations in
pairs with higher angular momentum are neglected, the scattering between pairs
with different shells is not constant in realistic forces, the coupling to the
continuum is not properly taken into account and the predictive power is limited.
Nonetheless this method is used in the constant gap approximation
in most of the large scale adjustments of CEDF's, in particular, also for
DD-ME2~\cite{DD-ME2} and DD-PC1~\cite{DD-PC1}. For each nucleus in the fit,
the gap parameter is determined directly from odd-even mass differences of neighboring
nuclei. In this case the occupation numbers $v_k^2$ in the neighborhood of the Fermi
surface, which depend crucially on the gap parameter,  have rather reasonable
values and in this way all quantities depending only on the $v_k^2$'s are not
influenced further neither by the value of $G$ nor by the pairing window.
Of course, the pairing energy~(\ref{Eq:pairing-energy}) depends on the constant $G$
and on the pairing window. The actual value of $G$ producing this experimental gap
parameter is determined after the self-consistent solution of the BCS equations
and depends on the nucleus under consideration and on the pairing window.
However, for a reasonable pairing window the total change in binding energy caused
by pairing, which is the difference between the gain in binding due to the
pairing energy~(\ref{Eq:pairing-energy}) and the loss in binding due to the
reoccupation of the single particle levels, is rather small.
Therefore, there is a clear separation of scales between the total binding energy, which is
of the order of 1000 MeV and more for heavy nuclei, and the additional binding
of a few MeV caused by pairing. By this reason the conventional procedure to adjust
the parameters of the Lagrangian in the constant gap approximation by RMF+BCS calculations
and to use for all further RHB calculations a more realistic pairing force is very successful.
In this way all the problems of the monopole pairing force are avoided.

In the present investigation two types of realistic effective pairing interaction
have been used. Both of them have finite range and, therefore, provide  an automatic
cutoff of high-momentum components. These are
\begin{itemize}
\item
the Brink-Booker part of phenomenological non-relativistic D1S
Gogny-type finite range interaction
\begin{eqnarray}
V^{pp}(1,2) &  = & f \sum_{i=1,2} e^{-[({\bm r}_1-{\bm r} _2)/\mu_i]^2}
\nonumber \\
&\times& (W_i+B_i P^{\sigma}- H_i P^{\tau} - M_i P^{\sigma} P^{\tau}).
\label{Eq:Vpp}
\end{eqnarray}
The motivation for such an approach to the description of pairing
is given in Refs.\ \cite{GonzalesLlarena1996_PLB379-13,CRHB}.
In Eq.\ (\ref{Eq:Vpp}), $\mu_i$, $W_i$, $B_i$,
$H_i$ and $M_i$ $(i=1,2)$ are the parameters of the force and
$P^{\sigma}$ and $P^{\tau}$ are the exchange operators for the spin
and isospin  variables. The D1S parametrization of the
Gogny force \cite{D1S,D1S-a} is used here. Note  that a scaling  factor
$f$ is introduced in Eq.\ (\ref{Eq:Vpp}). Its role is discussed below.

\item
a separable pairing interaction of finite range introduced by
Tian et al \ \cite{TMR.09}. Its matrix elements in $r$-space have the form
\begin{eqnarray}
\label{Eq:TMR}
V({\bm r}_1,{\bm r}_2,{\bm r}_1',{\bm r}_2') &=& \nonumber \\
= - f\,G \delta({\bm R}-&\bm{R'}&)P(r) P(r') \frac{1}{2}(1-P^{\sigma})
\end{eqnarray}
with ${\bm R}=({\bm r}_1+{\bm r}_2)/2$ and ${\bm r}={\bm r}_1-{\bm r}_2$
being the center of mass and relative coordinates.
The form factor $P(r)$ is of Gaussian shape
\begin{eqnarray}
P(r)=\frac{1}{(4 \pi a^2)^{3/2}}e^{-r^2/4a^2}
\end{eqnarray}
The parameters of this interaction have been derived
by a mapping of the $^1$S$_0$ pairing gap of infinite nuclear
matter to that of the Gogny force D1S. The resulting
parameters are: $G=738$ fm$^3$ and $a=0.636$ fm\ \cite{TMR.09}.
The scaling factor $f$ is the same as in Eq.\ (\ref{Eq:Vpp}).

\end{itemize}

\begin{figure*}[ht]
\centering
\includegraphics[width=16.0cm]{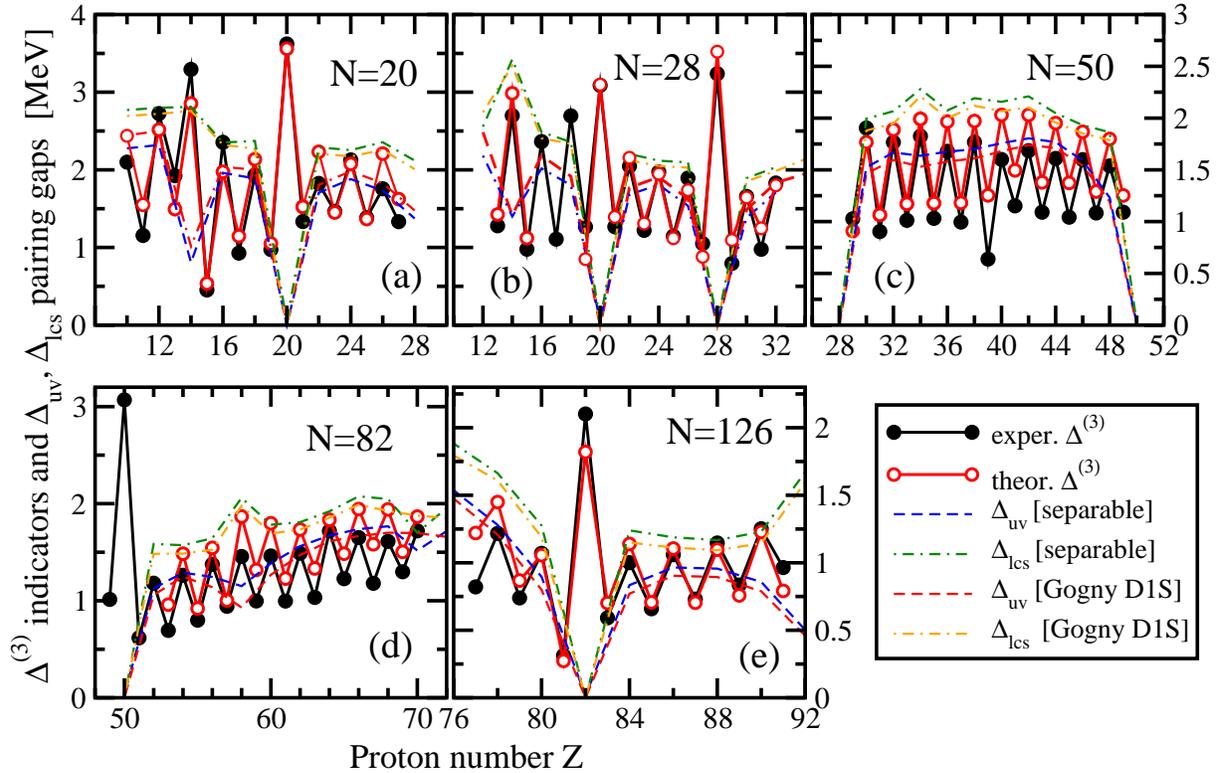}
\caption{(Color online) The same as Fig.\ \ref{Delta3N-NL3s} but
for proton three-point indicators $\Delta^{(3)}(Z)$ and
proton pairing gaps  $\Delta_{\rm uv}$ and $\Delta_{\rm lcs}$ as a
function of proton number $Z$. Note that it was not possible
to get a convergence for a few odd-mass nuclei in the $N=28$ and
$N=82$ isotone chains in the RHB calculations with Gogny D1S force
in pairing channel. This leads to the absence of theoretical
$\Delta^{(3)}$ values in some proton number range.
}
\label{Delta3P-NL3s}
\end{figure*}

Both in theory and in experiment the strength of pairing
correlations is usually accessed via the three-point indicator
\cite{DMNSS.01}
\begin{eqnarray}
\Delta ^{(3)}(N) = \frac{\pi_N}{2} \left[ B(N-1) + B(N+1) - 2 B(N)
\right],
\label{neut-OES}
\end{eqnarray}
which quantifies the odd-even staggering (OES) of  binding energies.
Here $\pi_N=(-1)^N$ is the number parity and $B(N)$ is the (negative)
binding energy of a system with $N$ particles. In Eq.\
(\ref{neut-OES}), the number of protons $Z$ is fixed, and $N$ denotes
the number of neutrons, i.e. this indicator gives the neutron OES. The
factor depending on the number parity $\pi_N$ is chosen so that the
OES centered on even and odd neutron number $N$ will both be positive.
An analogous proton OES indicator $\Delta ^{(3)}(Z)$ is obtained by
fixing the neutron number $N$ and replacing $N$ by $Z$ in Eq.\
(\ref{neut-OES}).

 As discussed in Ref.\ \cite{AO.13}, in many applications of RHB theory
with the pairing force D1S the same scaling factor $f$ has been used
across the nuclear chart. However, it was found a decade ago that
a proper description of rotational properties in actinides
\cite{A250} requires weaker pairing as compared with the rare-earth
region \cite{J1Rare,CRHB}. Subsequent systematic studies of pairing
(via the three-point indicator $\Delta^{(3)}$) and rotational
properties of actinides confirmed this observation in Refs.\
\cite{AO.13,A.14}. The investigation of odd-even mass
staggerings in spherical nuclei in Ref.\ \cite{WSDL.13} also
confirms the need for a scaling factor $f$ which depends on the
region in the nuclear chart. The studies of Refs.\ \cite{A250,AO.13,WSDL.13}
show also a weak dependence of the scaling factor $f$
on the CDFT parametrization. We therefore introduce in
Eqs.\ (\ref{Eq:Vpp}) and (\ref{Eq:TMR}) a scaling factor $f$ for a
fine tuning of the effective pairing force.

The scaling factor $f$ used in the present investigation
has been selected based on the results of a comparison between experimental
moments of inertia and those obtained in cranked RHB calculations with the
CEDF NL3*. As verified in the actinides in Ref.\ \cite{AO.13}, the strengths
of pairing defined by means of the moments of inertia and by
the three-point indicators $\Delta^{(3)}$ strongly correlate in deformed nuclei.
Following the results obtained in Ref.\ \cite{AO.13}, the scaling factor
has been fixed at $f=1.0$ in the $Z\geq 88$ actinides and superheavy nuclei.
The analysis of the moments of inertia in the rare-earth region \cite{RA.14}
leads to a scaling factor of $f=1.075$ for the $56\leq Z \leq 76$ rare-earth nuclei.
For $Z\leq 44$ nuclei,  the scaling factor was fixed at $f=1.12$
\cite{RA.14}. The scaling factor gradually changes with $Z$ in
between of these regions. Since the strength parameter $G$ of the separable
force has been determined in Ref.\ \cite{TMR.09} by a direct mapping to the
Gogny force D1S, the same scaling factors are also used in the
following RHB calculations with separable pairing.

Figs.\ \ref{Delta3N-NL3s} and \ref{Delta3P-NL3s} compare calculated
(open red circles) and experimental (solid black circles) three-point
indicators $\Delta^{(3)}$ for different chains of spherical nuclei.
Both in theory and experiment, these quantities have been obtained
from binding energies. The calculations have been performed within the
RHB formalism of Refs.\ \cite{CRHB,A250} which allows a fully
self-consistent treatment of even-even and odd-mass nuclei. Blocking
and time-odd mean fields have been taken into account in the case of
odd-mass nuclei.  The Gogny force D1S of Eq. (\ref{Eq:Vpp}) with the scaling
factors $f$ has been used in these calculations.
As shown in Ref.\ \cite{AA.10} the impact of the
time-odd mean fields on the $\Delta^{(3)}$ indicators cannot be ignored.
Large peaks appear in the experimental $\Delta^{(3)}$ indicators at shell
closures. This is connected with the fact, that pairing correlations
disappear in these cases and the peaks are not produced by pairing, but by
the increasing shell gap for closed shell configurations. Therefore they
are not relevant for the present discussions.

One can see that on average the RHB calculations reproduce the experimental
data and the magnitude of the observed staggering in $\Delta^{(3)}$ rather
well. However, in some nuclei the calculations somewhat overestimate
experimental $\Delta^{(3)}$ indicators. There are two possible reasons for
that.
First, particle-vibration coupling in odd-mass nuclei is neglected in these
calculations. Extra correlations induced by this coupling increases
the binding energy in odd mass nuclei. According to Eq.\ (\ref{neut-OES}),
this will lead to smaller $\Delta^{(3)}$ values. Thus, the agreement with
experiments could improve if we would take into account the additional
correlations due to particle-vibrational coupling in odd-mass nuclei.
The analysis of Ref.\ \cite{DG.80} suggests that this effect is
non-negligible and that it can reach up to 300 keV. In addition, we have
to keep in mind, that the effects of particle-vibration coupling are
state-dependent \cite{LA.11}. The second reason for the deviations
between theory and experiment in Fig.\ \ref{Delta3N-NL3s} has to do
with the deficiencies in the underlying single-particle structure
produced by the CEDF NL3* \cite{AS.11,LA.11}.

Fig.\ \ref{Delta3N-NL3s} also shows that the accuracy of
the description of the $\Delta^{(3)}$ indicators depends on
the structure of underlying single-particle states. For example,
reasonable agreement between theory and experiment is obtained
in the Ni isotopes between the $N=28$ shell and the $N=40$ subshell
closures where the active neutrons occupy the spherical $2p_{3/2}$,
$1f_{5/2}$ and $2p_{1/2}$ orbits. However, the calculations
systematically overestimate the experiment between the $N=40$
subshell and the $N=50$ shell closure where the active neutron
occupy the $1g_{9/2}$ orbit. A similar situation and a reduced
accuracy in the description of experimental data can be seen
in the chain of Sn (Fig.\ \ref{Delta3N-NL3s}c) and
Pb (Fig.\ \ref{Delta3N-NL3s}d) isotopes when crossing the $N=82$
and $N=126$ shell closures. We do not have a clear explanation for
these features but two factors may contribute: first, the state-dependence
of particle-vibration coupling mentioned above, and second, a
deficiency of the Gogny force D1S to reproduce a possible
state-dependence of pairing correlations.

\begin{figure*}[ht]
\centering
\includegraphics[width=16.0cm]{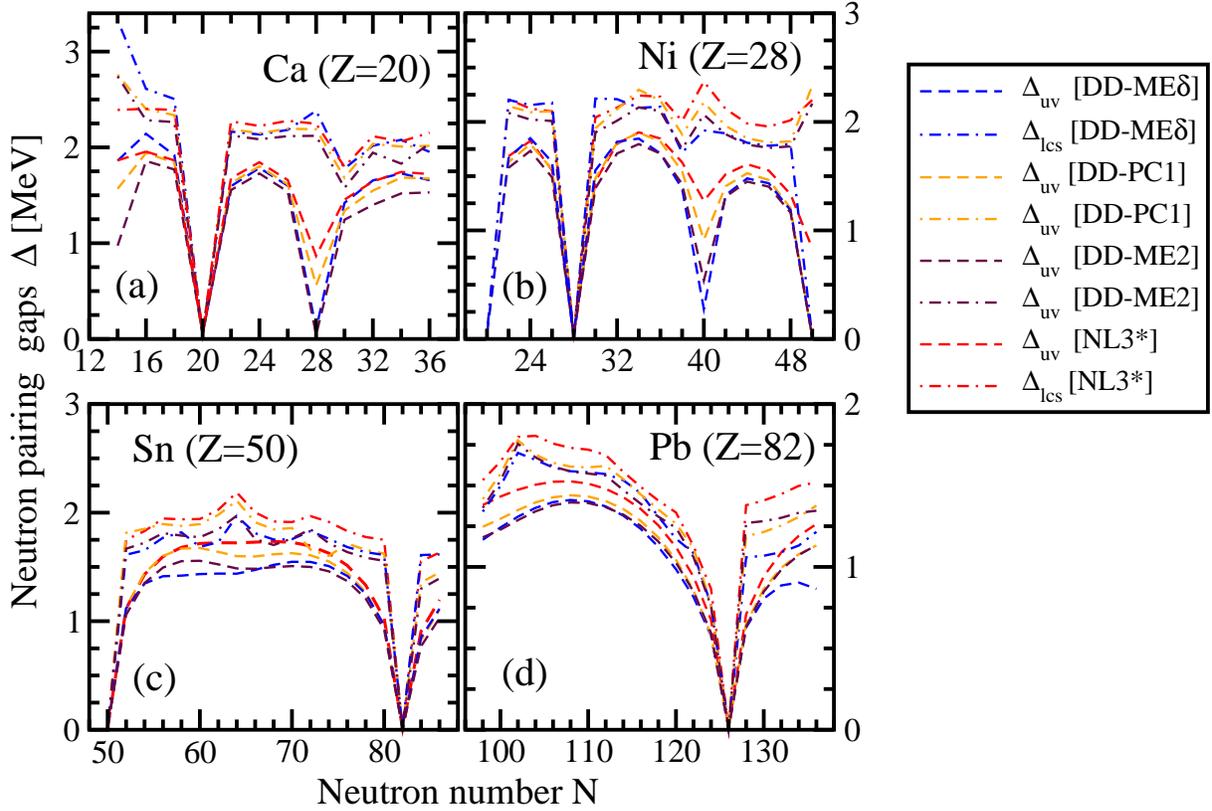}
\caption{(Color online) Calculated neutron pairing gaps
$\Delta_{\rm uv}$ and $\Delta_{\rm lcs}$ as a function of the neutron number
$N$ for different isotonic chains. The results of RHB calculations with
the separable pairing force (\ref{Eq:TMR}) are presented for the
indicated CEDF's.}
\label{Ne-gaps}
\end{figure*}

  There are clearly some differences in the approaches based on
fixing the pairing strength to the data in deformed and in spherical
nuclei. This is also seen in the Skyrme EDF \cite{DBHM.01} where similar
to our case the pairing strength adjusted to rotational structures
leads to too high $\Delta^{(3)}$ indicators in spherical nuclei. As
illustrated in Ref.\ \cite{AO.13}, deformed nuclei offer the
opportunity to fix the strength of pairing to two independent
physical observables, the rotational moments of inertia and the
$\Delta^{(3)}$ indicators. The accuracy of the description of the latter
quantity in deformed nuclei depends on the correctness of the reproduction
of the ground state configuration in the odd-mass nuclei and the impact
of particle-vibration coupling (see Sec.\ III.E of Ref.\ \cite{AO.13}).
However, these factors have less influence on the calculated
moments of inertia. Particle-vibration coupling is expected to be more
pronounced in spherical nuclei as compared with deformed ones (see
discussion in Sec.\ VI.B of Ref.\ \cite{A250}). Thus, we believe that
the experimental data in deformed nuclei allows a better and more
reliable estimate of pairing strength as compared with the one
in spherical nuclei.

However, it is too time-consuming to perform the analysis presented
in Refs.\ \cite{AO.13,RA.14} for the remaining three functionals. Thus,
we looked on alternative indicators for the strength of the effective
pairing force. It is well known that the connection between the $\Delta^{(3)}$
(or $\Delta^{(5)}$) indicators and theoretical pairing gaps is not
straightforward. Thus, several expressions for pairing gaps aimed on
circumventing this problem have been proposed. On the one hand, they have
the advantage of being calculated in even-even nuclei, thus avoiding the
complicated problem of calculating the blocked states in odd-mass nuclei
(see Refs.\ \cite{AS.11,AO.13}). On the other hand, their
validity for the comparison with experimental $\Delta^{(3)}$ indicators
is not clear.

In the literature the following definitions for the  average pairing gap
have been used:
\begin{itemize}

\item
The pairing gap
\begin{equation}
\Delta_{\rm vv}=\frac{\sum_k v^2_k\Delta_k}{\sum_k v^2_k}
\end{equation}
has been introduced in Ref.\ \cite{Dobaczewski1996_PRC53-2809}.
The sum runs over the states $k$ in the {\it canonical} basis
(for details see Ref.~\cite{RS.80}). $v^2_k$ are the corresponding
occupation probabilities and $\Delta_k$ is the diagonal
matrix element of the pairing field in this basis.

\item
The pairing gap
\begin{equation}
\Delta_{\rm uv}=\frac{\sum_k u_kv_k\Delta_k}{\sum_k u_kv_k}
\end{equation}
is related to the average of the state dependent gaps over the pairing tensor.
\item
The pairing gap $\Delta_{\rm lcs}$ (lcs stands for lowest
canonical state) \cite{DBHM.01} is defined by the smallest quasi-particle energy
\begin{eqnarray}
E_k=\sqrt{(\varepsilon_k-\lambda)^2+\Delta^2_k},
\end{eqnarray}
which is approximately equal to the gap $\Delta_k$ of the orbit closest to the
Fermi surface. Here $\varepsilon_k$ is the diagonal matrix element
of the single-particle field $\hat{h}$ in the canonical basis.
\end{itemize}

All these definitions have advantages and disadvantages. $\Delta_{\rm vv}$
averages over the occupation numbers $v^2_k$. For heavy nuclei with many
fully occupied states most of the contributions are therefore determined
by deeply bound states far from the Fermi surface, which have little to do
with the pairing phenomenon and the scattering of Cooper pairs around the
Fermi surface. $\Delta_{\rm lcs}$ considers only the canonical orbit closest
to the Fermi surface and, therefore, it is  more connected to the
pairing phenomenon. However, it has the disadvantage, that it depends on
a specific orbit and that it is not really an average. $\Delta_{\rm uv}$
finally averages over $u_kv_k$, a quantity which is concentrated around
the Fermi surface  However, because of the fact that $\kappa\sim \sum_k u_kv_k$
diverges for the seniority force and for zero range forces, $\Delta_{\rm uv}$
turns out to depend on the pairing window. This is, however, no problem for the
finite range pairing forces used in this investigation.
In addition, in the majority of the cases the $\Delta_{\rm vv}$ values are larger
than the $\Delta_{\rm lcs}$ ones, which as follows from the discussion below
overestimate experimental data. Therefore, in the current manuscript, we will
consider only $\Delta_{\rm uv}$ and $\Delta_{\rm lcs}$.

The calculated quantities are presented in Figs.\
\ref{Delta3N-NL3s} and \ref{Delta3P-NL3s} both for the
Gogny force D1S in Eq. (\ref{Eq:Vpp}) and for its separable
approximation (\ref{Eq:TMR}).
It is interesting to compare them with the five-point indicator
$\Delta^{(5)}$ discussed in Refs.\ \cite{BRRM.00,DBHM.01},
which is a better measure of pairing correlations since
it is less polluted by mean field effects as compared with
the $\Delta^{(3)}$ indicator. This quantity represents
a smooth curve and the $\Delta^{(3)}$
indicator oscillates around it (see Fig.\ 2 of Ref.\ \cite{DBHM.01} for the
graphical example of the relation between the $\Delta^{(3)}$ and
$\Delta^{(5)}$ indicators). It turns out that far from spherical shell
closures, the $\Delta_{\rm uv}$ values come close to the calculated
$\Delta^{(5)}$ indicators. On the other hand, the $\Delta_{\rm lcs}$ values
always overestimate the $\Delta^{(5)}$ indicators. This result is
contrary to the conclusions of Ref.\ \cite{DBHM.01} which concludes
that the $\Delta_{\rm lcs}$ value is a better measure of pairing
correlations. The difference maybe due to the zero-range pairing forces
in Ref.\ \cite{DBHM.01}, while finite-range pairing forces are used in our manuscript.

Figs.\ \ref{Delta3N-NL3s} and \ref{Delta3P-NL3s} also show that the
pairing gaps $\Delta_{\rm lcs}$ and $\Delta_{\rm uv}$ calculated with
the D1S Gogny force D1S and its separable limit are very close to each
other. Thus, all systematic calculations in this manuscript are performed
with the separable form of the Gogny force D1S. This reduces the
computational time considerably.

\begin{figure*}[ht]
\centering
\includegraphics[width=14.0cm]{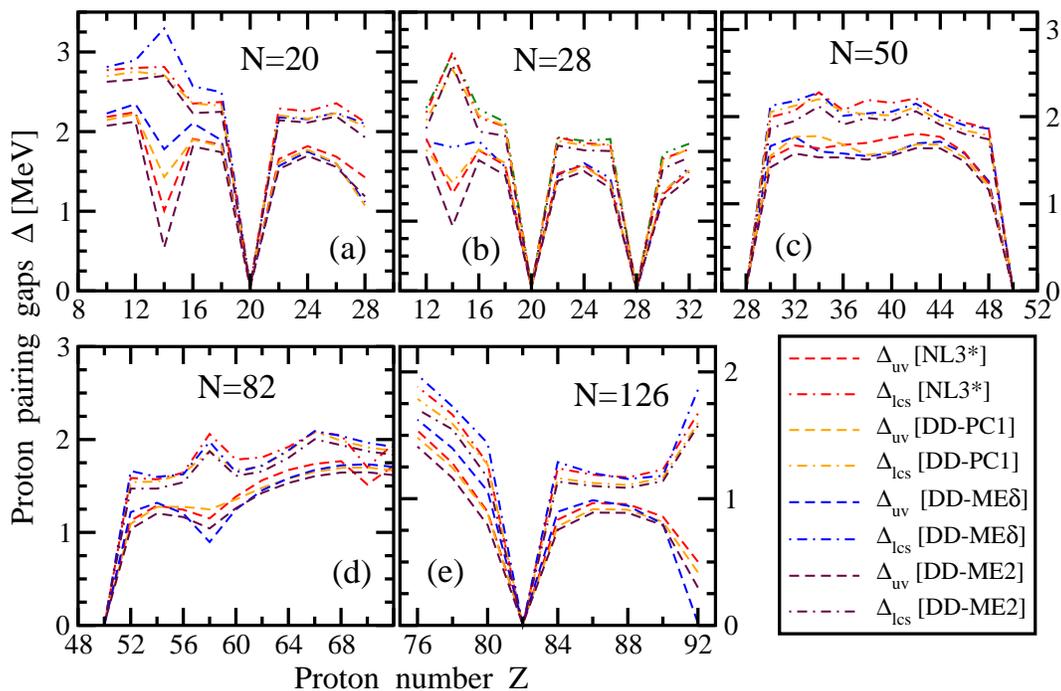}
\caption{(Color online) The same as Fig.\ \ref{Ne-gaps} but
for calculated proton pairing gaps $\Delta_{\rm uv}$ and
$\Delta_{\rm lcs}$ as a function of the proton number $Z$ for different
isotopic chains.}
\label{Pr-gaps}
\end{figure*}

Figs.\ \ref{Ne-gaps} and \ref{Pr-gaps} compare the pairing gaps
$\Delta_{\rm lcs}$ and $\Delta_{\rm uv}$ obtained in the calculations with
different CEDF's. Apart from proton number $Z=14$ in the $N=20$ and
the $N=28$ isotope chains (see Fig.\ \ref{Pr-gaps})
and from the proton subshell closure at $Z=40$ in the Ni isotopes (see
Fig.\ \ref{Ne-gaps}), the calculated gaps are similar for the different
parameterizations. The spread in the calculated values indicates that
scaling factors $f$ used here are reasonable within the limits of a few \%.
For example, the change of scaling factor $f$ by 4\% in $^{182}$Pb leads
to a change of the pairing gaps $\Delta_{\rm lcs}$ and $\Delta_{\rm uv}$
by $\sim 0.14$ MeV. The weak dependence of the scaling factor $f$ on
the CEDF has already been seen in the studies of pairing and rotational
properties in the actinides~\cite{A250,AO.13}. Thus, the same scaling
factor $f$ as defined above for the CEDF NL3* is used in the calculations
with DD-PC1, DD-ME2 and DD-ME$\delta$. Considering the global character
of this study, this is a reasonable choice. Definitely there are
nuclei in which the choice of the scaling factor $f$ is not optimal.
However, the change of scaling factor by 1\% changes the binding energy
only by approximately 100 keV. The impact on physical observables such
as two-particle separation energies, the position of two-proton and two-neutron
drip-lines is even smaller since they are sensitive to the differences of
the binding energies. Changes of scaling factor by a few \% will only
marginally affect the deformations, radii and neutron skin thicknesses.

\section{Binding energies}
\label{B-energies}

In Table  \ref{deviat} we list the rms-deviations $\Delta E_{\rm rms}$
between theoretical and experimental binding energies for the global
RHB calculations
with the different CEDF's investigated in this manuscript.
The masses given in the AME2012 mass evaluation \cite{AME2012} can
be separated into two groups; one represents
nuclei with masses defined only from experimental data, the other
contains nuclei with masses depending in addition on either
interpolation or extrapolation procedures. For simplicity, we
call the masses of the nuclei in the first and second groups as
measured  and estimated. There are 640 measured and 195 estimated
masses of even-even nuclei in the AME2012 mass evaluation.  One can
see in Table \ref{deviat} that the extension to include also estimated masses
leads only to a slight decrease of the accuracy in the description
of  experimental data.

\begin{figure*}[ht]
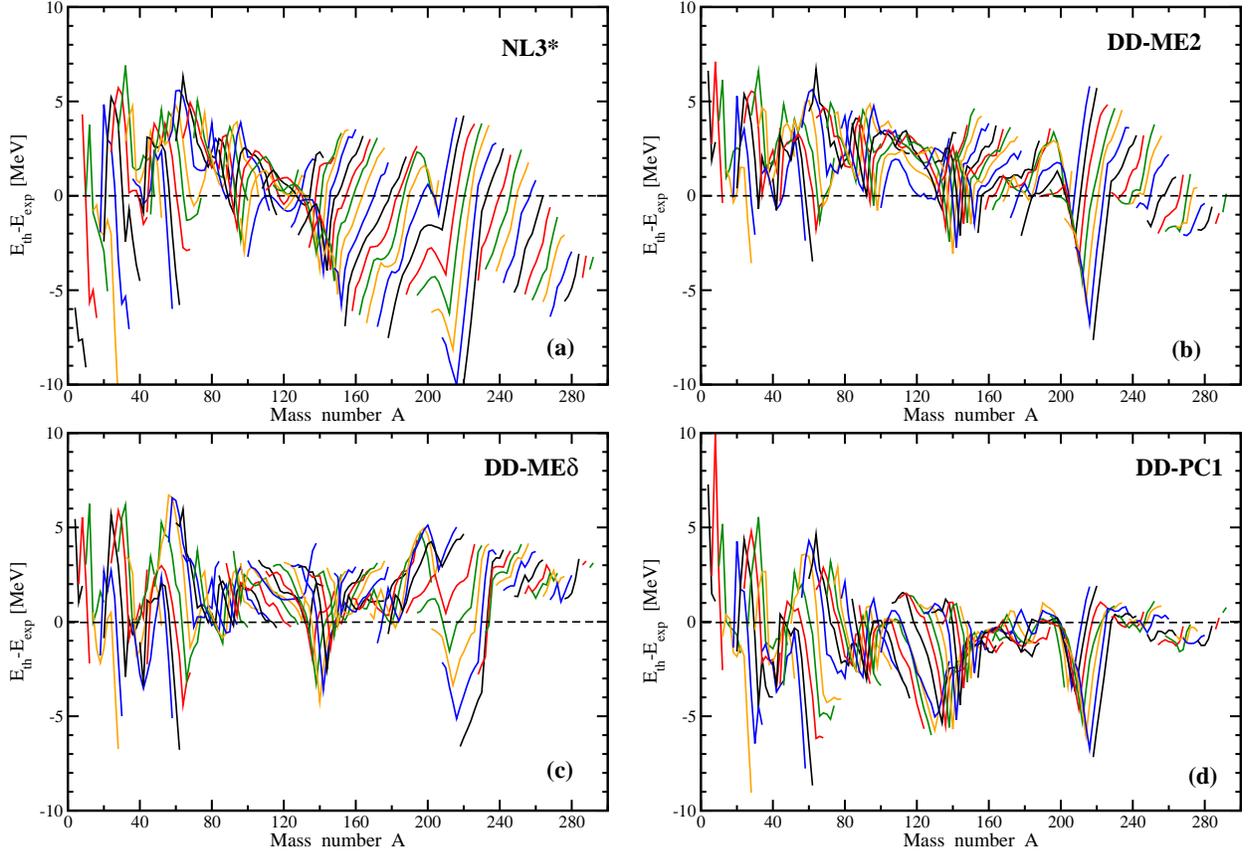

\begin{center}
\includegraphics[width=8.0cm,angle=0]{fig-6a.eps}
\hspace{0.2cm}
\includegraphics[width=8.0cm,angle=0]{fig-6b.eps}
\vspace{0.5cm}
\includegraphics[width=8.0cm,angle=0]{fig-6c.eps}
\hspace{0.2cm}
\includegraphics[width=8.0cm,angle=0]{fig-6d.eps}
\end{center}
\caption{(Color online) The difference between theoretical
and experimental masses of 835 even-even nuclei investigated
in RHB calculations with indicated CEDF's. If $E_{th}-E_{exp}<0$,
the nucleus is more bound in the calculations than in experiment.}
\label{Edif}
\end{figure*}

To our knowledge, for relativistic density functionals,
reliable\footnote{The masses were globally studied earlier in
the RMF \cite{HSet.97} or RMF+BCS \cite{LRR.99,GTM.05} formalisms.
However, the pairing correlations have been completely ignored
in the studies of Ref.\ \cite{HSet.97}. The treatment of pairing
via the BCS approximation in Refs.\ \cite{LRR.99,GTM.05} has to
be taken with care in the region of the drip line since this
approximation does not take into account the continuum properly
and leads to the formation of a neutron gas \cite{DFT.84} in nuclei
near neutron drip line. In addition, these calculations use at most
14 fermionic shells for the harmonic oscillator basis, which
according to our study and the one of Ref.\ \cite{RA.11} is not
sufficient for a correct description of binding energies of actinides
and superheavy nuclei and the nuclei in the vicinity of neutron drip line.}
global comparisons of experimental and theoretical masses have been
performed  so far only for the parametrizations NL3~\cite{NL3},
FSUGold~\cite{FSUGold}, BSR4~\cite{Agrawal2010_PRC81-034323} and
TM1~\cite{Sugahara1994_NPA579-557} in the RMF+BCS approach using
the constant gap approximation in Ref.\ \cite{RA.11} and for
PC-PK1 \cite{PC-PK1} in the RMF+BCS approach with density-dependent
pairing in Ref.\ \cite{ZNLYM.14}. Apart of BSR4 and PC-PK1 these
CEDF's were fitted more than ten years ago. The rms-errors for the
masses found for these CEDF's are 3.8 MeV for NL3, 6.5 MeV for FSUGold,
2.6 MeV for BSR4, 5.9 MeV for TM1 and 2.6 MeV for PC-PK1 (at the
mean field level).

One can see that the CEDF's NL3*, DD-ME2, DD-ME$\delta$, and DD-PC1
investigated in the present manuscript provide an improved
description of masses across the nuclear chart. The rms-deviations
for the binding energies presented in Table \ref{deviat} are more
statistically significant than those of Refs.\ \cite{RA.11}
and \cite{ZNLYM.14} since they are defined for 835 even-even nuclei.
On the contrary, rms-deviations for binding energies for the NL3,
FSUGold, BSR4 and TM1 CEDF's are defined only for 513 (575 for PC-PK1)
even-even nuclei in Refs.\ \cite{RA.11} and \cite{ZNLYM.14}. The
extension of the experimental database to 835 nuclei may lead to
further deterioration of the rms-deviations for these CEDF's.

\begin{figure*}[ht]
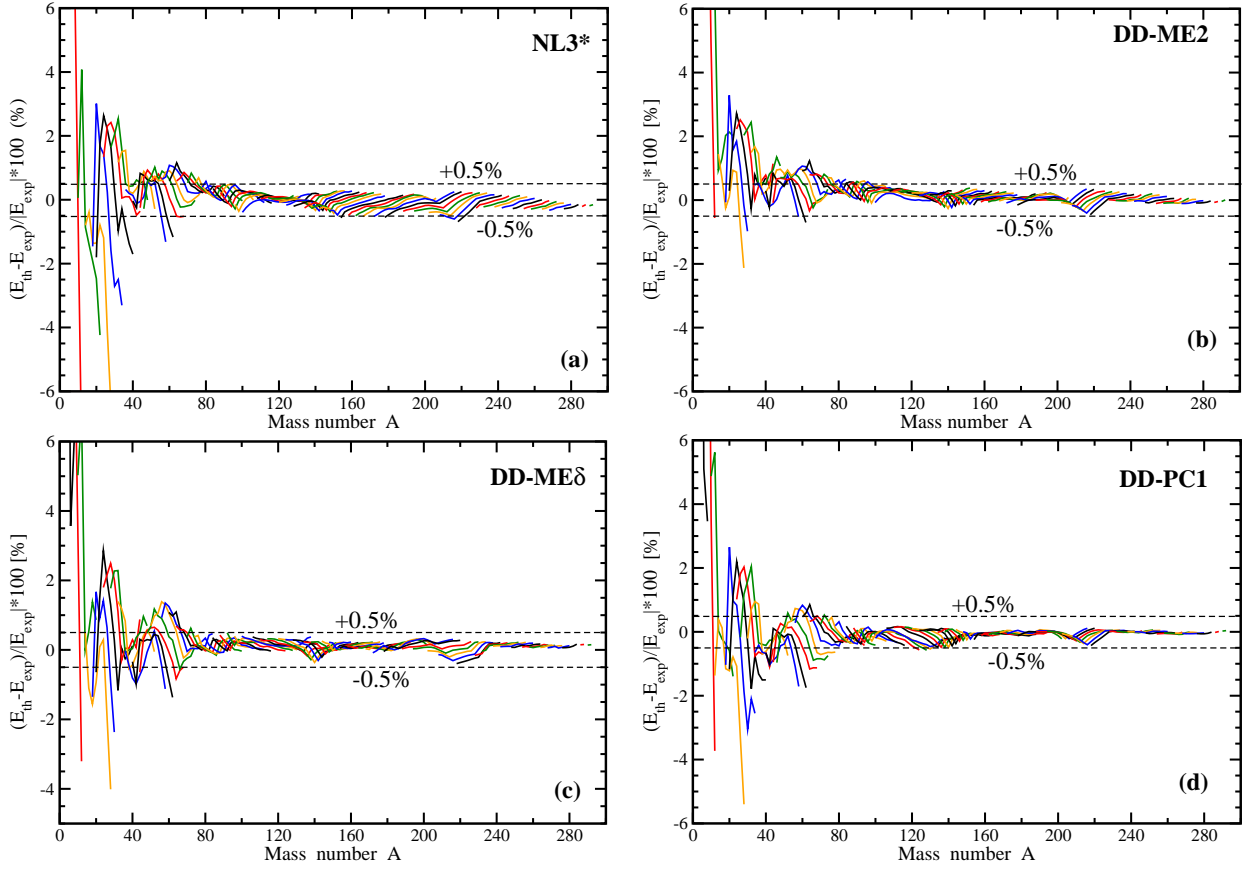

\begin{center}
\includegraphics[width=8.0cm,angle=0]{fig-7a.eps}
\hspace{0.2cm}
\includegraphics[width=8.0cm,angle=0]{fig-7b.eps}
\vspace{0.5cm}
\includegraphics[width=8.0cm,angle=0]{fig-7c.eps}
\hspace{0.2cm}
\includegraphics[width=8.0cm,angle=0]{fig-7d.eps}
\end{center}
\caption{(Color online) The relative accuracy of the description
of experimental masses in our model calculations. The same set of data
as in Fig.\  \ref{Edif} is used. Dashed lines show the $\pm 0.5\%$
error band.}
\label{Edif-percent}
\end{figure*}

In Fig.\ \ref{Edif-percent}, the errors in binding energies
are summarized for all experimentally known even-even nuclei.
This figure is prepared in the same style as Fig.\ 3 of Ref.\
\cite{RA.11}. This allows to compare the gross trends for the
binding energy errors of the current and previous generations
of the CEDF's. In particular, old CEDF's
show in all cases a growing deviation from the
zero line with increasing mass number (Fig.\ 3 in Ref.\ \cite{RA.11}).
These deviatiations are especially pronounced for
FSUGold and TM1, for which they reach 15 MeV for the highest measured
masses. The deviations are smaller for the NL3 CEDF for
which they reach 10 MeV for the highest measured masses, and quite
moderate for the BSR4 parametrization. On the contrary, no such problems
exist in the current generation of the CEDF's.
The accuracy of the description of the masses of heavy
nuclei is comparable with or even better (as in the case
of DD-PC1) than that of medium-mass and light nuclei
(Fig.\ \ref{Edif}). The large deviation peaks seen in Fig.\
\ref{Edif} are located in the vicinity of the doubly magic
shell closures. For such nuclei, medium polarization effects
associated with surface and pairing vibrations have a substantial
effect on the binding energies \cite{BBBBCV.06}.

Previous estimates of the rms-deviations for binding energies
with these CEDF's have been obtained only with restricted sets
of experimental data. For example, the RHB(NL3*) results were compared
with experiment only for approximately 180 even-even nuclei in Ref.\
\cite{NL3*}. However, no rms-deviations for binding energies were
presented for this set. An rms-deviation of 2.4 MeV has been obtained
in the analysis of 161 nuclei in the RMF+BCS calculations with
DD-ME$\delta$ using monopole pairing \cite{DD-MEdelta}. Note,
however, that the binding energies of these nuclei were used in the
fit of DD-ME$\delta$. 93 deformed nuclei calculated in the RMF+BCS
approach with DD-PC1 CEDF were compared with experiment in Ref.\
\cite{DD-PC1}. The binding energies of the most of these nuclei
deviate from experiment by less than 1 MeV, which is not surprising
considering that 64 of these nuclei were used in the fit of the corresponding CEDF.
However, much larger deviations have been reported for this CEDF
in spherical nuclei \cite{DD-PC1}. Note that, so far, DD-PC1 is the
only CEDF exclusively fitted to deformed nuclei. Theoretical binding
energies of approximately 200 nuclei calculated in the RHB framework
with DD-ME2 CEDF and the Gogny D1S interaction in the pairing channel
show rms-deviation of less than 0.90 MeV from experiment
\cite{DD-ME2}.

\begin{figure*}[ht]
\begin{center}
\includegraphics[width=18.0cm,angle=0]{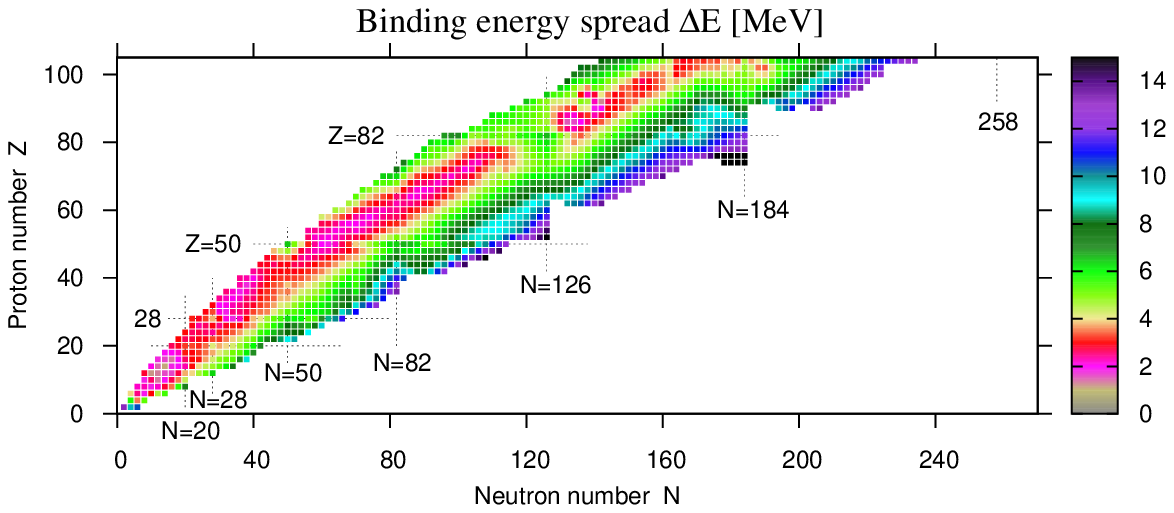}
\end{center}
\caption{(Color online) The binding energy spreads $\Delta E(Z,N)$
as a function of proton and neutron number.
$\Delta E(Z,N)=|E_{\rm max}(Z,N)-E_{\rm min}(Z,N)|$, where
$E_{\rm max}(Z,N)$ and $E_{\rm min}(Z,N)$ are the largest and
the smallest binding energies for each ($N,Z$)-nucleus
obtained with the four CEDF's used in this
investigation.}
\label{Ebin-error}
\end{figure*}

\begin{figure*}[ht]
\begin{center}
\includegraphics[width=18.0cm,angle=0]{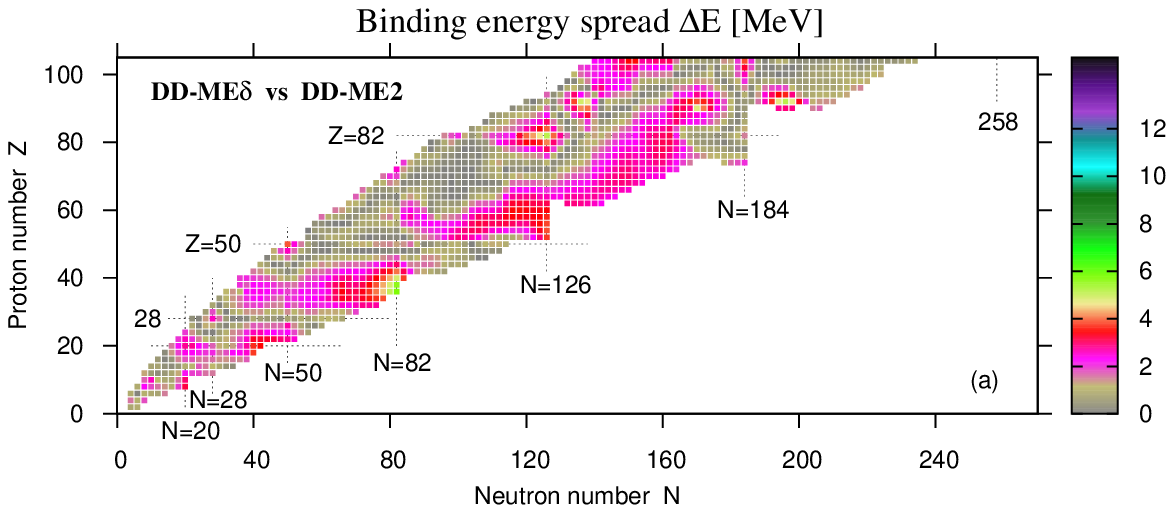}
\end{center}
\caption{(Color online) The same as Fig.\ \ref{Ebin-error},
but only for DD-ME2 and DD-ME$\delta$.}
\label{Ebin-error-med-vs-me2}
\end{figure*}

Comparing these rms-deviations with the ones presented in Table \ref{deviat}
one can see that the increase of the size of
experimental data set leads to a deterioration of the average
description of the binding energies. This clearly suggests that
the experimental data sets used in the fits of the CEDF's
(see Sect.\ \ref{CEDF} for details) are not sufficiently large to provide
an optimal localization of the model parameters in the parameter
space and reliable extrapolation properties of the CEDF's with respect
to binding energies. To our knowledge, so far, no attempt to create a ``mass
table'' quality CEDF based on a fit to the full set of available
experimental masses has been undertaken in CDFT.
This is contrary to non-relativistic models where mass tables based on
an extensive use of experimental data were generated in the
macroscopic+microscopic model~\cite{MNMS.95}, the Skyrme \cite{GCP.09}
and the Gogny \cite{GHGP.09} DFT. We have to keep in mind,
however, that the number of free parameters in such fits
to thousands of experimental masses is considerable larger
than that used in the CEDF's investigated in this manusrcipt.
In particular, many of these fitS include more or less
phenomenological terms for the Wigner energy
\cite{Wigner1937_PR51-947,Myers1997_NPA612-249} in close to
$N\approx Z$ nuclei and for the rotational corrections in deformed
nuclei.

One should also recognize the limitations of the description of
masses at the mean field level. This is clearly visible in
Fig.\ \ref{Edif-percent} where the relative errors are plotted as
a function of mass number $A$. One can see that these errors
are especially pronounced in light $A\leq 80$ nuclei for which the
configuration mixing effects (which go beyond mean field) are
important \cite{BBH.06,HNMM.09,FMXLYM.13}. In very light nuclei
the clusterization effects can also be important \cite{OFE.06}
and for the nuclei in the $N=Z$ region the Wigner term
~\cite{Wigner1937_PR51-947,Myers1997_NPA612-249}. Such effects
are not taken into account in these density functionals. For the
heavier $A\geq 80$ nuclei, the relative error in the description of
masses stays safely within $\pm 0.5\%$ error band. In this context,
it is interesting to mention that a similar level of error ($\sim 0.3\%$)
in the description of binding energies is achieved in the DFT local
density approximation in condensed matter physics  \cite{KK.08}.

In Fig.\ \ref{Ebin-error} we show the map of theoretical uncertainties 
$\Delta E(Z,N)$ defined in Eq.~(\ref{eq:TSUC})
for the description of binding energies. The comparison
of this figure with Fig.\ 1 in Ref.\ \cite{AARR.13} (which presents
experimentally known nuclei in the nuclear chart), shows that the 
spreads in the predictions of binding energies stay within 5-6 MeV for
the known nuclei. These spreads are even smaller (typically around 3 MeV)
for the nuclei in the valley of beta-stability. However, the theoretical
systematic uncertainties (\ref{eq:TSUC}) for the masses increase 
drastically when approaching
the neutron-drip line and in some nuclei they reach 15 MeV. This is a
consequence of poorly defined isovector properties of many CEDF's.
Comparing different pairs of CEDF's one can conclude that the smallest
difference in the predictions of binding energies exists for the
DD-ME2/DD-ME$\delta$ pair of CEDF's (Fig.\ \ref{Ebin-error-med-vs-me2}).
The next smallest difference in terms of $\Delta E(Z,N)$ exist for the
DD-PC1/NL3* pair of CEDF's.

\section{Separation energies}
\label{Sep-energies}

Since our investigation is restricted to even-even nuclei, we
consider two-neutron $S_{2n}=B(Z,N-2)-B(Z,N)$ and two-proton
$S_{2p}=B(Z-2,N)-B(Z,N)$ separation energies. Here $B(Z,N)$ stands
for the binding energy of a nucleus with $Z$ protons and $N$
neutrons. Two-neutron $S_{2n}$ and two-proton $S_{2p}$ separation
energies are described with a typical accuracy of 1 MeV (Table
\ref{deviat}). The accuracy of the description of separation
energies depends on the accuracy of the description of mass
differences. As a result, not always the functional
which provides the best description of masses gives the best
description of two-particle separation energies.

\begin{figure*}[ht]
\includegraphics[width=16.5cm,angle=0]{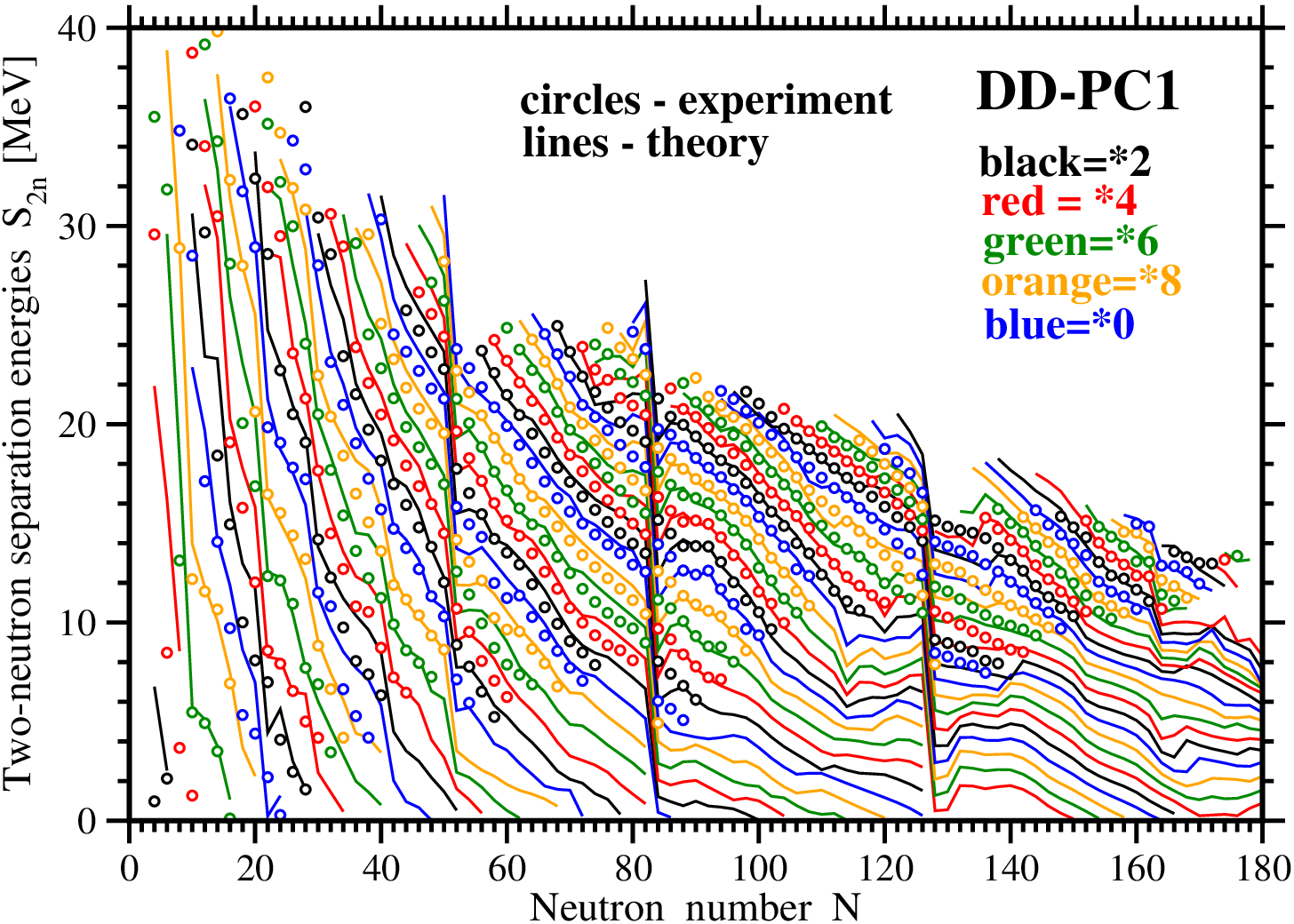}
\caption{(Color online) Two-neutron separation energies $S_{2n}(Z,N)$
given for different isotopic chains as a function of neutron number.
To facilitate the comparison between theory and experiment, five
different colors are used periodically as a function of neutron number.
Black, red, green, orange and blue colors are used for isotope chains
with proton numbers ending with 2, 4, 6, 8 and 0, respectively.}
\label{Sep-2-neut}
\end{figure*}

\begin{figure*}[ht]
\includegraphics[width=16.5cm,angle=0]{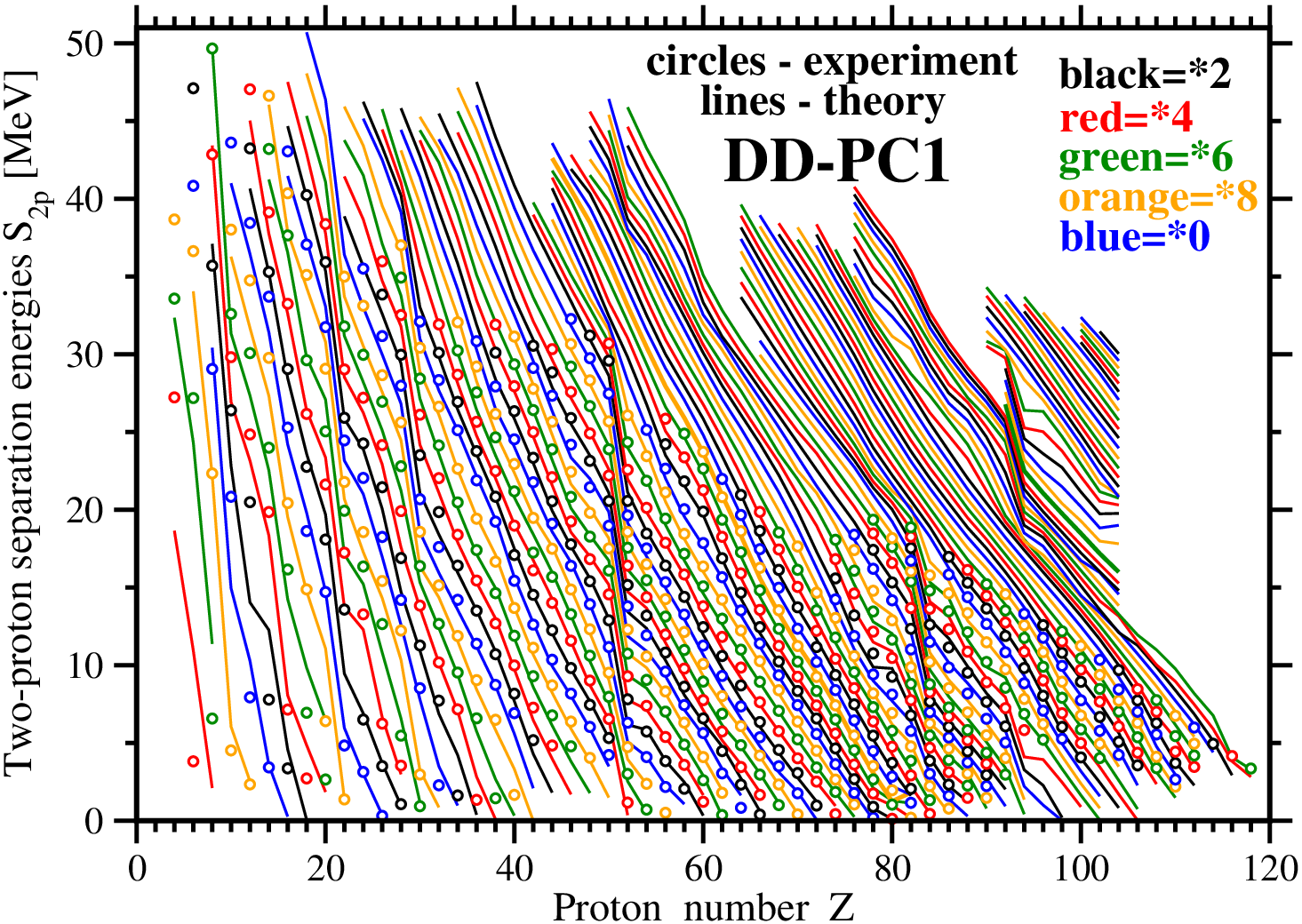}
\caption{(Color online) Two-proton separation energies $S_{2p}(Z,N)$
given for different isotonic chains as a function of proton number.
To facilitate the comparison between theory and experiment, five different
colors are used periodically as a function of proton number. Black, red,
green, orange and blue colors are used for isotonic chains with neutron
numbers ending with 2, 4, 6, 8 and 0, respectively.}
\label{Sep-2-prot}
\end{figure*}

The accuracy of the description of two-neutron and two-proton
separation energies is illustrated for different isotopic and isotonic
chains on the example of RHB calculations with DD-PC1
in Figs.\ \ref{Sep-2-neut} and \ref{Sep-2-prot}.
Similar results were obtained also in the calculations with NL3*,
DD-ME2 and DD-ME$\delta$. One can see that two-proton
separation energies are better described than two-neutron
separation energies (see also Table \ref{deviat}). In
part, this is a consequence of the behavior of the calculated
$S_{2n}$ curves in the vicinity of spherical shell gaps. The
experimental $S_{2n}$ curves are smooth (frequently almost
straight) as a function of neutron number between shell gaps
(Fig.\ \ref{Sep-2-neut}). For a given isotope
chain, the calculations rather well reproduce this behavior
of experimental $S_{2n}$ curves in the regions of a few neutrons
away from shell closures. However, the situation is different
in the vicinity of the $N=82$ and 126 shell closures. Here,
the calculations overestimate (underestimate) experimental
$S_{2n}$ values for a few nuclei before (after) the shell closure
in a number of isotopic chains with $Z\geq 40$.

It is interesting that such problems do not exist for two-proton
separation energies (Fig.\ \ref{Sep-2-prot}). The origin of these
problems is most likely related to the relative impact of proton and
neutron shell closures. Fig.\ \ref{Charge-deform} shows that the
band of nuclei with spherical or near-spherical deformations (gray
area in the figure) is wider around $N=82$ and $N=126$ as compared
with the one around $Z=50$ and $Z=82$. Thus, the transition from
spherical shapes to well-deformed shapes (where the mean field description
is justified) proceeds faster (in terms of particle number) for the
proton subsystem than for the neutron subsystem.
In contrast, the transitional shapes requiring a beyond mean field
description are expected for a wider range of nuclei around the $N=82$
and $N=126$ shell closures. Neglecting these beyond mean field correlations
is most likely the source for the above mentioned discrepancies between
experimental and calculated $S_{2n}$ values in the vicinity of the
$N=82$ and $N=126$ shell closures.

This analysis leads to a more critical look on the reappearance
of two-neutron binding with increasing neutron number beyond
the primary two-neutron drip line which exists in a number of DFT
calculations \cite{Eet.12,AARR.13,ZPX.13}. This reappearance shows
itself in the nuclear chart via the peninsulas emerging from the nuclear
mainland. For example, as we see in Fig.\ \ref{Charge-deform}, such
peninsulas exist at $(Z=62,N=132-146)$ and $(Z=88,N=194-206)$ for DD-PC1,
at $(Z=74,N=176-184)$ and $(Z=90,N=194-206)$ for DD-ME2,
and at $(Z=62,N=132-142)$,
$(Z=74,N=178-184)$, and $(Z=90,N=204-206)$ for DD-ME$\delta$,
but they are absent in NL3*. The physical mechanism for
their appearance was discussed in Ref.\ \cite{AARR.13}. Its
basic is the following: the two-neutron separation
energy $S_{2n}$ is slightly negative immediately after the
large shell gap at the neutron number $N^{(1)}_{\rm drip}$
that defines the primary neutron drip line, but then with
increasing neutron number it becomes slightly positive at
a higher neutron number $N_{\rm penin}$ and remains like that
for a range of neutron numbers up to $N^{(2)}_{\rm drip}$.
A further increase of $N$ beyond $N^{(2)}_{\rm drip}$
leads to two-neutron unbound nuclei. For example, these features
are visible in Fig.\ 3 of Ref.\ \cite{AARR.13}. However, the
present analysis clearly shows that immediately after the large
neutron shell closure CDFT calculations (and very likely also  SDFT
calculations since the shapes of calculated $S_{2n}$ curves
(see Fig.\ 8 in Ref.\ \cite{TOV-min} and Fig.\ 2 in \cite{Eet.12})
indicate the possibility of such a scenario) underestimate the experimental
$S_{2n}$ values. For some isotope chains, this underestimate may lead to
negative $S_{2n}$ values, and, thus, to the formation of peninsula in the
nuclear chart. Therefore, the calculated peninsulas may in some cases be
an artifact of the mean field approximation. The inclusion of correlations
beyond mean field may increase the two-neutron separation energies $S_{2n}$
and make them positive for neutron numbers from $N^{(1)}_{\rm drip}$ up to
$N_{\rm penin}$. As a consequence, the peninsula will disappear and the
two-neutron drip line will be located at $N^{(2)}_{\rm drip}$.

\begin{table}[ht]
\caption{The rms-deviations $\Delta E_{\rm rms}$, $\Delta (S_{2n})_{\rm rms}$
($\Delta (S_{2p})_{\rm rms}$) between calculated and experimental binding
energies $E$ and two-neutron(-proton) separation energies $S_{2n}$
($S_{2p}$). They are given in MeV for the indicated CDFT
parameterizations with respect to ``measured'' and ``measured+estimated''
sets of experimental masses.}
\begin{tabular}{|c|c|c|c|c|} \hline
EDF   & measured  & \multicolumn{3}{|c|}{measured+estimated}   \\ \hline
  & $\Delta E_{\rm rms}$  & $\Delta E_{\rm rms}$ & $\Delta (S_{2n})_{\rm rms}$ & $\Delta (S_{2p})_{\rm rms}$ \\ \hline
   NL3*            &    2.96  &  3.00 & 1.23 & 1.29 \\
   DD-ME2          &    2.39  &  2.45 & 1.05 & 0.95 \\
   DD-ME$\delta$   &    2.29  &  2.40 & 1.09 & 1.09 \\
   DD-PC1          &    2.01  &   2.15 & 1.16 & 1.03 \\ \hline
\end{tabular}
\label{deviat}
\end{table}

\begin{figure*}[ht]
\includegraphics[width=16.5cm,angle=0]{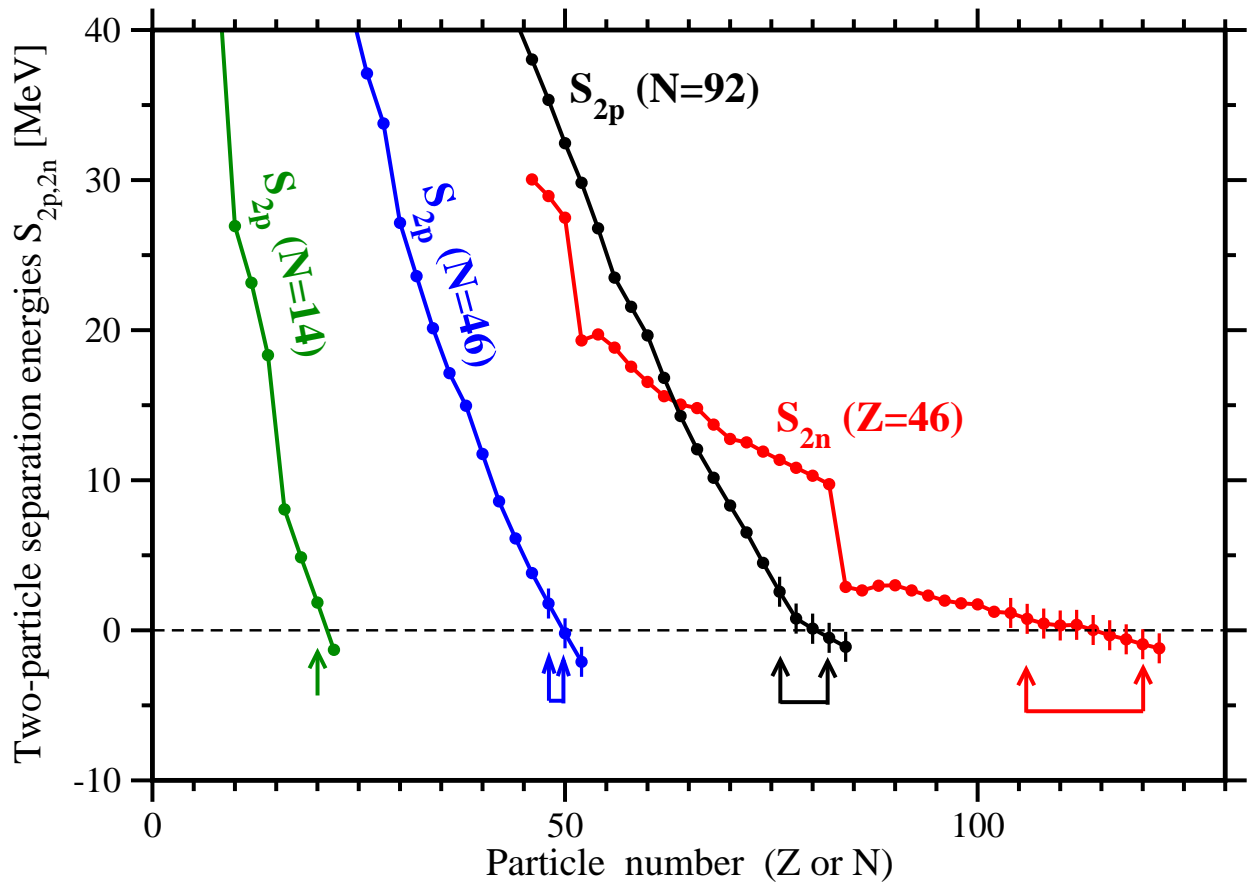}
\caption{(Color online) Schematic illustration of the dependence of the accuracy of the
prediction for the position of the two-particle drip line on the slope of the two-particle
separation energy curve as a function of the relevant particle number.  The error bars for
the calculated results show typical rms-deviations (1 MeV) between theory and experiment
(Table III). If these error bars would be taken into account (as it is effectively done
when different CEDF's are compared), they would lead to the possible ranges of particle
numbers corresponding to the two-particle drip line shown by arrows. For particle-bound
nuclei the results for DD-PC1 are used. The separation energies for particle unbound
nuclei ($S_{2n,2p}<0$) represent extrapolations. They are used here only for illustration
purposes.}
\label{accuracy}
\end{figure*}

\section{The two-proton drip line.}
\label{proton-drip}

The particle stability (and, as a consequence, a drip line)
of a nuclide is specified by its separation energy, namely,
the amount of energy needed to remove particle(s). If the two-neutron
and the two-proton separation energies are positive, the nucleus is
stable against two-nucleon emission. Conversely, if one of these separation energies is
negative, the nucleus is unstable. Thus, the two-neutron or the two-proton
drip line is reached when $S_{2n}\leq 0$ or $S_{2p}\leq 0$,
respectively.

\begin{figure*}[ht]
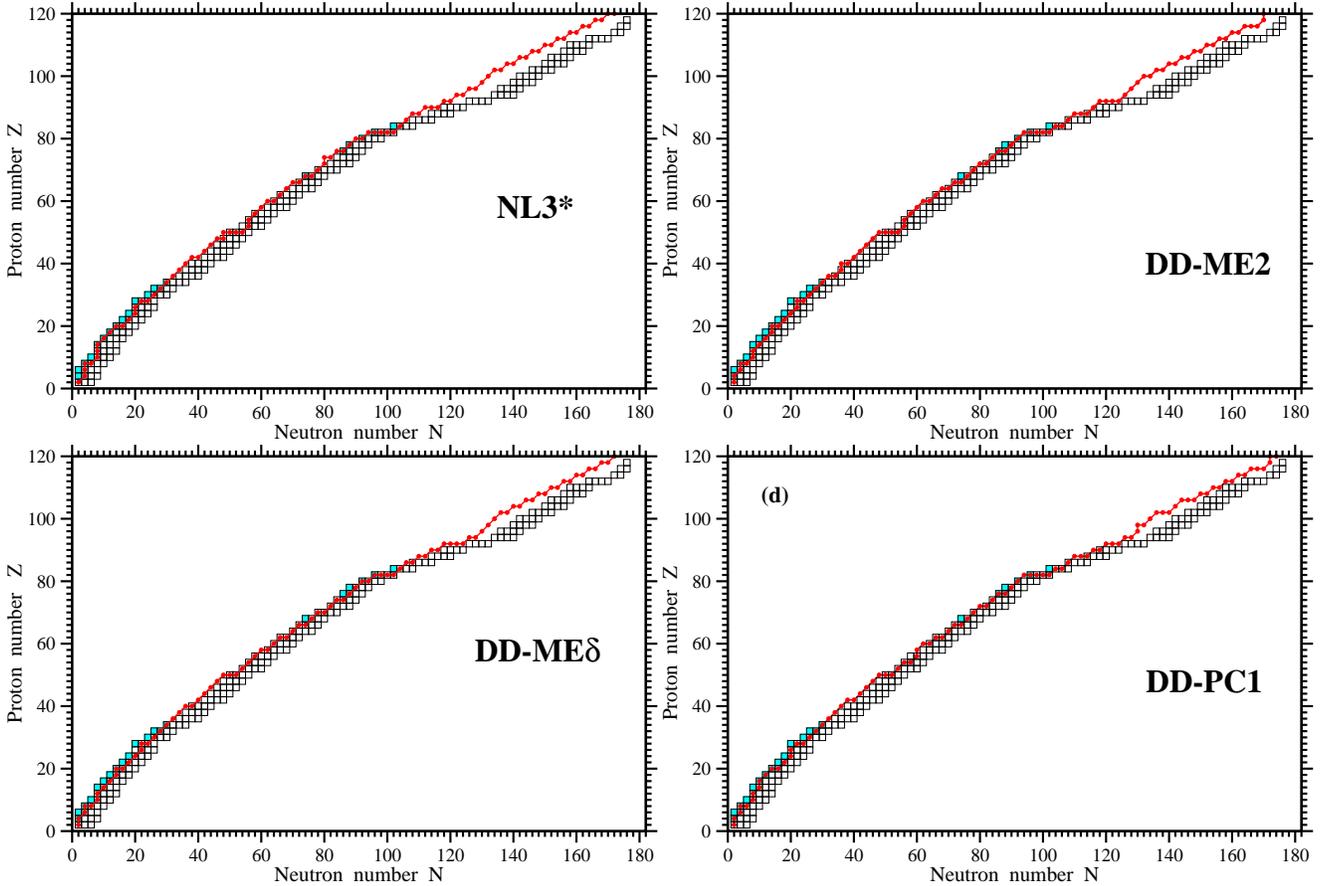

\begin{center}
%
\includegraphics[width=8.7cm,angle=0]{fig-13a.eps}
\hspace{-0.2cm}
\includegraphics[width=8.7cm,angle=0]{fig-13b.eps}
\hspace{-0.2cm}
\includegraphics[width=8.7cm,angle=0]{fig-13c.eps}
\hspace{-0.2cm}
\includegraphics[width=8.7cm,angle=0]{fig-13d.eps}
\end{center}
\caption{(Color online) The calculated two-proton drip lines versus
experimental data. For each isotope chain, the four experimentally known
most proton-rich nuclei are shown by squares. Cyan shading of the squares
is used for the nuclei located beyond the two-proton drip line ($S_{2p}<0)$.
The experimental data are from Ref.\ \protect\cite{AME2012}.  The
borderline between shaded and open squares delineates the known two-proton
drip lines. Only in the case of the $Z=4$, $6$, $8$, $80$, $82$, and $84$
isotope chains, the location of two-proton drip line is firmly established
since the masses of the nuclei on both sides of the drip line are
directly and accurately measured. The two-proton drip line is only
tentatively delineated for
other isotope chains since either
the  masses of beyond the drip line nuclei are only estimated in Ref.\
\protect\cite{AME2012} or beyond the drip line nuclei are not known
experimentally. The red lines with small symbols show the calculated
two-proton drip lines which go along the last two-proton bound nuclei.}
\label{Fig-2-prot-drip}
\end{figure*}

The proton drip line has been studied extensively more than
a decade ago in the  RHB framework with the finite range Gogny
pairing force D1S in
Refs.~\cite{LR.98,VLR.98,VLR.99,Lalazissis1999_NPA650-133,Lalazissis1999_PRC60-051302,LVR.01,LVR.04}.
However, the main emphasis was put on the one-proton drip line, for
which, at the time of these studies, experimental data was more
available than that for the two-proton drip line. In addition, only
the NL3 parametrization \cite{NL3} has been used in these studies.
Therefore, no estimate of theoretical errors in the prediction of one-
and two-proton drip lines are available. These gaps in our
knowledge of the CDFT performance have been filled in
Ref.~\cite{AARR.13}, where the two-proton drip lines were studied with
NL3*, DD-ME2, DD-PC1 and DD-ME$\delta$. Theoretical uncertainties
in the definition of two-proton drip line have been deduced.

In this chapter, we present a more detailed comparison of
RHB results with the experiment. Fig.\ \ref{Fig-2-prot-drip}
compares experimental data with calculated two-proton drip lines
obtained with NL3*, DD-ME2, DD-PC1, and DD-ME$\delta$. Note that the experimental two-proton
drip line is delineated firmly or tentatively up to $Z=84$ (see caption
of Fig.\ \ref{Fig-2-prot-drip} for details).
The red line with small solid circles shows the calculated two-proton
drip line. Nuclei to the left of this line are proton unstable
in the calculations. Nuclei which are proton unstable in experiment
are shown by solid cyan squares. In the following discussion we concentrate
on isotope chains containing proton unstable nuclei since this
provides the most reliable experimental information on the position
of two-proton drip line. One can see that NL3* tends to predict
the two-proton drip line at too low values of the neutron number $N$.
Indeed, experimentally known proton unstable nuclei at $Z=8$, $14$, $16$, $18$, $20$, $32$, $34$, $68$, $76$,
$78$, $80$, and $82$ (shown by cyan squares in Fig.\ \ref{Fig-2-prot-drip})
are predicted to be proton bound by NL3*. On the other side, the
two-proton drip line is predicted too early for the $Z=52$ chain.
Similar problems with the description of the proton unstable
$Z=4$, $8$, $20$, $32$, $34$, $76$, $80$, and $82$ nuclei exist for DD-ME2.
Note also that the two-proton drip line is predicted too early in the
$Z=26$ and $52$ isotope chains in this CDFT parametrization. Also for
DD-ME$\delta$, proton unbound $Z=4$, $8$, $20$, $30$, $32$, $80$, $82$ nuclei
are predicted to be proton bound, and the two-proton drip line is predicted too
early for $Z=26$. A similar situation is observed with DD-PC1 for which $Z=4$, $8$, $16$, $18$, $20$, $32$, $34$, $76$, $80$, and $82$ proton unbound
nuclei are bound in the calculations. In addition, the two-proton drip
line is predicted too early for this parametrization for the $Z=56$
isotopes.

The best reproduction of the two-proton drip line is achieved with
DD-ME2 and DD-ME$\delta$, which are characterized by the best residuals
for two-proton separation energies $S_{2p}$ (Table \ref{deviat}). In general,
the results of the calculations are very close to experimental data.
This is because the proton-drip line lies close to the valley of stability,
so that extrapolation errors towards it are small. Another reason is the fact
the Coulomb barrier provides a rather steep potential reducing considerably
the coupling to the proton continuum. This leads to a relatively low density of
the single-particle states in the vicinity of the Fermi level.

Since this density is comparable with the one for the nuclei away
from two-proton drip line, the slope of the two-proton separation
energy $S_{2p}$ as a function of proton number for a given isotonic
chain remains almost unchanged on approaching the two-proton drip line
(Fig.\ \ref{Sep-2-prot}). This slope is directly related to the
uncertainties in the prediction of the position of two-proton drip line.
For a given accuracy of the description of two-proton separation energies
these uncertainties in the definition of position of the two-particle drip
line increase with the decrease of the slope of $S_{2p,2n}$ 
(see Fig. \ref{accuracy}).
As a consequence, theoretical uncertainties for the two-proton drip line
are rather small for $Z\leq 86$ but somewhat larger for higher $Z$
(see Fig.\ 2 in Ref.\ \cite{AARR.13}) due to the increase of the
single-particle level density and the related decrease of the slope
of $S_{2p}$ as a function of proton number (Fig.\ \ref{Sep-2-prot}).

According to Fig.\ 2 of Ref.\ \cite{AARR.13},
theoretical uncertainties in the predictions of the position of
two-proton drip line are either very small (2 neutrons) or
non-existent for isotope chains with $Z\leq 86$. These small
uncertainties may be a source of observed discrepancies between
calculations and experiment for a number of isotope chains
(for example, the ones with $Z=4$, $14$, $16$, $18$, $20$, $26$, $68$, $76$, $78$, and $80$
in Fig.\ 2 of Ref.\ \cite{AARR.13}). However, in a number of
the cases (for example, in the $Z=32$ and $34$ isotopes chains) there
is no uncertainty in the predicted position of two-proton drip
line (Fig.\ 2 in Ref.\ \cite{AARR.13}). Thus, the observed
discrepancies between theory and experiment may be due to the
limitations of the model description on the mean field level.
Indeed, it is well known that the Ge ($Z=32$) \cite{Ge-As} and Se ($Z=34$)
\cite{Se.oblate,HNMM.09} isotopes show prolate-oblate shape
coexistence and/or $\gamma$-softness near the proton-drip line.
A similar shape coexistence is also observed in heavier Kr
\cite{Krexp,FMXLYM.13,AF.05,BBH.06} and Rb \cite{Rb74} nuclei as
well as in the $Z\sim 82$ proton-drip line nuclei \cite{YBH.13,Pb.prot.drip}.
By ignoring the correlations beyond mean field, which are expected
to be most pronounced in light nuclei, we may introduce an error
in the predicted position of two-proton drip line.

\section{The two-neutron drip line.}
\label{Two-neu-drip-sec}

\begin{figure*}[ht]
\includegraphics[width=16.5cm,angle=0]{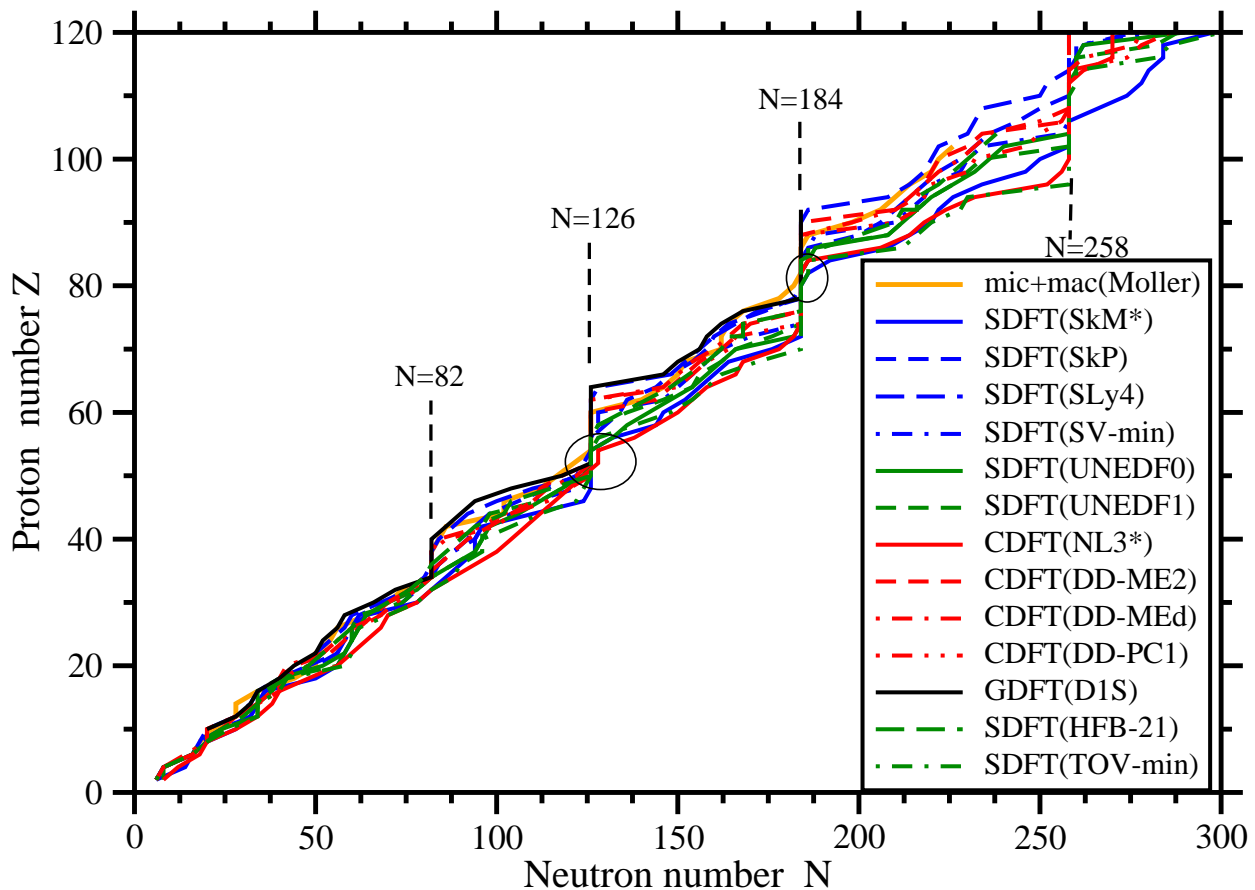}
\caption{(Color online) Two-neutron drip-lines obtained in
state-of-the-art DFT calculations. The regions of well defined
localization of the two-neutron drip-line are encircled.}
\label{systematics}
\end{figure*}

As discussed in Refs.\ \cite{Eet.12,AARR.13}, the situation
is different for the two-neutron drip line. Fig.\ \ref{systematics}
presents the compilation of known calculated two-neutron drip lines
obtained with the state-of-the-art relativistic and non-relativistic
EDF's. They include four two-neutron drip lines obtained in the CDFT
calculations of Ref.\ \cite{AARR.13}, which are tabulated in Table
\ref{drip-lines}. Non-relativistic results are represented by
two-neutron drip lines obtained with the Gogny functional D1S
\cite{DGLGHPPB.10}
and with eight functionals of Skyrme type~\cite{Eet.12,TOV-min}. In
addition, the two-neutron drip line from the microscopic+macroscopic
calculations of Ref.\ \cite{MNMS.95} is shown. One can see that
with the exception of two encircled regions, the  theoretical differences
in the location of two-neutron drip line are much larger than the
ones for the two-proton drip line. They are generally growing
with increasing proton number.

\begin{figure*}[ht]
\includegraphics[width=16.5cm,angle=0]{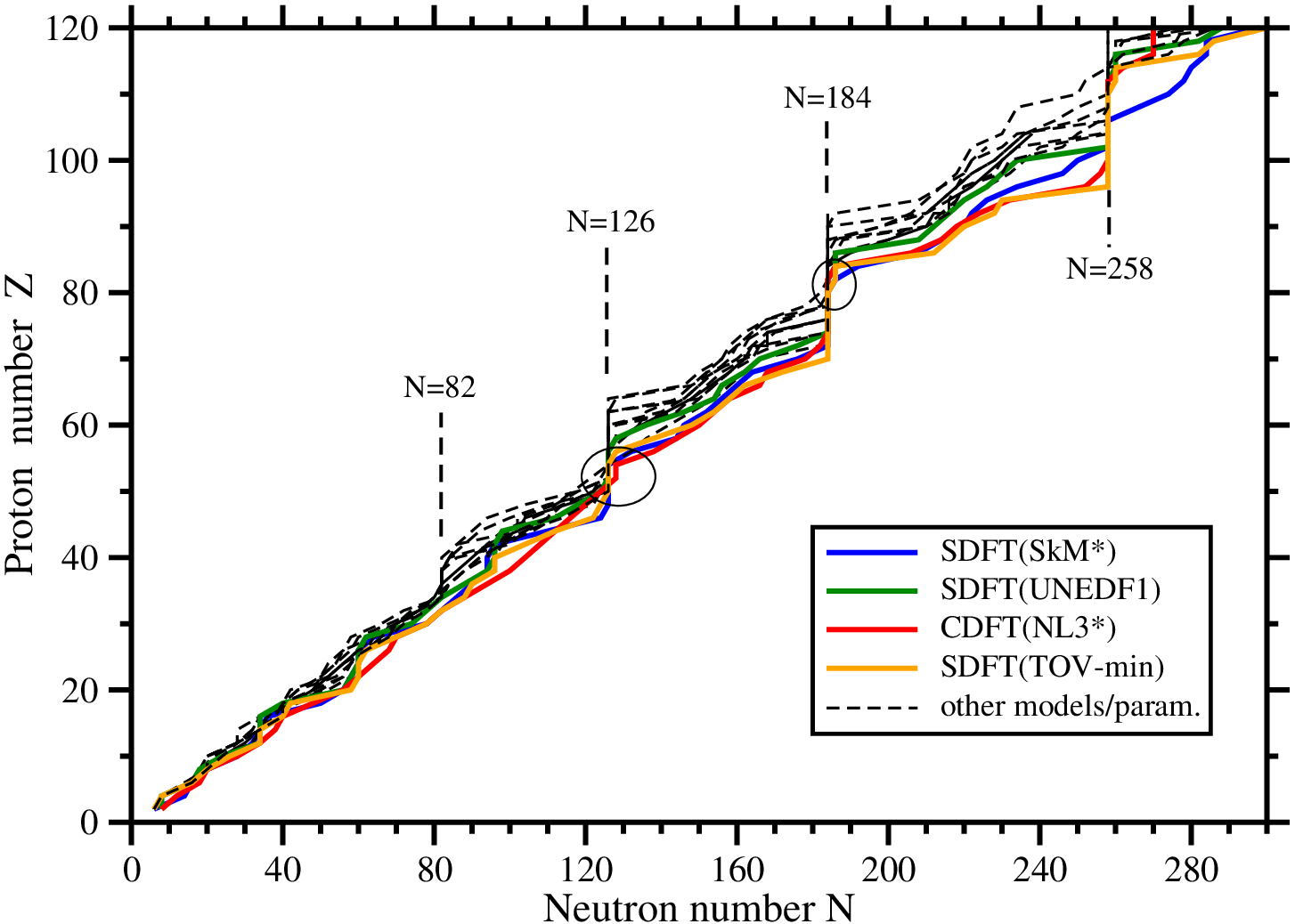}
\caption{(Color online) The same as in Fig.\ \ref{systematics} but with
the four most neutron-rich two-neutron drip lines shown in color
and the rest in  black.}
\label{system-4-most-rich}
\end{figure*}

\begin{figure*}[ht]
\includegraphics[width=16.5cm,angle=0]{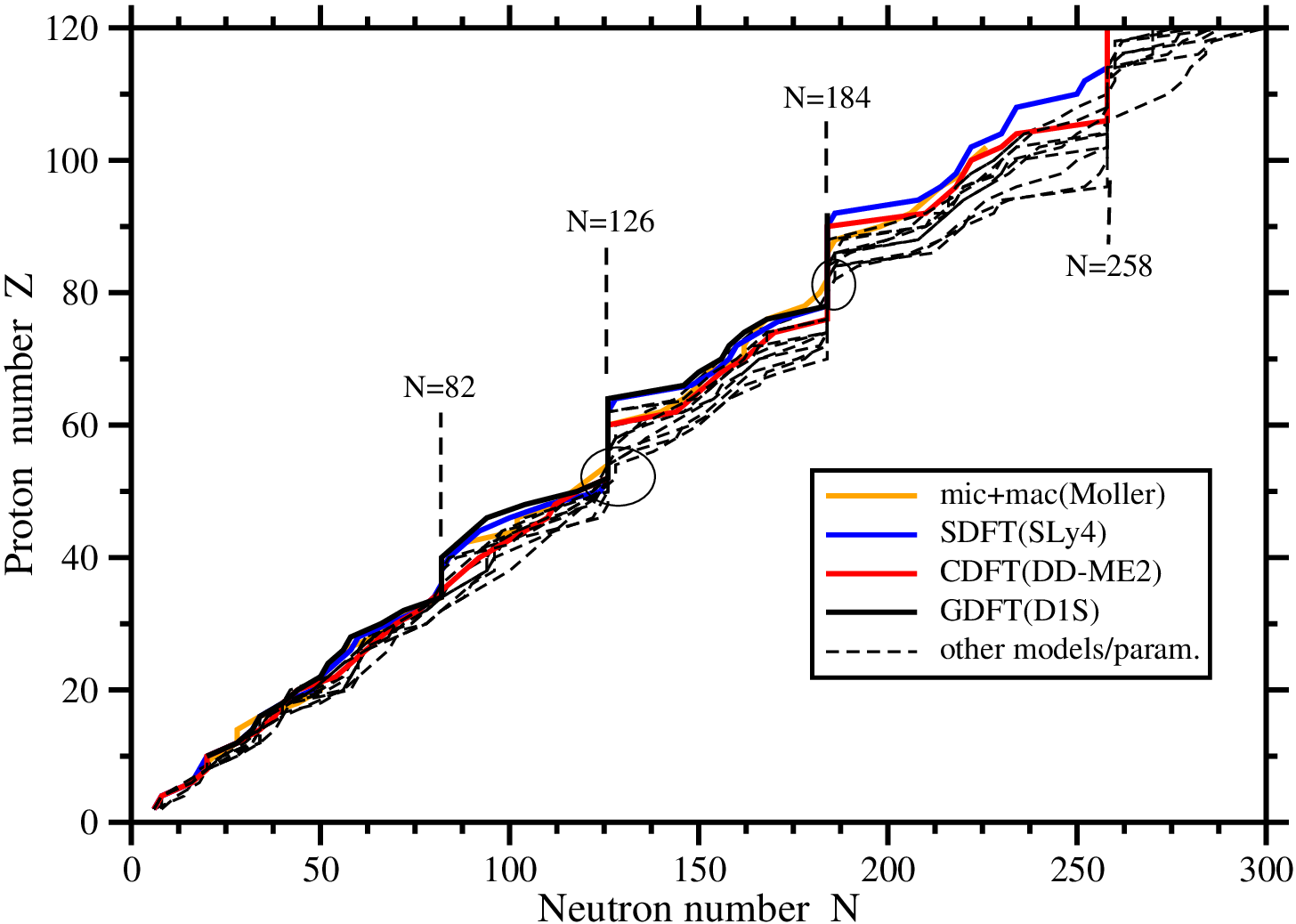}
\caption{(Color online) The same as in Fig.\ \ref{systematics} but with
the four least neutron-rich two-neutron drip lines shown in color and the
rest in  black.}
\label{system-4-least-rich}
\end{figure*}

One could ask the question whether there exist correlations
between the position of two-neutron drip line for a given EDF
and its nuclear matter properties. With that goal Figs.\
\ref{system-4-most-rich} and  \ref{system-4-least-rich} show
the four most neutron-rich and the four least neutron-rich
two-neutron drip lines amongst the 14 compiled lines.
The nuclear matter properties of the corresponding EDF's are shown in
Table~\ref{tab-nuclear-matter}. Let us consider the EDF's NL3* and
DD-ME2 leading to the most and the least neutron-rich two-neutron
drip lines amongst the relativistic functionals. It is tempting to
associate the difference in the position of two-neutron drip lines
with different symmetry energies $J$  ($J=32.30$ MeV for
DD-ME2 and $J=38.68$ MeV for NL3*) and the slope parameter
$L$ of the symmetry energy at saturation density ($L=51.26$ MeV
for DD-ME2 and $L=123$ MeV for NL3*). However, a detailed
comparison of the position of the 14 two-neutron drip lines
presented in Figs.\ \ref{systematics}, \ref{system-4-most-rich} and
\ref{system-4-least-rich} with nuclear matter properties of their
EDF's (Table \ref{tab-nuclear-matter}) does not reveal clear
correlations between the location of two-neutron drip line and the
nuclear matter properties of the corresponding functional.
In fact, for nuclei close to the neutron drip line
the Fermi surface is very small and negative close to the
continuum limit and it changes only slowly with the
neutron number. The precise position of the drip line
therefore depends very much on the behavior of the
tail of the neutron density. At these very low densities
the properties $J$ and $L$ of nuclear matter at saturation
is not really relevant.

 Possible sources of the uncertainties in the position of the
two-neutron drip line have been discussed in Ref.\ \cite{AARR.13}.
They include the isovector properties of the EDF's \cite{Eet.12} 
and the underlying shell structure connected with inevitable
inaccuracies of the single particle energies in the DFT
description~\cite{AARR.13}.

The isovector properties of an EDF define the depth of the nucleonic
potential with respect to the continuum and may thus affect the
location of two-neutron drip line.
However, such uncertainties in the depth of the nucleonic potential exist
also in known nuclei (see discussion in Sect. IVC of Ref.\ \cite{LA.11}).
They cannot describe the observed features completely.

The shell structure effects are clearly visible in the fact that
for some combinations of $Z$ and $N$ there is basically
no (or very little) dependence of the predicted location
of the two-neutron drip line on the CDFT parameterization. Such a weak
(or vanishing) dependence, seen in all model calculations,
is especially pronounced at spherical neutron
shell closures with $N=126$ and $184$ around the proton numbers $Z=54$
and $80$, respectively. In addition, a similar situation
is seen in the CDFT calculations at $N=258$ and $Z\sim 110$.
This fact is easy to understand because of the large neutron shell gap
at the magic neutron numbers in all DFT's.

Inevitable inaccuracies in the DFT description of single particle
energies \cite{AS.11,LA.11} also contribute to increasing uncertainties
in the prediction of two-neutron drip line position on moving away
from these spherical shell closures. This move induces deformation.
The comparison of Figs.\ \ref{systematics} and \ref{Charge-deform} shows
that there is a close correlation between the nuclear deformation at
the neutron-drip line and the uncertainties in their prediction.
The regions of large uncertainties corresponds to transitional and
deformed nuclei. Again this is caused by the underlying level densities
of the single-particle states. The spherical nuclei under discussion are
characterized by large shell gaps and a clustering of highly degenerate
single-particle states around them. Deformation removes this high degeneracy
of single-particle states and leads to a more equal distribution of the
single-particle states with energy. Moreover, the density of bound neutron
single-particle states close to the neutron continuum is substantially larger
than that on the proton-drip line which leads to a small slope of two-neutron
separation energies $S_{2n}$ as a function of neutron number in the vicinity
of two-neutron drip line for medium and heavy mass nuclei
(see Fig.\ \ref{Sep-2-neut}). This slope is smaller than the slope of
two-proton
separation energies $S_{2p}$ as a function of proton number
in the vicinity of two-proton drip line (Fig.\ \ref{Sep-2-prot}).
Note that the $S_{2n}$ and $S_{2p}$ values are described with a similar
accuracy in the various parameterizations (Table \ref{deviat}). However, the difference
in the slope of $S_{2n}$ and $S_{2p}$ as a function of proton and neutron
numbers
translates into much larger uncertainties in the definition of the position of
two-neutron drip line as compared with two-proton drip line. This also
indicates that the predictions for the two-neutron drip line depend more
sensitively on the
single-particle energies than those for two-proton drip line.

\begin{table*}[ht]
\caption{Properties of symmetric nuclear matter at saturation for the energy density functionals used in Fig.\
\ref{systematics}: the density $\rho_0$ , the energy per particle $(E/A_{\infty})$, the incompressibility $K_{\infty}$,
the symmetry energy $J$ and its slope $L$, and the isoscalar effective masses $m^*/m$ of a nucleon at the Fermi surface.
In the relativistic cases we show the Lorentz effective masses \cite{Jaminon1989_PRC40-354}.
The results of the compilation \cite{Sk-nm} is used for the Skyrme functionals when possible.}
\label{tab-nuclear-matter}
\begin{center}
\begin{tabular}{|c|c|c|c|c|c|c|}\hline
Parameter              & $\rho_0$ [fm$^{-3}$] & $(E/A)_{\infty}$ [MeV] & $K$ [MeV] & $J$ [MeV] & $L$ [MeV] &  m*/m \\ \hline
                         \multicolumn{7}{|c|}{four most neutron-rich two-neutron drip lines}                  \\ \hline
NL3* \cite{NL3*}       &   0.150        &  -16.31           & 258    & 38.68   & 122.6  &  0.67  \\
SkM* \cite{SkM*,Sk-nm} &   0.160        &  -15.77           & 217    & 30.03   &  45.8  &  0.79   \\
UNEDF1  \cite{UNEDF2}  &   0.159        &  -15.80           & 220    & 28.99   &  40.0  &  0.99   \\
TOV-min \cite{TOV-min} &   0.161        &  -15.93           & 222    & 32.30   &  76.0  &  0.94   \\ \hline
                         \multicolumn{7}{|c|}{four least neutron-rich two-neutron drip lines}                  \\ \hline
mic+mac [FRDM] \cite{MNMS.95}  &        & -16.25            & 240    & 32.73   &        & 1.00       \\
DD-ME2 \cite{DD-ME2}   &  0.152         & -16.14            & 251    & 32.40   &  49.4  & 0.66  \\
SLy4 \cite{SLy4,Sk-nm} &  0.160         & -15.97            & 230    & 32.00   &  45.9  & 0.69   \\
D1S [Gogny] \cite{CGH.08}  &  0.160     & -15.90            & 210    & 32.00   &        & 0.70   \\ \hline
                         \multicolumn{7}{|c|}{remaining parametrizations (drip-lines in the middle) }                \\ \hline
UNEDF0  \cite{UNEDF2}  & 0.161         & -16.06            & 230    & 30.54   &  45.1  & 0.90   \\
DD-ME$\delta$ \cite{DD-MEdelta} & 0.152 & -16.12            & 219    & 32.35   &  52.9  & 0.61  \\
SkP \cite{SkP,Sk-nm}   &  0.163         & -15.95            & 201    & 30.00   &  19.7  & 1.00   \\
SV-min \cite{SV-min,Sk-nm} &   0.161    & -15.91            & 222    & 30.66   &  44.8  & 0.95   \\
DD-PC1 \cite{DD-PC1,PC-PK1}&   0.152    & -16.06            & 230    & 33.00   &  68.4  & 0.66   \\
HFB-21 [BSk21] \cite{GCP.10} & 0.158    & -16.05            & 246    & 30.00   &  46.6  & 0.80  \\ \hline
\end{tabular}
\end{center}
\end{table*}

\begin{table*}[ht]
\caption{Two-proton and two-neutron drip lines predicted by the CEDF's
used in this work. Neutron numbers $N$ (columns 2-9) corresponding
to these drip lines are given for each even proton number
$Z$ (column 1). An asterisk at a neutron number at the two-neutron
drip line indicates isotope chains with additional two-neutron
binding at higher $N$-values (peninsulas).}
\begin{tabular}{|c|c|c|c|c|c|c|c|c|} \hline
Proton     & \multicolumn{4}{|c|}{Two proton drip-line} &\multicolumn{4}{|c|}{Two neutron drip-line}   \\
number $Z$ & NL3* & DD-ME2 &DD-ME$\delta$& DD-PC1 & NL3* & DD-ME2 & DD-ME$\delta$ & DD-PC1 \\ \hline
   1       &  2   &  3     &  4          &  5     &  6   &  7     &   8           &  9     \\ \hline
   2       &  2   &  2     &  2          &  2     &  8   &  6     &   6           &  6     \\
   4       &  4   &  2     &  2          &  2     & 12   &  8     &   8           &  8     \\
   6       &  4   &  4     &  4          &  4     & 18   &  16    &  14           & 16     \\
   8       &  4   &  4     &  4          &  4     & 20   &  20    &  20           & 20     \\
  10       &  8   &  8     &  8          &  8     & 28   &  20    &  20           & 24     \\
  12       &  8   &  8     &  8          &  8     & 34   &  28    &  28           & 28     \\
  14       &  8   &  10    &  10         & 10     & 38   &  34    &  34           & 34     \\
  16       & 10   &  12    &  12         & 10     & 40   &  38    &  40           & 40     \\
  18       & 12   &  14    &  14         & 12     & 48   &  40    &  40           & 40     \\
  20       & 14   &  14    &  14         & 14     & 56   &  44    &  42           & 48     \\
  22       & 18   &  18    &  18         & 18     & 60   &  54    &  52           & 52     \\
  24       & 20   &  20    &  20         & 20     & 64   &  58    &  56           & 56     \\
  26       & 20   &  22    &  22         & 20     & 68   &  62    &  60           & 62     \\
  28       & 22   &  22    &  22         & 22     & 70   &  66    &  68           & 68     \\
  30       & 26   &  26    &  26         & 26     & 78   &  70    &  70           & 72     \\
  32       & 28   &  28    &  28         & 28     & 82   &  76    &  76           & 78     \\
  34       & 30   &  30    &  30         & 30     & 88   &  80    &  82           & 82     \\
  36       & 32   &  32    &  32         & 32     & 94   &  84    &  82           & 82     \\
  38       & 34   &  36    &  34         & 34     & 100  &  88    &  82           & 82     \\
  40       & 36   &  36    &  36         & 36     & 104  &  92    &  84           & 86     \\
  42       & 38   &  40    &  40         & 38     & 108  &  98    &  96           & 100    \\
  44       & 42   &  42    &  42         & 42     & 112  &  104   & 102           & 104    \\
  46       & 44   &  44    &  44         & 44     & 116  &  110   & 110           & 114    \\
  48       & 46   &  46    &  46         & 46     & 120  &  112   & 114           & 120    \\
  50       & 48   &  48    &  48         & 48     & 124  &  118   & 122           & 126    \\
  52       & 56   &  56    &  54         & 54     & 128  &  126   & 126           & 126    \\
  54       & 56   &  56    &  56         & 56     & 128  &  126   & 126           & 126    \\
  56       & 58   &  58    &  58         & 60     & 138  &  126   & 126           & 126    \\
  58       & 60   &  60    &  60         & 60     & 144  &  126   & 126           & 126    \\
  60       & 62   &  62    &  64         & 62     & 150  &  126   & 126           & 126    \\
  62       & 66   &  66    &  66         & 66     & 154  &  144   & 126*          & 126*   \\
  64       & 68   &  68    &  70         & 70     & 158  &  148   & 146           & 150    \\
  66       & 70   &  72    &  72         & 72     & 166  &  152   & 150           & 154    \\
  68       & 74   &  76    &  76         & 76     & 168  &  156   & 154           & 158    \\
  70       & 78   &  78    &  78         & 78     & 178  &  162   & 160           & 164    \\
  72       & 80   &  80    &  82         & 80     & 182  &  166   & 164           & 166    \\
  74       & 80   &  84    &  84         & 84     & 184  &  170*  & 168*          & 184    \\
  76       & 84   &  86    &  88         & 86     & 184  &  184   & 184           & 184    \\
  78       & 88   &  90    &  90         & 90     & 184  &  184   & 184           & 184    \\
  80       & 90   &  92    &  92         & 92     & 184  &  184   & 184           & 184    \\
  82       & 94   &  94    &  96         & 94     & 184  &  184   & 184           & 184    \\
  84       & 104  & 104    & 104         & 104    & 186  &  184   & 184           & 184    \\
  86       & 106  & 108    & 106         & 108    & 206  &  184   & 184           & 184    \\
  88       & 108  & 110    & 110         & 110    & 214  &  184   & 184           & 184*   \\
  90       & 112  & 116    & 114         & 116    & 218  &  184*  & 198*          & 210    \\
  92       & 118  & 120    & 118         & 120    & 224  &  210   & 210           & 216    \\
  94       & 122  & 126    & 126         & 126    & 232  &  214   & 216           & 218    \\
  96       & 126  & 128    & 130         & 130    & 252  &  218   & 218           & 220    \\
  98       & 130  & 130    & 132         & 130    & 256  &  220   & 222           & 230    \\
 100       & 132  & 132    & 134         & 134    & 258  &  222   & 228           & 232    \\
 102       & 134  & 136    & 136         & 136    & 258  &  230   & 232           & 246    \\
 104       & 138  & 140    & 140         & 142    & 258  &  234*  & 236           & 250    \\
 106       & 142  & 144    & 144         & 144    & 258  &  258   & 250           & 256    \\
 108       & 146  & 148    & 148         & 150    & 258  &  258   & 258           & 258    \\
 110       & 150  & 152    & 152         & 154    & 258  &  258   & 258           & 258    \\
 112       & 154  & 156    & 156         & 158    & 258  &  258   & 258           & 258    \\
 114       & 158  & 160    & 160         & 162    & 262  &  258   & 258           & 258    \\
 116       & 162  & 164    & 164         & 166    & 270  &  258   & 262           & 274    \\
 118       & 166  & 170    & 168         & 172    & 270  &  258   & 276           & 278    \\
 120       & 170  & 170    & 172         & 172    & 270  &  258*  & 278           & 286    \\ \hline
\end{tabular}
\label{drip-lines}
\end{table*}

\section{Deformations}
\label{Sec-def}

\begin{figure*}[ht]
\includegraphics[width=14.0cm,angle=0]{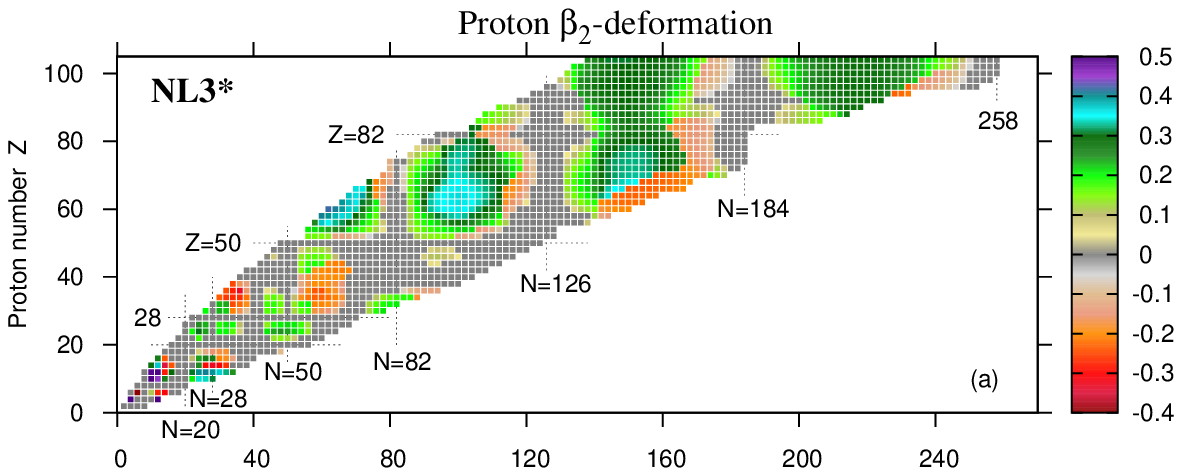}
\includegraphics[width=14.0cm,angle=0]{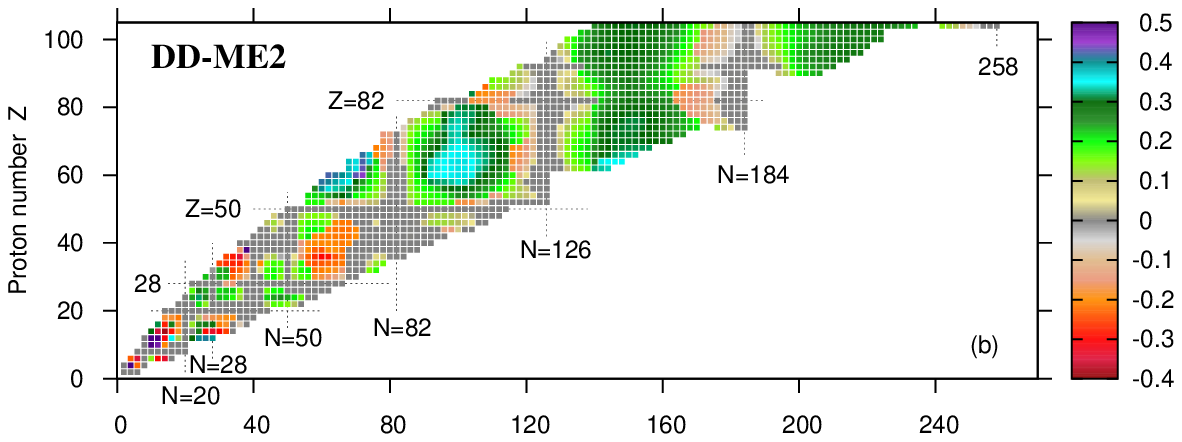}
\includegraphics[width=14.0cm,angle=0]{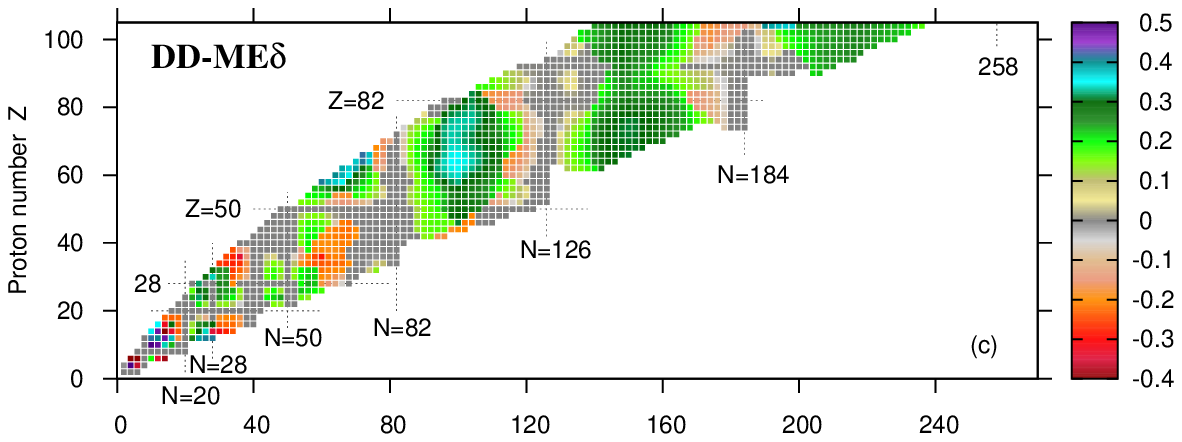}
\includegraphics[width=15.0cm,angle=0]{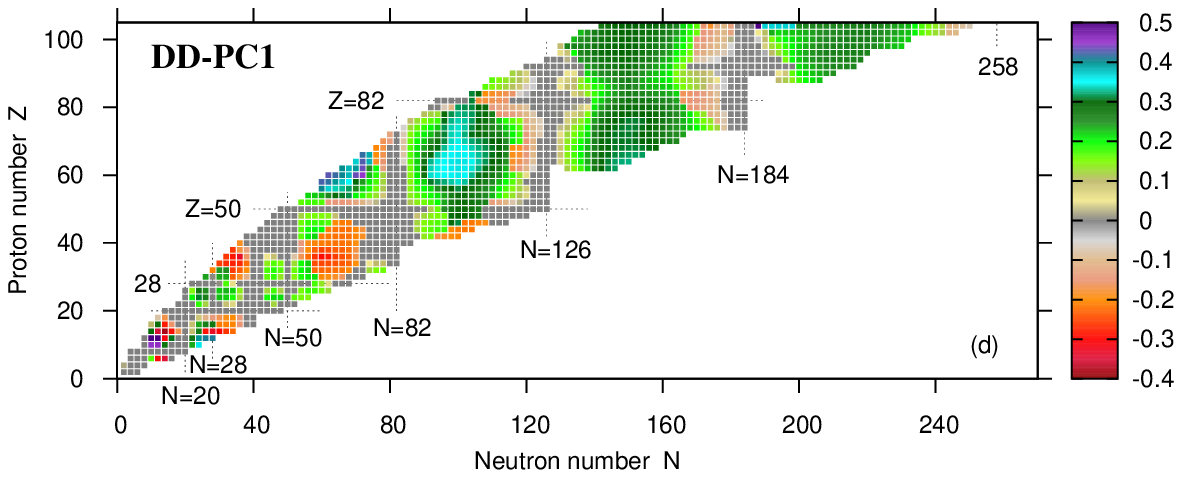}
\caption{(Color online) Charge quadrupole deformations $\beta_2$
obtained in the RHB calculations with indicated CEDF's.}
\label{Charge-deform}
\end{figure*}

\begin{figure*}[ht]
\includegraphics[width=14.0cm,angle=0]{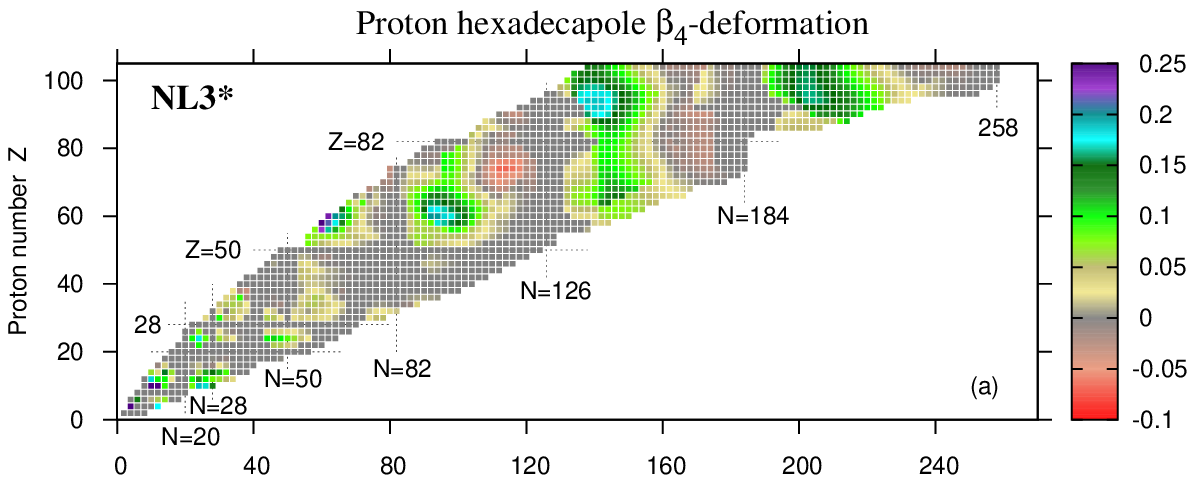}
\includegraphics[width=14.0cm,angle=0]{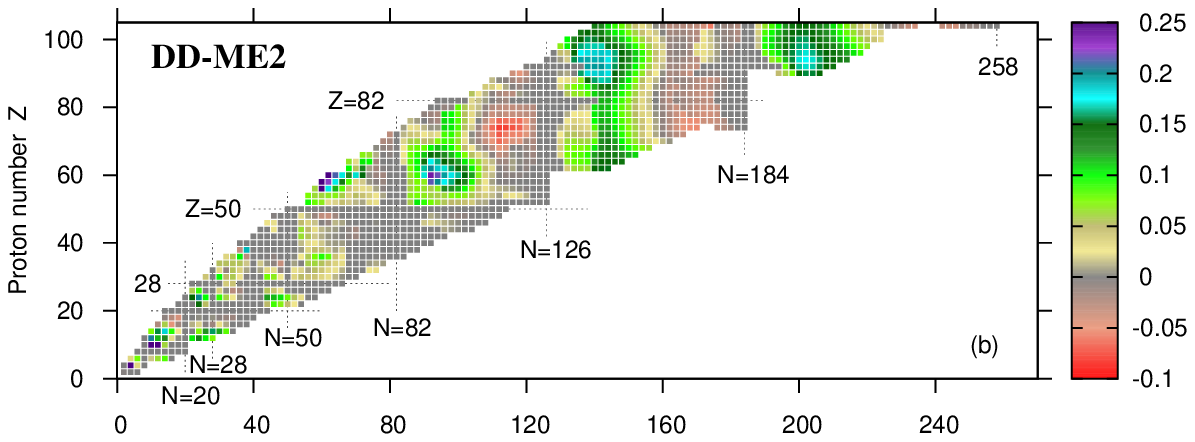}
\includegraphics[width=14.0cm,angle=0]{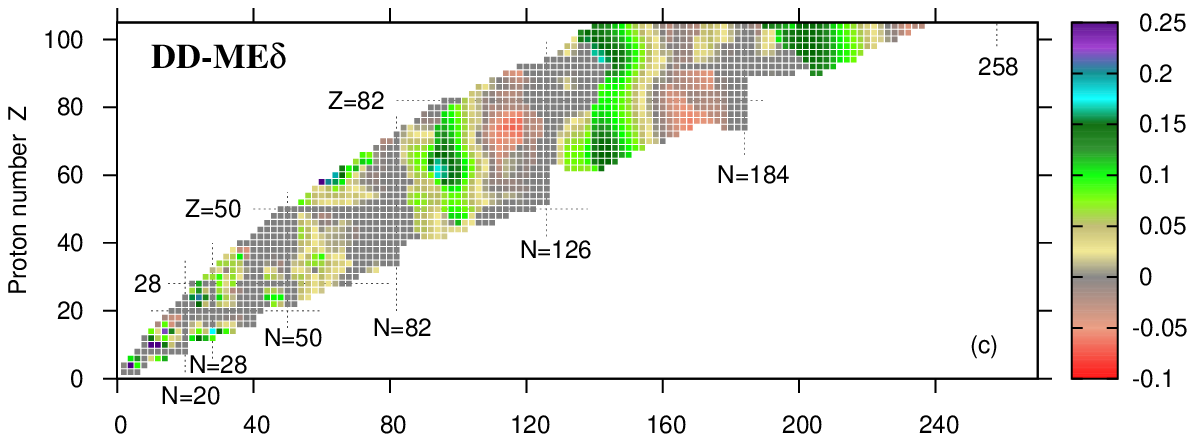}
\includegraphics[width=14.0cm,angle=0]{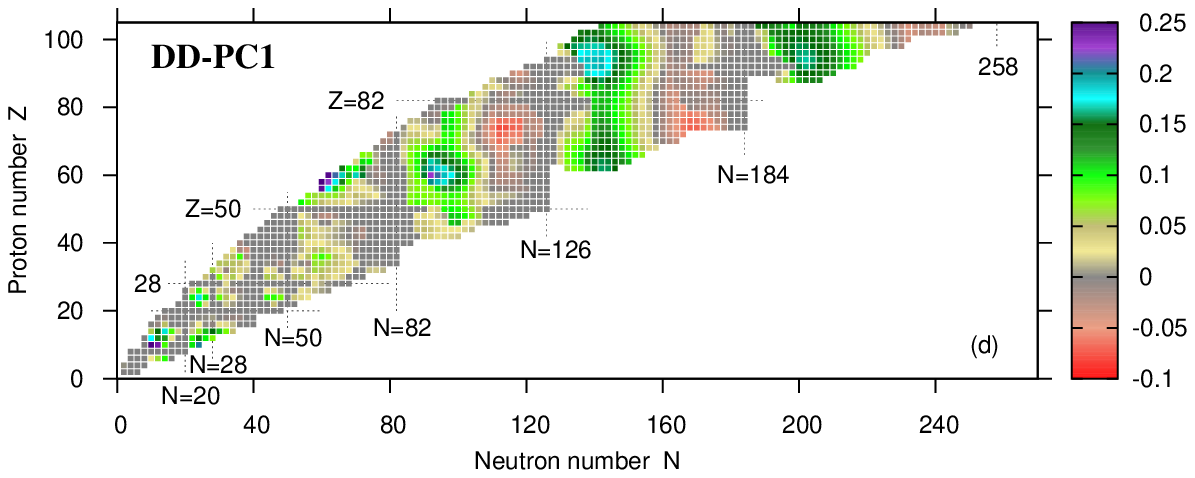}
\caption{(Color online) Proton hexadecapole deformations $\beta_4$
obtained in the RHB calculations with the indicated CEDF's.}
\label{Charge-deform-hex}
\end{figure*}

The solution of the variational equations of density functional theory
yields values for the single particle density $\rho({\bm r})$. Therefore
density functional theory not only allows us to derive the binding energies
of the system but in addition all quantities depending on $\rho({\bm r})$.
In this section we consider the charge quadrupole and hexadecupole
moments:
\begin{eqnarray}
Q_{20} &=& \int d^3r \rho({\bm r})\,(2z^2-r^2_\perp),\\
Q_{40} &=& \int d^3r \rho({\bm r})\,(8z^4-24z^2r^2_\perp+3r^4_\perp).
\end{eqnarray}
with $r^2_\perp=x^2+y^2$. In principle these values can be directly
compared with experimental data. However, it is more convenient to
transform these quantities into dimensionless deformation
parameters $\beta_2$ and $\beta_4$:
\begin{eqnarray}
Q_{20}&=&2\sqrt{\frac{4\pi}{5}} \frac{3}{4\pi} Z R_0^2 \beta_2,
\label{beta2_def} \\
Q_{40}&=&8\sqrt{\frac{4\pi}{9}}\frac{3}{4\pi} Z R_0^4 \beta_4,
\label{beta4_def}
\end{eqnarray}
where $R_0=1.2 A^{1/3}$. Eq.\ (\ref{beta2_def}) is used also
in the extraction of experimental $\beta_2$ deformation from
measured data \cite{RMMNS.87}. This justifies its application
despite the fact that this simple linear expression
ignores the contributions of higher power/multipolarity
deformations to the charge quadrupole moment. Including higher powers of
$\beta_2$, as in Ref.~\cite{NZ-def}, yields values of $\beta_2$
that are $\approx 10$\% lower. In Figs.~\ref{Charge-deform} and
\ref{Charge-deform-hex} we show the distribution of proton quadrupole $\beta_2$ and
hexadecapole $\beta_4$ deformations in the $(N,Z)$ plane
for the CEDF's NL3*, DD-ME2, DD-ME$\delta$ and DD-PC1.

Direct experimental information on the deformations of nuclei can be
obtained from Coulomb excitation and lifetime measurements
\cite{RMMNS.87}. An alternative method is to derive a quadrupole
moment from the $2^+ \rightarrow 0^+$ transition energy by
using the Grodzins relation~\cite{G.62} or its later
refinements~\cite{No252}. However, these prescriptions are applicable
only to well deformed nuclei. In general, it is estimated that
experimental methods give an accuracy of around $10\%$ \cite{No252}
for the static charge quadrupole deformation $\beta_2$ in the
case of well deformed nuclei. The error can be larger in transitional
nuclei since in this case the deformation extracted from experimental
data will contain also dynamic deformation resulting from zero-point
oscillations of the nuclear surface in the ground state \cite{BVIKO.07}.

These considerations basically limit the possibilities of
a comparison between calculated and experimental $\beta_2$
deformations to the well-deformed nuclei in the rare-earth and
actinide regions. Although deformation exists also in
the ground states of nuclei in many other regions, the potential energy
surfaces of these nuclei are, in general, soft in $\beta_2$
or $\gamma$-deformation, leading to the phenomena
of shape fluctuations, shape coexistence \cite{HW.11} and quantum phase
transitions~\cite{Niksic2007_PRL99-092502}.
For such situations, the mean field description is not completely adequate,
and, thus, a comparison between theoretical and experimental deformation
properties is not conclusive.

\begin{figure*}[ht]
\begin{center}
\includegraphics[width=16.0cm,angle=0]{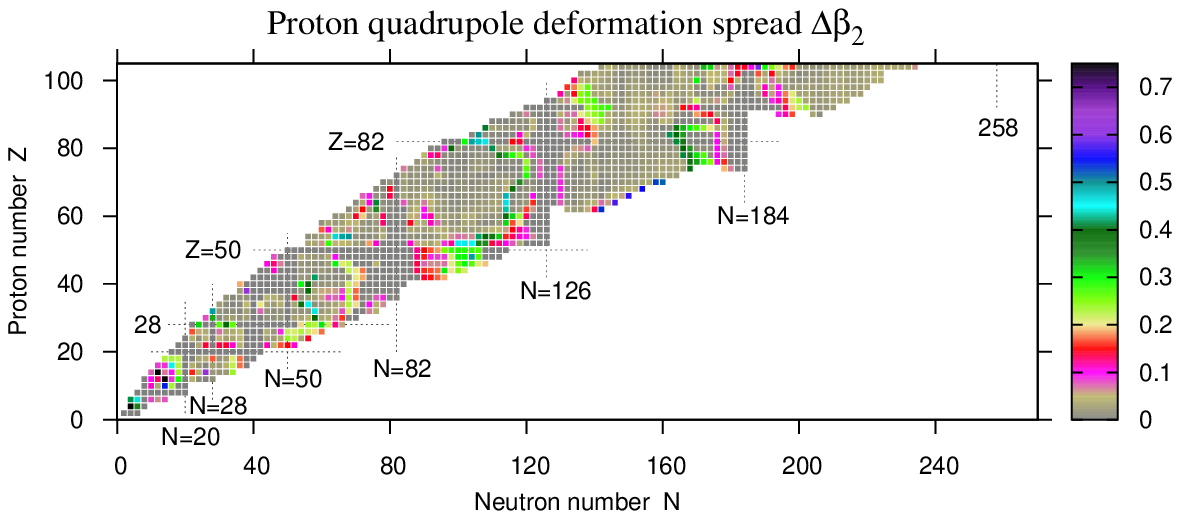}
\end{center}
\caption{(Color online) Proton quadrupole deformation spreads
$\Delta \beta_2(Z,N)$ as a function of proton and neutron number.
$\Delta \beta_2(Z,N)=|\beta_2^{\rm max}(Z,N)-\beta_2^{\rm min}(Z,N)|$,
where $\beta_2^{\rm max}(Z,N)$ and $\beta_2^{\rm min}(Z,N)$ are the
largest and smallest proton quadrupole deformations obtained
with four employed CEDF for  the $(Z,N)$
nucleus.}
\label{Edif-charge-deform}
\end{figure*}

A systematic comparison between calculated and experimental static
charge quadrupole deformations $\beta_2$ has already been performed
in each of these regions (with NL3* \cite{AO.13} in the actinides
and with DD-ME2 and DD-PC1 \cite{DD-PC1} in the rare-earth region).
They describe the experimental data well, typically within
the experimental uncertainties. Fig.\ \ref{Edif-charge-deform} shows that
in these regions of well deformed nuclei the spread of the theoretical
predictions, i.e. the difference between results obtained with various
CEDF's, is rather small for static quadrupole deformations $\beta_2$.
Thus, we do not repeat such a comparison here.

The distribution of calculated static quadrupole deformations
$\beta_2$ is similar in all four CEDF's under consideration
(see Fig.~\ref{Charge-deform}). The biggest difference between these results
is related to the presence of two regions of oblate deformation at
$(Z\sim 70, N\sim 160)$ and $(Z\sim 95, N\sim 230$) in the
calculations with NL3*. These regions are absent in the other
CEDF's.  However, this is a consequence of the fact that the two-neutron
drip line is located at higher $N$ values in NL3* as compared with other
CEDF's. As a result, these regions are neutron-unbound for DD-ME2,
DD-ME$\delta$, and DD-PC1.

The width of the gray region in Fig.\ \ref{Charge-deform} (the
gray color corresponds to spherical and near-spherical shapes)
along a specific magic number corresponding to a shell closure
indicates the impact of this shell closure on the structure of
the neighboring nuclei. Note that proton and neutron shell gaps act simultaneously
in the vicinity of doubly magic spherical nuclei. Thus, the effect
of a single gap is more quantifiable away from these nuclei.
One can see in Fig.\ \ref{Charge-deform} that the neutron $N=82$,
126 and 184 shell gaps have a more pronounced effect on the nuclear
deformations as compared with the proton shell gaps at $Z=50$ and $Z=82$.
This feature is common for all the CEDF's under investigation in this
manuscript.

It is interesting to compare the RHB results with those obtained
in non-relativistic models. The comparison
of Fig.\ \ref{Charge-deform} in the present manuscript
with HFB results based on the Gogny D1S force in Fig.\ 3a of Ref.\ \cite{DGLGHPPB.10},
with HFB results based on six Skyrme EDF's in Fig.\ 2 of the Supplement to Ref.\ \cite{Eet.12},
and with the microscopic+macroscopic model in Fig.\ 9 of Ref.\ \cite{MNMS.95}
show that the general structure of the distribution of charge quadrupole
deformations $\beta_2$ in the nuclear chart is similar in all model calculations.
Differences between models emerge mostly at the boundaries between the
regions of different types of deformation, i.e. in the transitional regions,
where the energy surfaces are rather flat and static deformations are
not well defined. There are boundaries between the regions of prolate and
oblate shapes and between the regions of deformed and spherical shapes. This
comparison also reveals that, similar to our relativistic results,
also in non-relativistic calculations the neutron shell gaps with $N=82$, 126 and 184
have a more pronounced effect on the nuclear deformations than the proton shell gaps
with $Z=50$ and $Z=82$.

Fig.\ \ref{Edif-charge-deform} shows the spreads
$\Delta\beta_2(Z,N)$ among four CEDF's for the predicted
charge quadrupole deformations. One can see that this spread is
either non-existent or very small for spherical or nearly spherical
nuclei as well as for well-deformed nuclei in the rare-earth
and actinide region. The largest uncertainties for predicting the
equilibrium quadrupole deformations exist at the boundaries
between regions of different deformations. They are extremely
high in the regions of the prolate-oblate shape coexistence, indicating
that the ground state in a given nucleus can be prolate (oblate) in one
CEDF and oblate (prolate) in another CEDF. These uncertainties are more
modest on the boundaries of the regions of spherical and deformed (oblate
or prolate) shapes. It is well known that such nuclei are difficult
to describe precisely at the mean field level \cite{Niksic2011_PPNP66-519,HW.11,Mo-Ru.13}.
Correlations going beyond mean field have to be taken into
account~\cite{BBH.06,LNVMLR.09,HNMM.09,YBH.13} and shape fluctuations
do not allow a precise definition of deformation parameters.
However, even if such correlations and fluctuations are taken into
account properly by methods based on density functional theory
and going beyond the mean field, there remain deficiencies of the
current generations of the DFT models with respect of the description of
single-particle energies \cite{BBH.06}. Indeed, when we compare
the profile of the potential energy surface (PES) as a function of the
deformation in spherical or well-deformed nuclei with that in transitional nuclei,
we find that this profile depends for transitional nuclei much more
sensitively on the underlying single-particle structure than in the
other two cases. However, it is well known that the single-particle
energies (both spherical and deformed) are not very accurately
described at the DFT level (see Refs.\ \cite{LA.11,AS.11}
and references quoted therein). Considering that the PES's obtained at the
mean field level form the starting points of many beyond mean field
calculations, further improvement in the description of the
single-particle energies is needed in order to describe
experimental data in transitional and shape-coexistent nuclei reliably
and consistently across the nuclear chart with a high level of predictive
power by the methods going beyond mean field.

\begin{figure*}[ht]
\begin{center}
\includegraphics[width=16.0cm,angle=0]{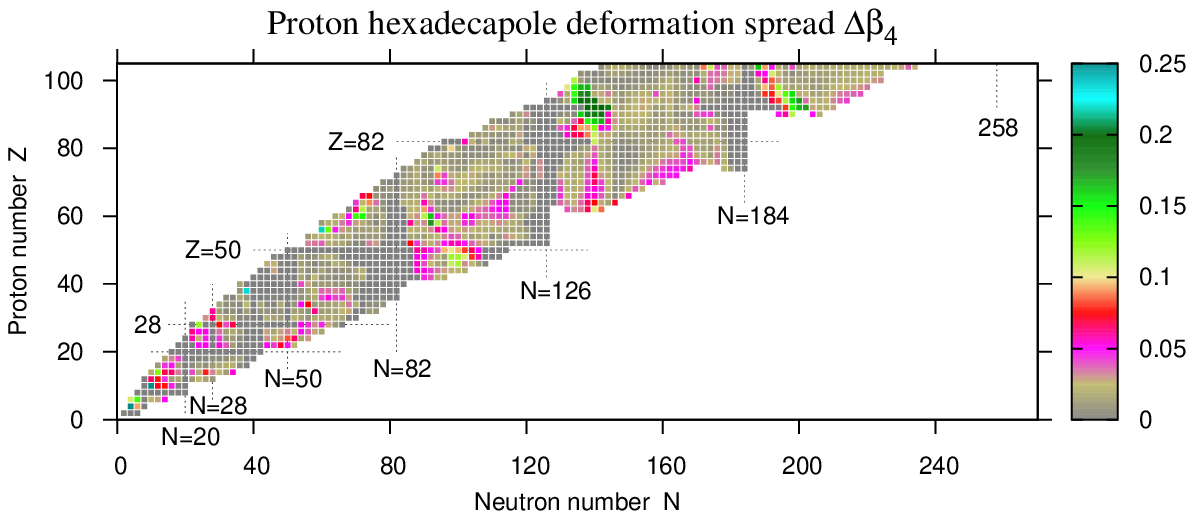}
\end{center}
\caption{(Color online) Proton hexadecapole deformation spreads
$\Delta \beta_4(Z,N)$ as a function of the proton and neutron numbers.
$\Delta \beta_4(Z,N)=|\beta_4^{\rm max}(Z,N)-\beta_4^{\rm min}(Z,N)|$,
where $\beta_4^{\rm max}(Z,N)$ and $\beta_4^{\rm min}(Z,N)$ are the
largest and smallest proton hexadecapol deformations obtained
with four employed CDFT parametrizations for the $(Z,N)$
nucleus.}
\label{Edif-percent-hex}
\end{figure*}

In Figs.\ \ref{Charge-deform-hex} and \ref{Edif-percent-hex} we present
the distribution of the calculated charge hexadecapole deformations
$\beta_4$ in the $(N,Z)$ plane and the spreads (\ref{eq:TSUC}) for 
this observable. The detailed comparison of Figs.\ \ref{Edif-percent-hex} 
and \ref{Charge-deform-hex} reveals a large degree of correlation 
between the uncertainties in the predictions of proton quadrupole and 
hexadecapole deformations.
Similar to quadrupole deformation (see discussion above), the largest
the spread of the calculated hexadecapole deformations exist near
the borderline separating the regions with different quadrupole deformations.
For non-relativistic theories, the distribution of hexadecapole deformations
of ground states in the nuclear chart has been published so far only in
the microscopic+macroscopic (MM) model (see Fig.\ 11 in Ref.\ \cite{MNMS.95}).
Although the general trends for hexadecapole deformations seems to be
similar with our results, the direct comparison between the two models is
very difficult.  In the MM model\ \cite{MNMS.95}, the deformation
parameters determine the shape of the potentials and the multipole moments
of the corresponding density distributions are complicated non-linear
functions of deformations: $Q_{L0} = Q_{L0}(\beta_2,\beta_4)$ for ($L=2,4$).
On the contrary, in the present investigation the deformation parameters
are defined from the $Q_{L0}$ moments via the linear expressions
(\ref{beta2_def}) and (\ref{beta4_def}) where all the  non-linear
coupling effects are neglected (see, for instance,
Ref. \cite{Libert1982_PRC25-571}).

\begin{figure*}[ht]
\includegraphics[width=14.0cm,angle=0]{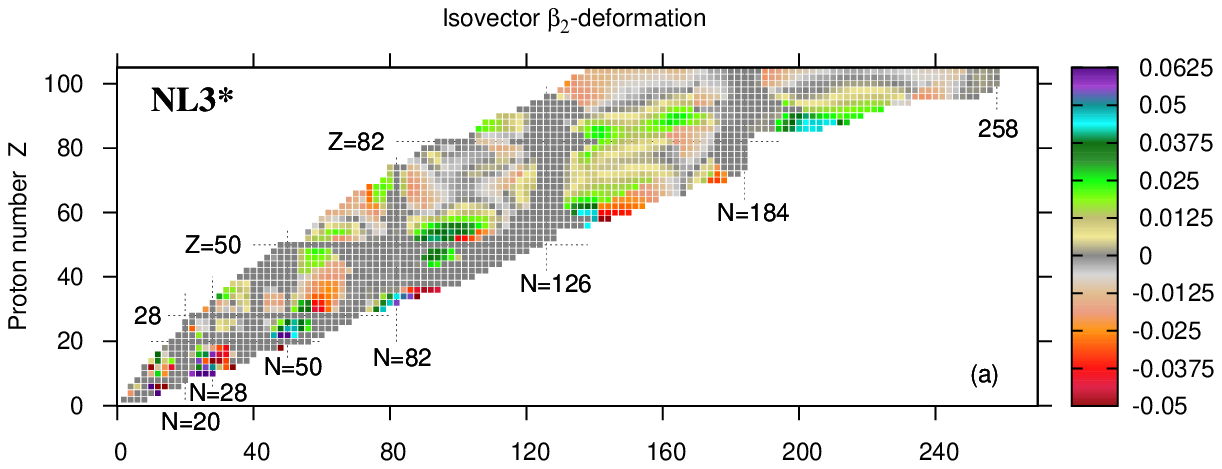}
\includegraphics[width=14.0cm,angle=0]{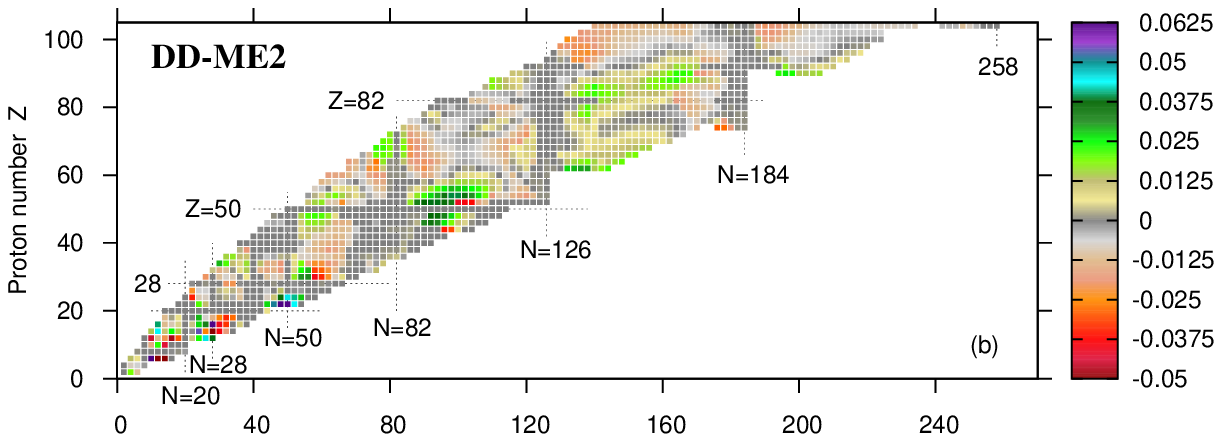}
\includegraphics[width=14.0cm,angle=0]{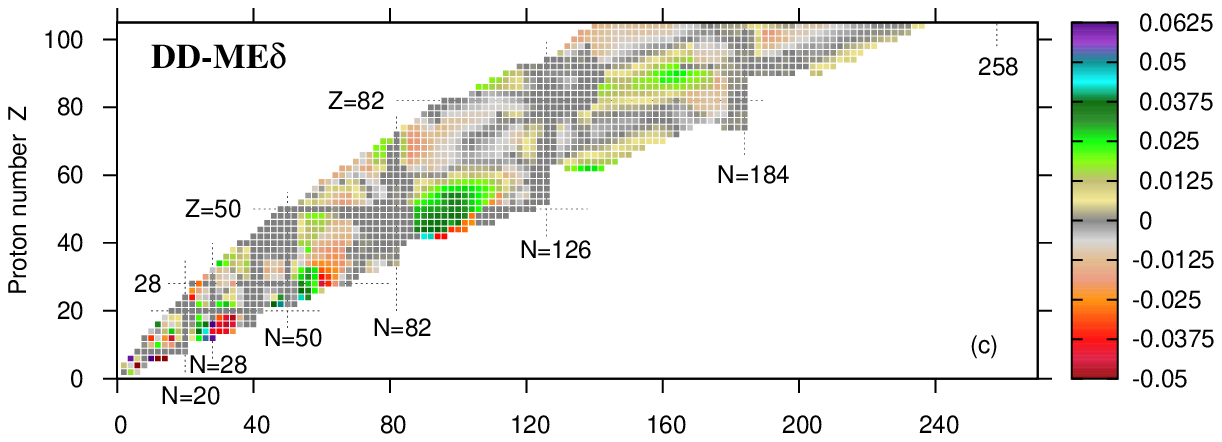}
\includegraphics[width=14.0cm,angle=0]{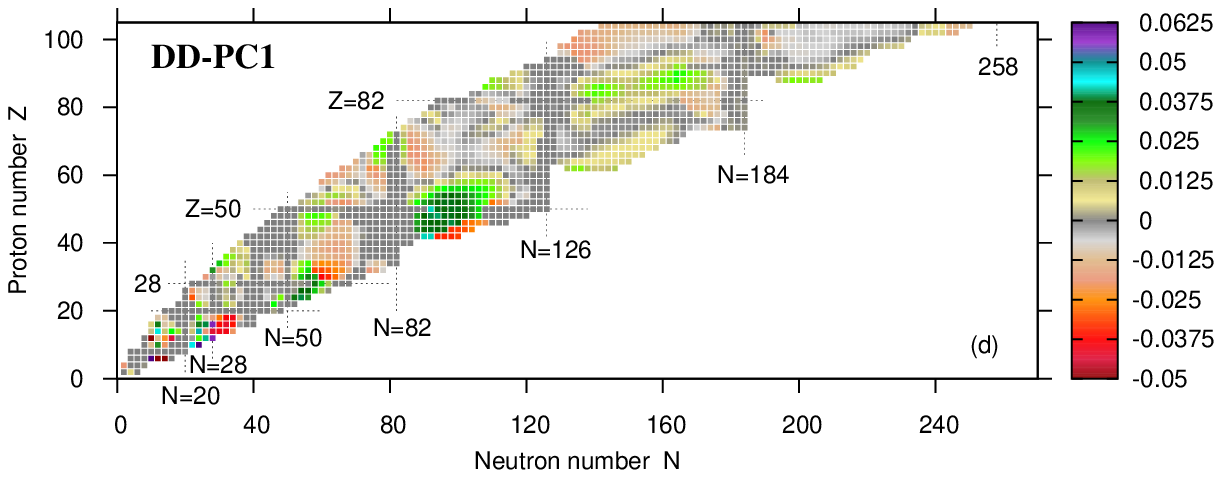}
\caption{(Color online) Isovector $\beta_2 ^{IV}=\beta_2(\nu)-\beta_2(\pi)$
deformations obtained in the RHB calculations with the indicated
CDFT parametrizations.}
\label{Isovec-charge-deform}
\end{figure*}

\begin{figure*}[ht]
\begin{center}
\includegraphics[width=18.0cm,angle=0]{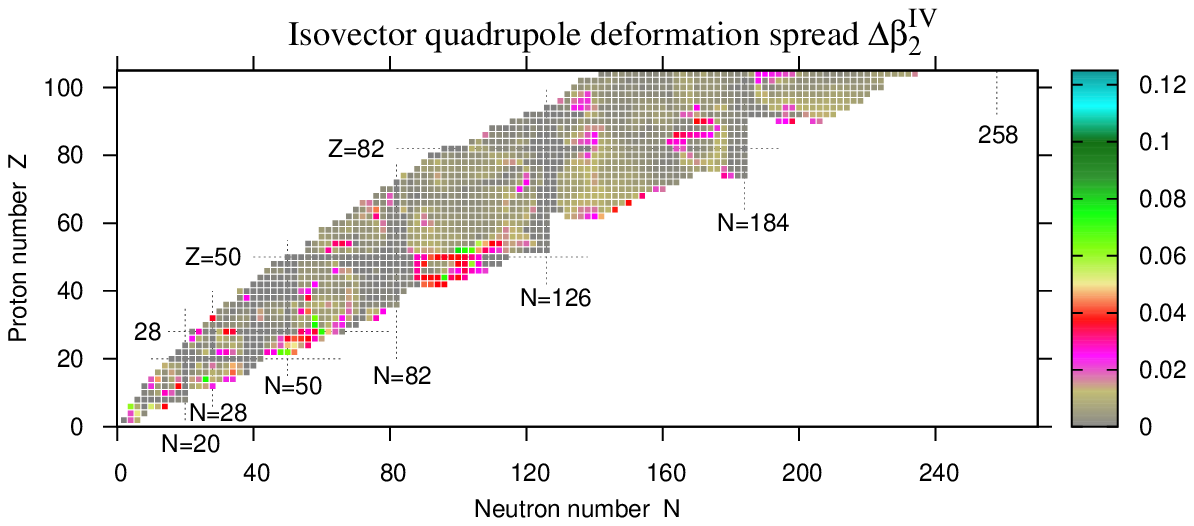}
\end{center}
\caption{(Color online) Isovector quadrupole deformation spreads
$\Delta \beta_2^{IV}(Z,N)$ as a function of proton and neutron number.
$\Delta \beta_2^{IV}(Z,N)=|\beta_{\rm 2, max}^{IV}(Z,N)-
\beta_{\rm 2,min}^{IV}(Z,N)|$,  where $\beta_{\rm 2, max}^{IV}(Z,N)$
and $\beta_{\rm 2,min}^{IV}(Z,N)$ are the largest and smallest
isovector quadrupole deformations obtained with four CDFT
parametrizations for the $(Z,N)$ nucleus.}
\label{Edif-isovec}
\end{figure*}

In Fig.\ \ref{Isovec-charge-deform} we present isovector deformations
$\beta_2^{IV}=\beta_2(\nu)-\beta_2(\pi)$. So far, there are no
experimental data on such a quantity. However, it is important to understand
how consistent are the predictions for $\beta_2^{IV}$ among modern DFT models.
The results presented in the Supplement of Ref.\ \cite{Eet.12} show that for
the absolute majority of nuclei Skyrme functionals favor smaller neutron $\beta_2$-deformations
as compared with the proton values. This result has been verified for 6 Skyrme
EDF's; although some differences between Skyrme parameterizations exist it
appears as a general rule. The situation is different in covariant functionals.
The neutron $\beta_2$-deformation is larger than the corresponding proton deformation
in approximately 2/3 of the nuclei, while in 1/3 of the nuclei the opposite
situation is seen. The absolute difference between proton and neutron $\beta_2$-deformations
is less than 0.0125 in approximately 70\% of the deformed nuclei. As
illustrated by Fig.\ \ref{Edif-isovec}, these results do not depend much on the
selection of the CEDF. On the contrary, the difference exceeds 0.02 for at least
half of the deformed nuclei in the Skyrme DFT (see Fig.\ 3 in Supplement of Ref.\ \cite{Eet.12}).
Thus, the MM model, which assumes the same deformations for protons
and neutrons, is better justified in CDFT than in Skyrme DFT. One also should note
that in CDFT there are several regions in the periodic chart, where the differences
of neutron and proton quadrupole deformations become substantial. These are $(Z\sim 16,
N\sim 34)$, $(Z\sim 28, N\sim 60)$ and $Z\sim 50, N\sim 100)$ regions located
in the vicinity of two-neutron drip line (Fig.\ \ref{Isovec-charge-deform}).
They are present in all CDFT parameterizations. At the moment we do not understand 
all these details, but we have to emphasize that most of the regions with a large 
differences between neutron and proton deformations are close to the neutron
drip line, where the neutron densities are more dilute than those of the protons.
In addition, the neutron densities are more deformed than the proton ones in 
these regions. Of course it would be interesting to investigate in future these 
facts in more detail.

\%
\section{Charge radii and neutron skin thickness.}
\label{Sec-radii}

 The charge radii were calculated from the corresponding point proton radii as
\begin{equation}
r_{ch} = \sqrt{<r^2>_p + 0.64}\,\,\,\, {\rm fm}
\end{equation}
where the factor 0.64 accounts for the finite-size effects of the
proton. Here we have neglected the small contributions to the charge
radius originating from the electric neutron form factor
and the electromagnetic spin-orbit coupling \cite{BFHN.72,NS.87}
as well as the corrections due to the center of mass motion.
Note that in the fits of the three density functionals NL3*~\cite{NL3*}, 
DD-ME2~\cite{DD-ME2}, and DD-ME$\delta$~\cite{DD-MEdelta} the same 
finite size of the proton of $0.8$ fm has been used and that the 
functional DD-PC1~\cite{DD-PC1} has been adjusted only to nuclear
binding energies.

\begin{figure*}[ht]
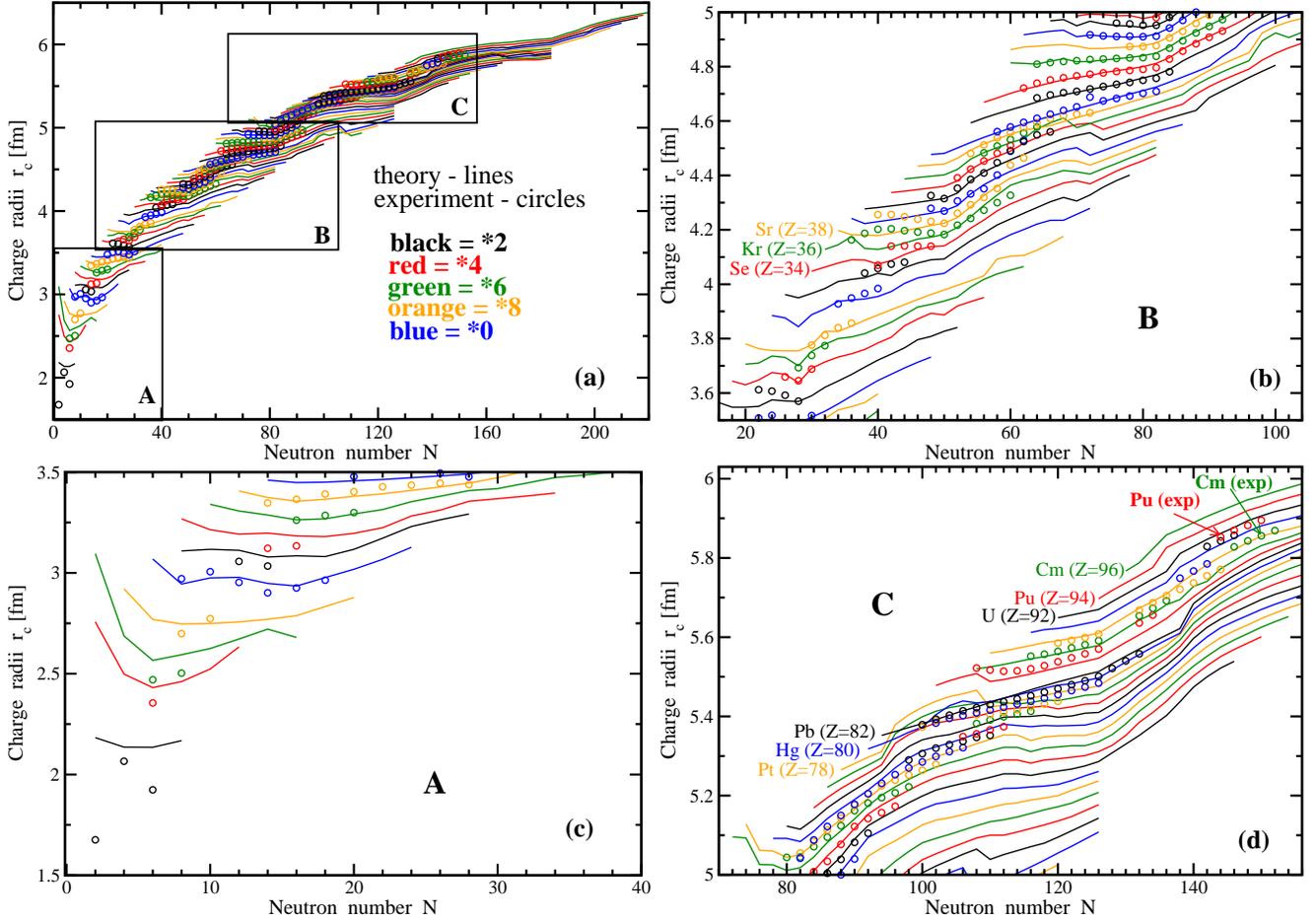

\includegraphics[width=8.7cm,angle=0]{fig-23a.eps}
\includegraphics[width=8.7cm,angle=0]{fig-23b.eps}
\includegraphics[width=8.7cm,angle=0]{fig-23c.eps}
\includegraphics[width=8.7cm,angle=0]{fig-23d.eps}
\caption{(Color online) Experimental and theoretical charge radii
as a function of neutron number. The calculations are performed with
DD-PC1. Black, red, green, orange and blue colors are used for
isotope chains with proton numbers ending with 2, 4, 6, 8 and 0,
respectively. The experimental data are taken from Ref.\ \cite{AM.13}.
Panels (b), (c) and (d) show the comparison in an enlarged scale.}
\label{Charge_radii}
\end{figure*}

\begin{figure*}[ht]
\begin{center}
\includegraphics[width=16.0cm,angle=0]{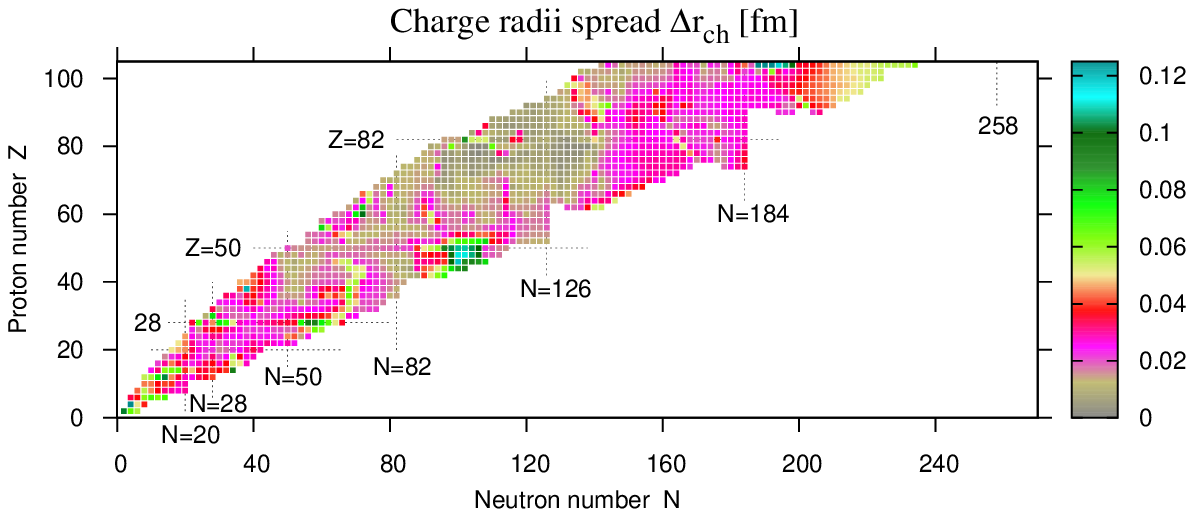}
\end{center}
\caption{(Color online)  Charge radii spread $\Delta r_{\rm ch}(Z,N)$
as a function of proton and neutron number.
$\Delta r_{\rm ch}(Z,N) = |r_{\rm ch}^{\rm max}(Z,N) - r_{\rm ch}^{\rm min}(Z,N)|$,
where $r_{\rm ch}^{\rm max}(Z,N)$ and $r_{\rm ch}^{\rm min}(Z,N)$ are the
largest and the smallest charge radii obtained with the four CDFT
parameterizations for the $(Z,N)$ nucleus.
}
\label{Edif-radii}
\end{figure*}

The accuracy of the description of charge radii is illustrated
on the example of the CEDF DD-PC1 in Fig.\ \ref{Charge_radii}. We do
not present such a comparison for the CEDF's NL3*, DD-ME2 and DD-ME$\delta$
because they show very similar results. This similarity
is clearly seen from Fig.\ \ref{Edif-radii}, which presents
the spreads (\ref{eq:TSUC}) in the theoretical results on charge radii, and
from Table \ref{Radii-deviat}, which presents the rms-deviations between
calculated and experimental radii. These comparisons are based on the latest
compilation of experimental charge radii in Ref.\ \cite{AM.13}, which includes
charge radii for 351 even-even nuclei,

One can see that the calculations provide in general a good description
of experimental data. However, there are four exceptions.
First, there are very light nuclei He, Be and C  (Fig.\
\ref{Charge_radii}a), where the mean field description has
obviously limitations. The discrepancy between theory and
experiment is especially pronounced in the case of the He nuclei.
Then, there is a substantial discrepancy
between theory and experiment for charge radii of Se, Kr
and Sr isotopes at neutron numbers $N=38-46$ (see Fig.\ \ref{Charge_radii}b).
The calculated ground state quadrupole deformations of these nuclei are
predicted to be either spherical or near-spherical (see Fig.\
\ref{Charge-deform}). However, the potential energy surfaces are soft.
This indicates that a proper description of their structure requires
the inclusion of beyond mean field correlations.
Next, the ground states of some
 proton-rich Hg and Pb isotopes are
predicted to be oblate (or prolate) in contradiction
with experiment. These earlier observed features \cite{NVRL.02} are in part
due to incorrect position of the proton $1h_{9/2}$ spherical subshell
\cite{NVRL.02,A250} and they are present in all the CEDF's used here
(see Fig.\ \ref{Charge-deform}).
When comparing theory with experiment we use for these nuclei the radii
from the minimum of the potential energy surface
corresponding to the experimental minimum, i.e. the spherical minimum for
the $N=104-114$ Pb isotopes and the oblate minimum for the $N=100-108$ Hg isotopes.
Finally, the last case is related to the unusual behavior of the charge
radii in the U-Pu-Cm isotopes (see Fig.\ \ref{Charge_radii}d).
For a fixed neutron number, the increase of proton number leads in these isotopes to an increase of the calculated charge radius.
Such a feature is seen not only for the CDFT results, but also for the 
results of the non-relativistic DFT calculations based on the
Gogny D1S force (see supplement to Ref.\ \cite{DGLGHPPB.10}).
However, in experiment the charge radii of the Cm $(Z=96)$ nuclei 
are lower than those of Pu $(Z=94)$ and U $(Z=92)$.
This is the only case in the nuclear chart where such an inversion exists. 
Considering that both the ground state quadrupole deformations are very 
stable in this region, i.e. their variations with particle number are 
much less pronounced than in the rare-earth region, and that covariant 
density functional theory describes the experimental deformations in the 
actinides well \cite{A250,AO.13} it is impossible based on the current 
CDFT's and on the Gogny functional D1S to understand this highly 
unusual behavior of experimental charge radii in the Cm isotopes.

\begin{table}
\caption{The rms-deviations $\Delta r_{ch}^{\rm rms}$ between calculated
and experimental charge radii. They are given in fm for the indicated CEDF's.
For the calculations of the rms-values, all experimental data are used
in column 2, while the data on radii of He ($Z=2$) and Cm ($Z=96$) isotopes
are excluded in column 3. See text for the discussion of these cases.}
\begin{tabular}{|c|c|c|} \hline
CEDF   & $\Delta r_{ch}^{\rm rms}$ [fm] & $\Delta r_{ch}^{\rm rms}$ [fm] \\ \hline
  1    &             2              &          3                 \\ \hline
   NL3*            & 0.0407  & 0.0283  \\
   DD-ME2          & 0.0376  & 0.0230  \\
   DD-ME$\delta$   & 0.0412  & 0.0329  \\
   DD-PC1          & 0.0402  & 0.0253  \\ \hline
\end{tabular}
\label{Radii-deviat}
\end{table}

In neutron-rich nuclei the excess of neutrons over protons
creates a neutron skin. The neutron skin thickness is commonly
defined as the difference of proton and neutron root-mean-square
(rms) radii
\begin{equation}
r_{\rm skin}=<r^2_n>^{1/2}-<r_p^2>^{1/2}.
\end{equation}
The neutron skin thickness is an important indicator of isovector
properties. It is closely related with a number of observables
in finite nuclei which are sensitive to isovector
properties \cite{B.00,RN.10,RCVW.11} and it affects the physics
of neutron stars \cite{HP.01,SPLE.05,RN.10,FP.12}.

\begin{table}
\caption{Neutron skin thicknesses $r_{\rm skin}$ in $^{48}$Ca and
$^{208}$Pb obtained in calculations with the indicated CEDF's.
The results of the calculations with FSUGold are taken from
Ref.\ \cite{FP.13}.}
\begin{tabular}{|c|c|c|} \hline
CEDF           & $r_{skin}(^{48}$Ca) [fm] & $r_{skin}(^{208}$Pb) [fm] \\ \hline
NL3*          & 0.236                    &  0.288 \\
DD-ME2        & 0.187                    &  0.193 \\
DD-ME$\delta$ & 0.177                    &  0.186 \\
DD-PC1        & 0.198                    &  0.201 \\ 
FSUGold       &                          &  0.21  \\ \hline
\end{tabular}
\label{Table-neu-skin}
\end{table}

\begin{table}
\caption{Neutron skin thicknesses $r_{\rm skin}$ [in fm] in selected
neutron-rich nuclei obtained in calculations with relativistic functionals
(CEDF) and Skyrme functionals (SEDF). The latter results are extracted from
Fig.\ 3 of Ref.\ \cite{KENBGO.13}.}
\begin{tabular}{|c|c|c|c|} \hline
EDF                 & Ca (N=42) & Zr (N=84) & Er (N=68) \\ \hline
CEDF(NL3*)          & 0.688     &  0.666    & 0.752     \\
CEDF(DD-ME2)        & 0.598     &  0.522    & 0.582     \\
CEDF(DD-ME$\delta$) & 0.542     &  0.495    & 0.529     \\
CEDF(DD-PC1)        & 0.539     &  0.509    & 0.532     \\
SEDF(SV-min)        & 0.55      &  0.470    & 0.490     \\
SEDF(UNEDF0)        & 0.55      &  0.510    & 0.560     \\ \hline
\end{tabular}
\label{Table-neu-skin-comp}
\end{table}

The experimental data on the neutron skin thickness in
$^{208}$Pb is contradictory. On the one hand, there is a large set of
experiments which suggests that the neutron skin is around 0.2 fm
or slightly smaller (see Table 1 in Ref.\  \cite{KPVH.13}). However, these
experimental data are extracted in model dependent ways (see
Ref.\ \cite{TSCD.12} and references quoted therein).
The neutron skin thicknesses $r_{\rm skin}=0.161\pm 0.042$ \cite{KPVH.13}
and $r_{\rm skin}=0.190\pm 0.028$ \cite{Anti-an.2013} obtained recently
from the energy of the anti-analogue giant dipole resonance rely on
relativistic proton-neutron quasiparticle random-phase
approximation calculations based on the RHB model. Another recent
value of the neutron skin thickness of
$r_{\rm skin}=0.15\pm 0.03({\rm stat})^{+0.00}_{-0.03}({\rm sys})$ fm has been
extracted from coherent pion photo-production cross sections \cite{Pion-rskin}.
However, the extraction of information on the nucleon density distribution
depends on the comparison of the measured $(\gamma,\pi^0)$ cross
sections with model calculations. On the other hand, a measurement
using an electro-weak probe has very recently been carried out in parity
violating electron scattering on nuclei (PREX)~\cite{PREX.12}. It utilizes the preferential
coupling of the exchanged weak boson to neutrons. The electro-weak
probe has the advantage over experiments using hadronic probes
that it allows a nearly model-independent extraction of the neutron
radius that is independent of most strong interaction uncertainties
\cite{H.98}. However, a first measurement at a single momentum transfer
gave $r_{\rm skin}=0.33\pm 0.17$ with a relatively large error bar \cite{PREX.12}.
A central value of 0.33 fm is particularly intriguing since it is around 0.13 fm
higher than central values obtained in other experiments (see Table 1 in Ref.\  
\cite{KPVH.13}). The analysis performed in Ref.\ \cite{FP.13} has found no 
compelling reason to rule out the models with large neutron skin in $^{208}$Pb. 
However, as indicated in Ref.\ \cite{FP.13}, the parameters of these models 
do not follow from a strict optimization procedure. All systematic fits with 
density dependent couplings in the isovector channel for 
DD-ME1~\cite{Niksic2002_PRC66-024306}, DD-ME2~\cite{DD-ME2}, 
DD-ME$\delta$~\cite{DD-MEdelta}, DD-PC1~\cite{DD-PC1}, or FSUGold~\cite{FSUGold}
find for the neutron skin thickness in $^{208}$Pb values close to 0.2 fm 
(see Table \ref{Table-neu-skin}). Only in the first two cases the small neutron 
skins have been used in the fit. For the CEDF's DD-ME$\delta$ and DD-PC1 the 
density dependence in the isovector channel has been determined from ab-initio 
calculations of nuclear matter.

It is clear that the already approved follow-up PREX measurement \cite{PREX-CREX} 
designed to achieve the original 1\% error in the neutron radius of $^{208}$Pb 
will provide useful constraints on the selection of the proper CEDF. 
Table \ref{Table-neu-skin} also provides the predictions for neutron skin thickness in
$^{48}$Ca. It will be measured in the approved CREX experiment
at JLab with an accuracy of around 0.02 fm \cite{PREX-CREX}.
Again the neutron skin thickness is the largest for the NL3* CEDF
and the density dependent (DD) CEDF's provide similar, but smaller predictions
for it. However, the difference between the NL3* and the DD CEDF's is
less pronounced in $^{48}$Ca as compared with $^{208}$Pb. Apart from NL3*,
the neutron skin thicknesses are only slightly (by $\sim 0.05$ fm) smaller in
$^{48}$Ca as compared with $^{208}$Pb.

\begin{figure*}[ht]
\includegraphics[width=8.7cm,angle=0]{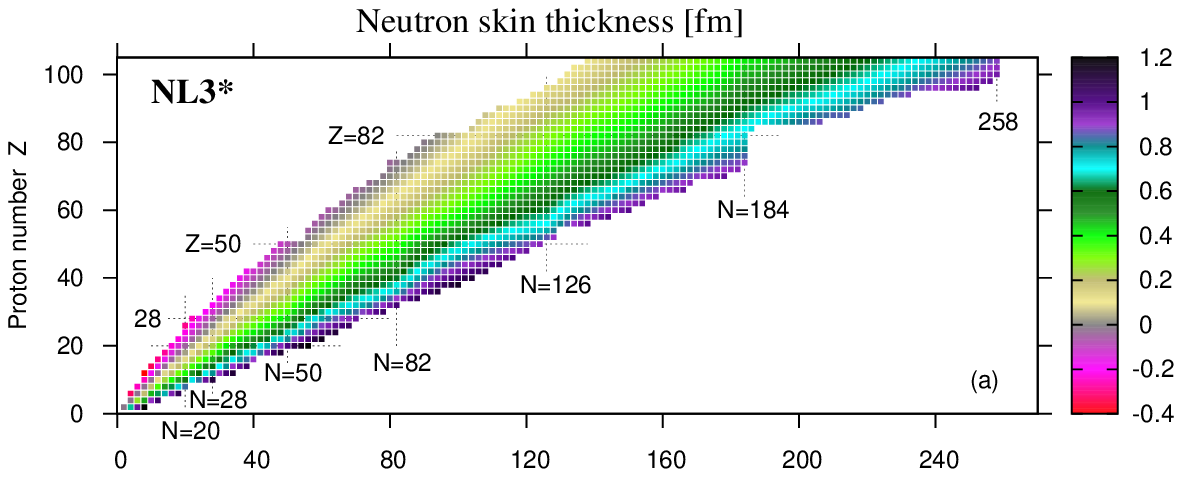}
\includegraphics[width=8.7cm,angle=0]{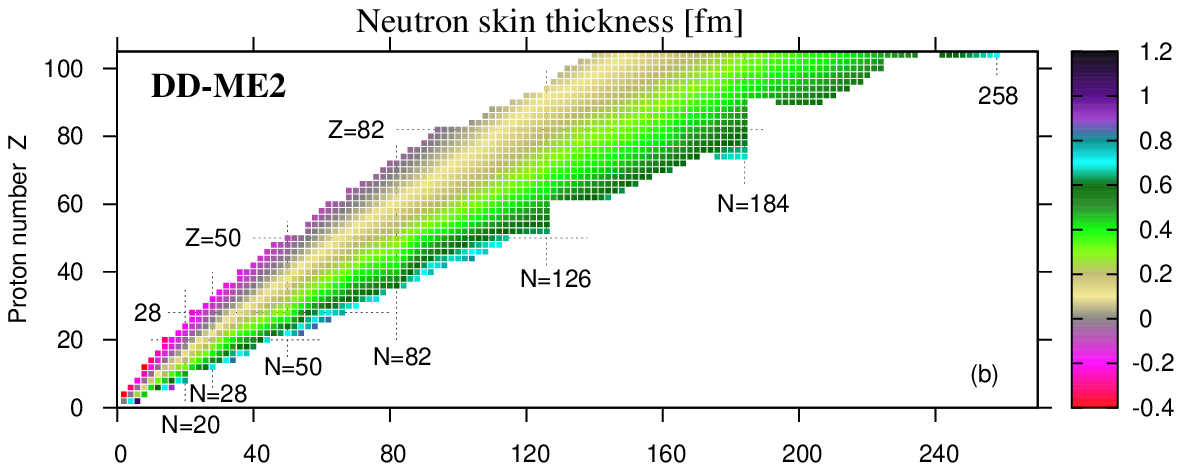}
\includegraphics[width=8.7cm,angle=0]{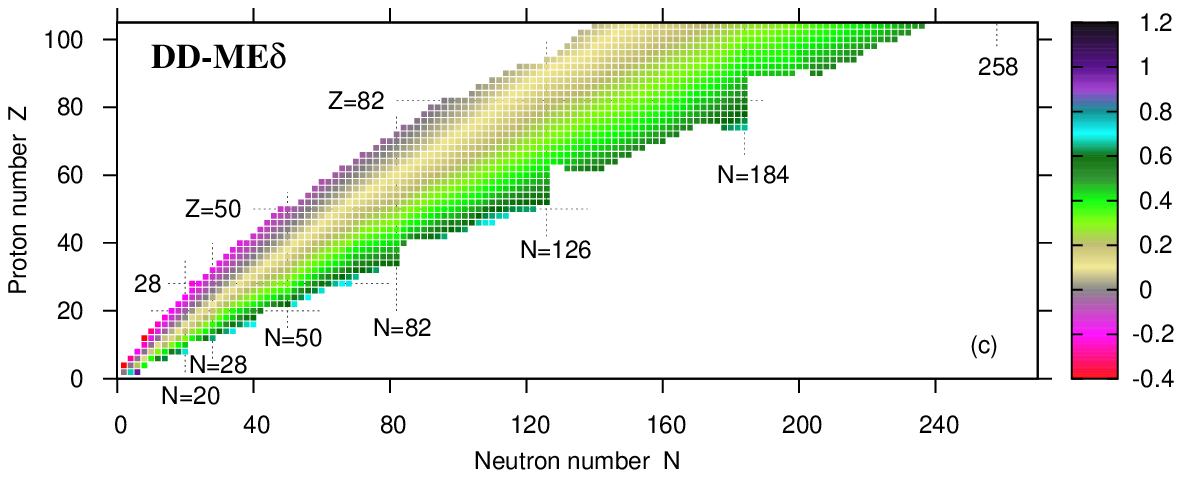}
\includegraphics[width=8.7cm,angle=0]{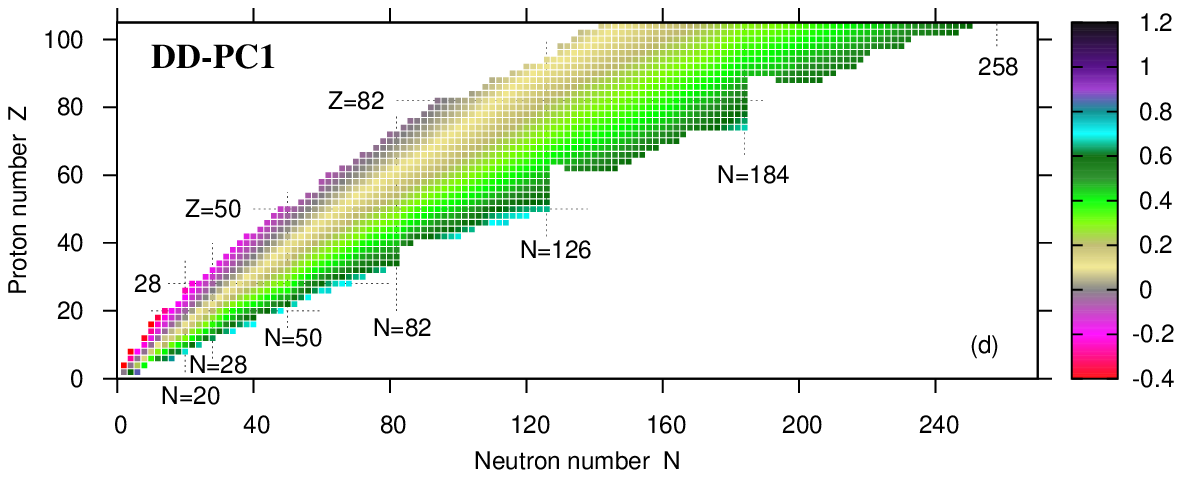}
\caption{(Color online) Neutron skin thicknesses obtained
in RHB calculations with several CEDF's.}
\label{Neutron-skin-NZ}
\end{figure*}

On going to the neutron-drip line we observe the same trends
which are already seen in $^{48}$Ca and $^{208}$Pb (see Table
\ref{Table-neu-skin-comp}). First, the neutron skin thicknesses
obtained with DD CEDF's cluster around the same value. Second, the
neutron skin thickness obtained with NL3* exceeds
substantially those found with DD CEDF's. It is interesting
that the neutron skin thicknesses obtained with DD CEDF's are
very close to those found in Skyrme DFT's calculations with
SV-min and UNEDF0 in Ref.\ \cite{KENBGO.13}.

In Fig.\ \ref{Neutron-skin-NZ} we present calculated distributions of
neutron skin thicknesses in the $(Z,N)$ chart.
One can see that they are very similar for the DD CEDF's. On
the other side, the neutron skin thickness is larger for NL3*.
In some nuclei it can reach 1.2 fm. This is a consequence
of two factors. First, the neutron skin is larger for NL3*
than for the DD CEDF's already in the valley of beta-stability
and the neutron skin thickness increases with isospin. Second,
the two-neutron drip line extends to more neutron-rich nuclei in
NL3* as compared with DD CEDF's (see Sec.\ \ref{Two-neu-drip-sec}) 
leading to these high values of $r_{\rm skin}$. The comparison of the 
results for DD CEDF's shows significant similarities with the 
results obtained for 6 Skyrme functionals in Ref.\ \cite{KENBGO.13}. 
In part, this is a consequence of the fact that similar to the DD 
CEDF's these Skyrme EDF (SEDF)'s favor smaller values for the 
neutron skin in $^{208}$Pb.

\begin{figure*}[ht]
\begin{center}
\includegraphics[width=16.0cm,angle=0]{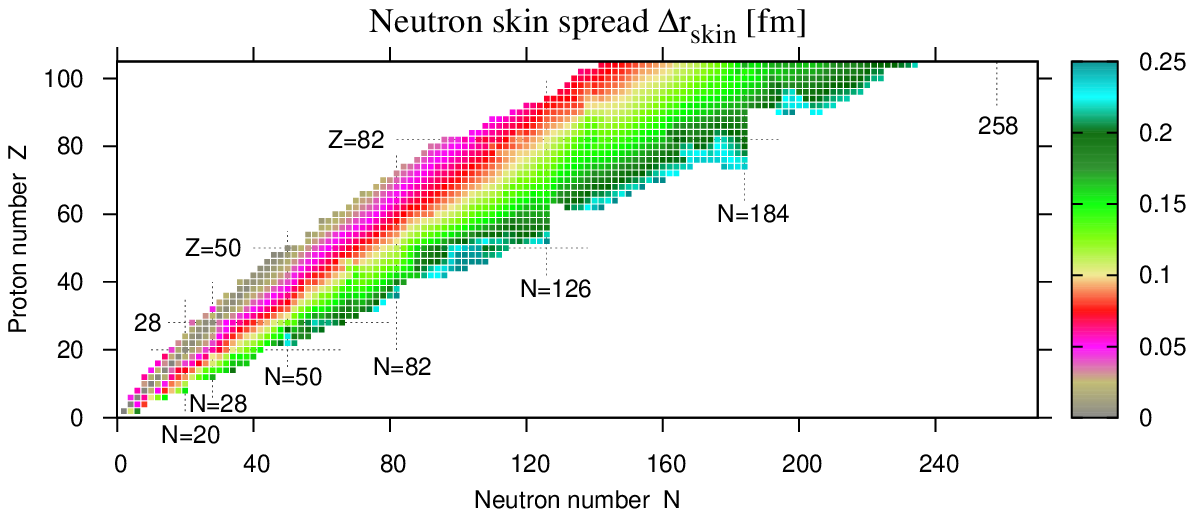}
\end{center}
\caption{(Color online) Neutron skin thickness spreads
$\Delta r_{\rm skin}(Z,N)$ as a function of proton and neutron number.
$\Delta r_{\rm skin}(Z,N)=|r_{\rm skin}^{\rm max}(Z,N)-r_{\rm skin}^{\rm min}(Z,N)|$,
where $r_{\rm skin}^{\rm max}(Z,N)$ and $r_{\rm skin}^{\rm min}(Z,N)$ are the
largest and smallest proton hexadecapol deformations obtained
with four CDFT parameterizations for the $(Z,N)$ nucleus.}
\label{Edif-neu_skin}
\end{figure*}

\begin{figure*}[ht]
\begin{center}
\includegraphics[width=16.0cm,angle=0]{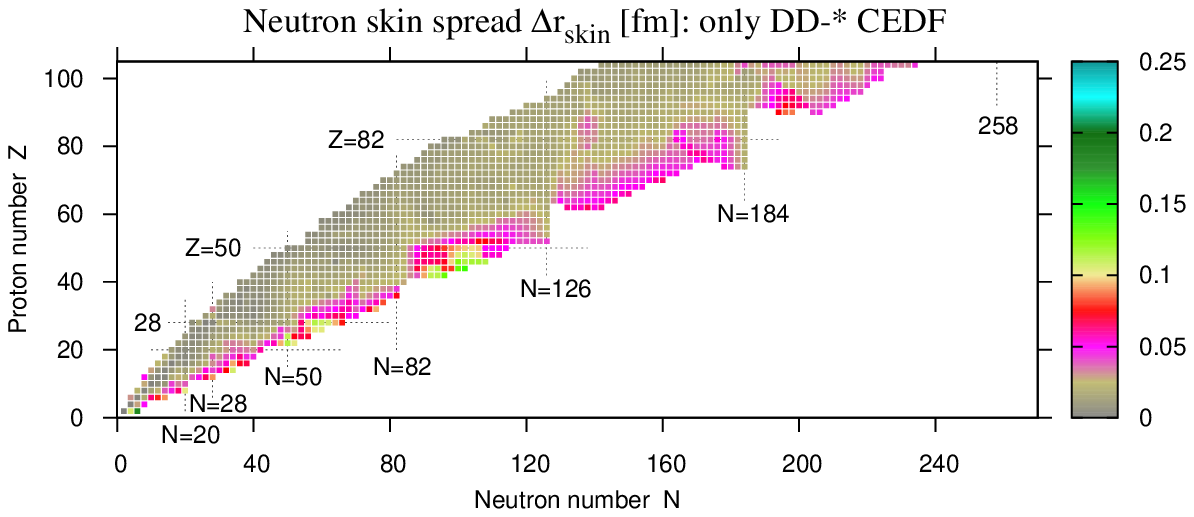}
\end{center}
\caption{(Color online) The same as Fig.\ \ref{Edif-neu_skin} but
for neutron skin thickness spreads obtained with exclusion of the
NL3*.}
\label{Edif-neu_skin_dd}
\end{figure*}

As shown in Fig.\ \ref{Edif-neu_skin} the spreads (\ref{eq:TSUC})
of theoretical predictions
in the neutron skin thickness increase with isospin 
and become rather large in neutron-rich nuclei (reaching 0.25 fm in some 
cases). They are larger than those found in Skyrme calculations in Ref.\ 
\cite{KENBGO.13}. This is a consequence of the use of NL3*, which contrary 
to DD CEDF's of the present manuscript and the Skyrme EDF's used in Ref.\ 
\cite{KENBGO.13}, favors large neutron skins. As illustrated in Fig.\ 
\ref{Edif-neu_skin_dd}, the spreads (\ref{eq:TSUC}) in the neutron skin 
thicknesses become substantially smaller if we exclude NL3* from our 
consideration. This again stresses the importance of future PREX-II and 
CREX experiments.
If PREX-II confirms the large neutron skin in $^{208}$Pb ($r_{\rm skin}\sim 0.33$ fm)
obtained in the first PREX experiment, this would also require to look for
density dependent CEDF's and Skyrme EDF's with larger neutron skins.
If this experiment will lead to a smaller neutron skin thickness $r_{\rm skin}\sim 0.2$ fm,
then the EDF's with large neutron skins (such as NL3*) should be excluded from
further consideration. In either case, this experiment will lead to a reduction 
of the uncertainty in the prediction of neutron skins in neutron-rich nuclei.

\newpage

\section{Conclusions.}
\label{Concl}

The global performance of covariant energy density
functionals has been assessed investigating the state-of-the-art
functionals NL3*, DD-ME2, DD-ME$\delta$, and DD-PC1. They
represent three classes of functionals which differ by basic
model assumptions and fitting protocols.  The available
experimental data on ground state properties of even-even
nuclei have been confronted with the results of the
calculations. For the first time, theoretical systematic 
uncertainties in the prediction of physical observables (as 
defined in Eq. (\ref{eq:TSUC})) have been investigated on a
global scale for relativistic functionals. Special
attention has been paid to the propagation of these
uncertainties towards the neutron-drip line. The main
results can be summarized as follows:

\begin{itemize}

\item
 The current generation of CEDF's investigated in
the present manuscript provides an improved
description of masses across the nuclear chart as compared with
the previous generation. This leads not only to reduced global
rms-deviations  but also to improved gross trends of the
deviations between theory and experiment as a function of
the mass number. The rms-deviations for the available experimental
masses of 835 even-even nuclei range from 2.16 MeV (DD-PC1) to 2.96 MeV
(NL3*). This is  achieved with a relatively small number of model
parameters fitted to a rather modest set of data on finite nuclei
ranging from 12 (for NL3* and DD-ME2) to 161 (for DD-ME$\delta$) nuclei.
The spread for binding energies increases
on going from the beta-stability valley towards the neutron-drip line.
This is a consequence of poorly defined isovector properties of the
current generation of CEDF's.  In the light of the model limitations
and the relatively narrow isospin range measured in experiment, it
still remains an open question whether the isovector properties of
EDF's can accurately be defined from masses alone.

\item
The analysis of discrepancies between theory and experiment
for two-neutron separation energies and their sources leads to a
more critical look on the reappearance of two-neutron binding with
increasing neutron number beyond  the primary two-neutron drip line.
This reappearance shows itself in the nuclear chart via peninsulas
emerging from the nuclear mainland and it is directly related to the
behavior of two-neutron separation $S_{2n}$ energies with neutron
number. This effect exists in a number of DFT calculations
\cite{Eet.12,AARR.13,ZPX.13} but it maybe an artifact of the mean
field approximation. These peninsulas usually appear above the $N=126$
and $N=184$ shell closures. However, the range of nuclei around these
shell closures, in which transitional shapes are expected, is wide.
Thus, the inclusion of correlations beyond mean field may lead to
the merging of these peninsulas with the nuclear mainland.

\item
The calculated two-proton drip lines are very close to experiment.
The best reproduction of the two-proton drip line is achieved
for the CEDF's DD-ME2 and DD-ME$\delta$, which are characterized by
the best residuals for the two-proton separation energies $S_{2p}$.
Since the proton-drip line lies close to the valley of stability, the
extrapolation errors towards it are small. In addition, the Coulomb
barrier provides a rather steep potential reducing considerably the
coupling to the proton continuum. This leads to a relatively low
density of the single-particle states in the vicinity of the Fermi
level, which helps to minimize the errors in the prediction of
two-proton drip line.

\item
A detailed analysis of the sources of the spread in the
predictions of the two-neutron drip lines existing in non-relativistic
and covariant DFT has been performed. Poorly known isovector
properties of the EDF's, the underlying shell structure and
inevitable inaccuracies in the DFT description of the single-particle
energies contribute to these uncertainties. However, no clear
correlations between the location of the two-neutron drip line
and the nuclear matter properties of the corresponding EDF have
been found.

\item
The spread between the different models in the definition of the
two-neutron drip line at $Z\sim 54, N=126$ and $Z\sim 82, N=184$
are very small due to the
impact of the spherical shell closures at $N=126$ and $184$. The
largest difference between covariant and Skyrme DFT exist in
superheavy nuclei, where the first model (contrary to the second)
consistently predicts a significant impact of the $N=258$ spherical
shell closure. The spread of the theoretical predictions grows on
moving away from these spherical closures. This is caused by the
increasing deformation.

\item
The experimental static $\beta_2$ deformations of well-deformed
nuclei are well described in these calculations. The difference
between the four CEDF's is small and within the experimental uncertainties.
As a result, such experimental data cannot be
used to differentiate between the functionals. Theoretical 
uncertainties for this physical observable are either non-existent 
or very small
for spherical or nearly spherical nuclei as well as for well-deformed
nuclei in the rare-earth and in the actinide regions. The largest
spreads for predicting the equilibrium quadrupole  deformations
exist at the boundaries between regions of different deformations.
They are extremely high in the regions of the prolate-oblate shape
coexistence, indicating that the ground state in a given nucleus
can be prolate (oblate) in one CEDF and oblate (prolate) in
another CEDF. These uncertainties are due to the deficiencies of
the current generations of the DFT models with respect of the
description of single-particle energies.

\item
The analysis of isovector deformations $\beta_2^{IV}$ reveals that
the neutron $\beta_2$-deformation is typically larger than the
corresponding proton deformation. However, in most of the nuclei
the absolute value of $\beta_2^{IV}$ is small. Only in  the
$(Z\sim 16, N\sim 34)$, $(Z\sim 28, N\sim 60)$ and
($Z\sim 50, N\sim 100)$ regions located in the vicinity of
two-neutron drip line the isovector deformation
is substantial in all relativistic functionals.
On the contrary, for Skyrme functionals  in the majority
of the nuclei the neutron $\beta_2$-deformations are smaller
than proton ones and the absolute values of $\beta_2^{IV}$ are larger.
Thus, the microscopic+macroscopic model, which assumes the same deformations
for protons and neutrons, is better justified in CDFT than in Skyrme DFT.

\item
A comparable level of accuracy (with a slightly better description
by DD-ME2) is achieved by all the functionals under investigation
for charge radii. Fig.\ \ref{Edif-radii} shows that the spread in predicting
charge radii are not necessarily larger near the neutron drip line as compared
with the valley of beta-stability.

\item
The experimental data on the neutron skin thickness $r_{\rm skin}$
in $^{208}$Pb is somewhat contradictory. Hadronic probes give $r_{\rm skin} \sim 0.2$
fm, whereas in the PREX experiment the electro-weak probe provides a central value
of $r_{\rm skin}=0.3$ fm, however with very large error bars. The NL3* results come
close to the central PREX value, while DD-ME2, DD-ME$\delta$ and DD-PC1
give much smaller neutron skins in the vicinity of $r_{\rm skin}=0.2$ fm.
This can be understood by the fact that the last three functionals have a
density dependence in the isovector channel, which leads to a smaller slope
$L$ of the symmetry energy at saturation and, therefore, to larger values of the
symmetry energy in the region of densities $\rho\sim 0.1$ fm below saturation
(see Refs.~\cite{B.00,Niksic2002_PRC66-024306}).
As a consequence, the neutrons are less bound to the protons in this region
of densities. Globally, the spreads in the neutron skin thickness increase
with isospin and become rather large in neutron-rich nuclei (reaching
$r_{\rm skin}=0.25$ fm in some cases) reflecting the difference between
NL3* and the DD CEDF's. There is hope that these uncertainties can 
be reduced, if future PREX-II and CREX experiments provide neutron skin
thicknesses in $^{208}$Pb and $^{48}$Ca with the required accuracy.

\end{itemize}

The current investigation shows that the biggest uncertainties in
theoretical description exist in transitional nuclei. On the
one hand, this is expected since these nuclei have usually flat
potential energy surfaces, often in the $\beta$- and $\gamma$-directions.
The minima are not well defined in these flat
energy surfaces and the fluctuations cannot be neglected.
These nuclei have to be treated by the methods going beyond mean
field~\cite{Niksic2011_PPNP66-519,FMXLYM.13,YBH.13}.
On the other hand, the mean field is the starting
point of these approaches. However, in some specific cases
we find a strong dependence of the equilibrium deformations
and the potential energy surfaces of transitional and shape-coexistent
nuclei on the employed EDF which originates from the deficiencies
of mean field methods in the description of single-particle energies.
These uncertainties will eventually affect the results of
beyond mean field calculations. The analysis indicates that
further  improvement in the description of the single-particle energies
is needed in order to describe experimental data in transitional
and shape-coexistent nuclei reliably and consistently
across the nuclear chart with a high level of predictive power
by the methods going beyond mean field.

Historically it was considered an advantage of the CDFT over
non-relativistic DFT that no single-particle information
has been used in the fit of CEDF's. However, it is clear from
the current investigation that such an approach has its own
limits since further improvement of CEDF's may require additional
terms, such as tensor terms, in the functional which cannot
be firmly constrained by only nuclear matter properties and by
the fit to masses and radii of finite nuclei~\cite{Lalazissis2009_PRC80-041301}.
The inclusion of experimental data on giant resonances in the spin-
and isospin degrees of freedom and/or on the energies of the 
single-particle states into the fitting protocol may offer such an 
extra tool and allow to define the functional with better 
single-particle properties.
However, we also have to consider that according to the concept of
density functional theory~\cite{KS.65,KS.65a} single-particle 
energies are only auxiliary quantities, which are not automatically 
reproduced well. As it is well known, very often, in particular in 
the relativistic case,
DFT theories suffer from low effective masses and the corresponding
low level densities at the Fermi surface. One has to go beyond mean
field and to take into account energy dependent self-energies~
\cite{Vretenar2002_PRC65-024321,Typel2005_PRC71-064301,Marketin2007_PRC75-024304},
as for instance particle-vibrational coupling,
to deal with this problem~\cite{Litvinova2006_PRC73-044328,LA.11}.

\section{Acknowledgements}

The authors would like to thank J.\ Erler for valuable discussions.
This work has been supported by the U.S. Department of Energy under
the grant DE-FG02-07ER41459 and by the DFG cluster of excellence
\textquotedblleft Origin and Structure of the Universe
\textquotedblright\ (www.universe-cluster.de). This work was also
supported partially through CUSTIPEN (China-U.S. Theory Institute
for Physics with Exotic Nuclei) under DOE grant number
DE-FG02-13ER42025 and by an allocation of advanced computing resources
provided by the National Science Foundation. The computations were
partially performed on Kraken at the National Institute for
Computational Sciences (http://www.nics.tennessee.edu/).

\bibliography{references}

\end{document}